\documentclass[amsmath,amssymb,prd, notitlepage,nofootinbib,twocolumn, superscriptaddress]{revtex4-2}
\usepackage{ulem}
\usepackage{orcidlink}

\usepackage{bm}

\newcommand{\aap}{{Astron.~Astrophys.}}

\newcommand{\apjl}{{Astrophys.~J.~Lett.}}
\newcommand{\apjs}{{Astrophys.~J.~Supp.}}

\usepackage{graphicx}
\usepackage{amssymb, amsmath}
\usepackage[amssymb]{SIunits}
\usepackage{aas_macros}
\usepackage{natbib}
\usepackage{xcolor}
\usepackage{booktabs}
\usepackage{multirow} 
\usepackage{verbatim}
\usepackage{subcaption}

\begin{document}

\preprint{APS/123-QED}

%\title{Constraints on ultralight axions from cosmic microwave background lensing}

%\title{Hints of ultralight axions from cosmic microwave background lensing \alex{Trying out flashier title} \Mat{Might be too flashy!}} 

\title{The Atacama Cosmology Telescope: Probing new signatures of ultralight axions with gravitational lensing}

\author{Alex Laguë \orcidlink{0000-0003-4642-6720}}
\affiliation{Department of Physics and Astronomy, University of
Pennsylvania, 209 South 33rd Street, Philadelphia, PA, 19104, United States of America}
\author{Keir K. Rogers}
\affiliation{Department of Physics, King's College London, Strand, London, WC2R 2LS, United Kingdom}
\affiliation{Department of Physics, Imperial College London, Blackett Laboratory, Prince Consort Road, London, SW7 2AZ, United Kingdom}
\affiliation{Dunlap Institute for Astronomy and Astrophysics, University of Toronto, 50 St George St, Toronto, ON, M5S 3H4, Canada}

\author{Mathew S. Madhavacheril}
\affiliation{Department of Physics and Astronomy, University of
Pennsylvania, 209 South 33rd Street, Philadelphia, PA, 19104, United States of America}

\author{J. Richard Bond~\orcidlink{0000-0003-2358-9949}}
\affiliation{Canadian Institute for Theoretical Astrophysics, University of Toronto, Toronto ON M5S 3H8 Canada}

\author{Erminia Calabrese \orcidlink{0000-0003-0837-0068}}
\affiliation{School of Physics and Astronomy, Cardiff University, The Parade, Cardiff, Wales CF24 3AA, UK}

\author{Mark J. Devlin~\orcidlink{0000-0002-3169-9761}}
\affiliation{Department of Physics and Astronomy, University of
Pennsylvania, 209 South 33rd Street, Philadelphia, PA, 19104, United States of America}

\author{Jo Dunkley}
\affiliation{Joseph Henry Laboratories of Physics, Jadwin Hall, Princeton University, Princeton, NJ, USA 08544}
\affiliation{Department of Astrophysical Sciences, Peyton Hall, Princeton University, Princeton, NJ USA 08544}

\author{Vera Gluscevic \orcidlink{0000-0002-3589-8637
}}
\affiliation{Department of Physics \& Astronomy, University of Southern California, Los Angeles, CA 90007, USA 7}
\affiliation{Institute for Advanced Study, 1 Einstein Drive, Princeton, NJ 08540, USA}
\affiliation{Center for Computational Astrophysics, Flatiron Institute, 162 5th Avenue, New York, NY, 10010, USA}

\author{Renée Hlo\v{z}ek~\orcidlink{0000-0002-0965-7864}}
\affiliation{Dunlap Institute for Astronomy and Astrophysics, University of Toronto, 50 St George St, Toronto, ON, M5S 3H4, Canada}
\affiliation{David A. Dunlap Department of Astronomy and Astrophysics, University of Toronto, 50 St George St, Toronto, ON, M5S 3H4, Canada}

\author{Hidde T. Jense \orcidlink{0000-0002-9429-0015}}
\affiliation{School of Physics and Astronomy, Cardiff University, The Parade, Cardiff, Wales CF24 3AA, UK}

\author{Thibaut Louis}
\affiliation{Université Paris-Saclay, CNRS/IN2P3, IJCLab, 91405 Orsay, France}

\author{Frank~J.~Qu \orcidlink{0000-0001-7805-1068}}
\affiliation{Kavli Institute for Particle Astrophysics and Cosmology, Stanford University, 452 Lomita Mall, Stanford, CA, 94305, USA}
\affiliation{Department of Physics, Stanford University, 382 Via Pueblo Mall, Stanford, CA, 94305, USA}
\affiliation{SLAC National Accelerator Laboratory, 2575 Sand Hill Road, Menlo Park, California 94025, USA}

\author{Bernardita Ried Guachalla \orcidlink{0000-0002-0418-6258}}
\affiliation{Department of Physics, Stanford University, Stanford, CA, USA 94305-4085}
\affiliation{Kavli Institute for Particle Astrophysics and Cosmology, 382 Via Pueblo Mall Stanford, CA 94305-4060, USA}
\affiliation{SLAC National Accelerator Laboratory 2575 Sand Hill Road Menlo Park, California 94025, USA}

\author{Neelima Sehgal \orcidlink{0000-0002-9674-4527}}
\affiliation{Physics and Astronomy Department, Stony Brook University, Stony Brook, NY 11794, USA}

\author{Blake D. Sherwin}
\affiliation{DAMTP, Centre for Mathematical Sciences, University of Cambridge, Wilberforce Road, Cambridge CB3 OWA, UK}
\affiliation{Kavli Institute for Cosmology Cambridge, Madingley Road, Cambridge, CB3 0HA, UK}

\author{Suzanne T. Staggs~\orcidlink{0000-0002-7020-7301}}
\affiliation{Joseph Henry Laboratories of Physics, Jadwin Hall, Princeton University, Princeton, NJ, USA 08544}

\author{Alexander van Engelen \orcidlink{0000-0002-3495-158X}}
\affiliation{School of Earth and Space Exploration, Arizona State University, Tempe, AZ 85287, USA}

%%%%%%%%%%%%%%%%%%%%%%%%%%%%%%%%%%%%%%%%%%%%%%%%%%%%%%%%%%%%%%%%%%%%%%%%%%%
\date{\today}
\smallskip

\begin{abstract}
Ultralight axions (ULAs) are well-motivated dark matter particle candidates that arise in many extensions of the Standard Model of particle physics. ULAs with mass $m_\mathrm{a} \lesssim 10^{-27}$ eV have been strongly constrained by cosmic microwave background (CMB) observations in temperature and polarization. We fit recent  measurements of gravitational lensing of the CMB from \textit{Planck}, the Atacama Cosmology Telescope (ACT) and the South Pole Telescope (SPT-3G) using a state-of-the-art simulation-calibrated nonlinear clustering model for ULAs. We derive the strongest constraints on ULAs in the mass range $10^{-26}\;\mathrm{eV}\leq m_\mathrm{a}\leq 10^{-24.5}\;\mathrm{eV}$. ULAs of this mass have been shown to alleviate tensions between inferences of the matter clustering if they compose a few percent of the total dark matter content of the Universe. We conclude that ULAs with a mass of $10^{-26}$ eV make up less than 1.5\% of the dark matter and $10^{-25}$ eV axions make less than 9\% (both at 95\% confidence level). We identify a slight preference for non-zero axion density at $10^{-24.5}$ eV at $2.1\sigma$. We find that the preference for ULAs is largely driven by a few data points and that further investigation of non-linear ULA physics is needed to confirm or rule out this signal.

\end{abstract}
\maketitle
%\keywords{cosmology:cosmic microwave background --- cosmology:observations}  

%section 1: Introduction
\section{Introduction}
Dark matter (DM) comprises most of the mass in the Universe, but its origin remains a mystery. Some of the most promising DM particle candidates are axions and axion-like particles. Axions were originally proposed as a solution to the strong CP problem of quantum chromodynamics (QCD)~\cite{Peccei1977CP,Preskill_1983,Abbott_1983,Dine_1983,Weinberg1978LightBoson,Wilczek1978ProblemOf}. Axion-like particles are ubiquitous in quantum gravity extensions to the Standard Model. Studies of specific compactifications in string theory predict that individual ultralight axions (ULAs; particle masses \(m_\mathrm{a} < 10^{-17}\,\mathrm{eV}\)), as part of a spectrum of particles, could compose a few percent of the total DM; in most cases, we expect multiple axion fields with different masses to form, leading to the \textit{string axiverse} scenario~\cite{Witten1984Superstrings,Svrcek_2006,Arvanitaki_2010,Acharya2010MTheory,Conlon2006ModuliStabilisation,Cicoli2012TypeIIB,Arza:2026rsl}. The exact number of axion species in the string axiverse depends on the topology of the Calabi-Yau manifold and can range from a few to over 400 with a log-distributed mass spectrum~\cite{Mehta_2021,Gendler2024GlimmersOf}. This wide mass range includes ULAs with masses between $(10^{-33}-10^{-17})$ eV. Recent developments in the field of computational geometry have allowed the exploration of specific realizations of the string axiverse~\cite{Demirtas2022CYTools,Bachlechner_2017,Sheridan2024FuzzyAxions,Demirtas2018Kreuzer-Skarke,Gendler2024GlimmersOf,Gendler2024QCDAxion,Mehta_2021,Demirtas2023PQAxiverse,Benabou2025QCDAxion,Cheng2025Universality,Yin2025AxiverseReio,MacFadden2026DNA,Vander2025FTheory,Jain2025BayesianCY}.

ULAs are pseudoscalar bosons with a periodic potential, which we expand assuming small-field displacements. The potential is then assumed to be quadratic, and the equation of motion for the homogeneous part of the axion field can be found using the Klein-Gordon equation for a Friedmann-Lema\^itre-Robertson-Walker metric to be
\begin{align}
    \ddot{\varphi}_0 + 3H \dot{\varphi}_0 + m_\mathrm{a}^2\varphi_0 = 0,
\end{align}
where the overdot denotes derivative with respect to cosmic time, $\varphi_0$ is the background axion field and $H$ is the Hubble expansion rate (the full cosine potential was considered in Ref.~\cite{Winch2024Extreme}). The equation of motion is that of a damped harmonic oscillator, and admits two types of solutions. For $m_\mathrm{a}\ll H$, the field is frozen, and the axion behaves as a quintessence field; for $m_\mathrm{a}\gg H$, the field oscillates rapidly and its density scales as matter. ULAs with mass below $10^{-33}$ eV will thus behave as dark energy until the present day, while axions with mass $m_\mathrm{a}\gtrsim 10^{-27}$ eV start behaving as DM before the onset of matter-radiation equality. We will consider only DM-like axions for this work and leave dark energy-like axions for future work.

DM-like ULAs do not cluster like cold DM (CDM) due to their macroscopic wavelike nature (de Broglie wavelength \(\sim\) kpc -- Mpc). Indeed, the ULA density perturbations obey a Jeans equation
\begin{align}
    \ddot{\delta}^\mathrm{L}_\mathrm{a} + 2H\dot{\delta}^\mathrm{L}_\mathrm{a} + \left(\frac{\hbar^2k^4}{4m_\mathrm{a}^2 a^4} -4\pi G\bar{\rho}_\mathrm{m} \right)\delta^\mathrm{L}_\mathrm{a} = 0, \label{eq:ax-jeans}
\end{align}
where $\delta_\mathrm{a}$ is the axion density fluctuation, $k$ is the comoving wavenumber, $\bar\rho_\mathrm{m}$ is the mean matter density of the Universe, and \(a\) is the scale factor. The superscript L indicates that Eq.~\eqref{eq:ax-jeans} describes the evolution of axions in the linear regime. On small scales (large $k$), the diffusion term dominates, and the axion perturbations do not grow. Conversely, on large scales or at high axion mass, Eq.~\eqref{eq:ax-jeans} reduces to the CDM limit. This is similar to the Jeans instability of massive neutrinos near their free streaming scale or baryons undergoing collapse.

In models with a large number of axions, many of them will have a sufficiently high mass to be indistinguishable from CDM on cosmological scales. Studying multi-axion cosmologies requires the inclusion of two free parameters (for relic densities and masses) for each axion species. To limit the number of additional free parameters beyond those of $\Lambda$CDM, we approximate multi-axion models as a combination of one ultralight species with relic density $\Omega_\mathrm{a}$ and many heavier axions with collective relic density $\Omega_{\rm c}$. It is useful to define the proportion of DM composed of ultralight particles as the axion fraction
\begin{align}
    \frac{\Omega_\mathrm{a}}{\Omega_\mathrm{d}}\equiv\frac{\Omega_\mathrm{a}}{\Omega_\mathrm{c}+\Omega_\mathrm{a}},
\end{align} 
where $\Omega_\mathrm{d}$ is the total DM relic density. This parametrization has been constrained by various astrophysical probes. If we assume that the ULA composes all of the DM, then the axion mass is strongly constrained by the Lyman-$\alpha$ forest~\cite{Irsic2017FirstConstraints,Armengaud2017ConstrainingThe,Rogers:2020cup,Rogers2021StrongBound}, galaxy weak lensing~\cite{Dentler2022FuzzyDark}, and the galaxy ultra-violet luminosity function (UVLF)~\cite{Sipple2025FuzzyDark}. Studies of stellar kinematics in ultra-faint dwarf galaxies also disfavor $m_\mathrm{a}\lesssim 10^{-19}$ eV at the 95\% confidence level (assuming $\Omega_\mathrm{a}/\Omega_\mathrm{d}=1$)~\cite{2025ApJ...986..127N,2025ApJ...986..127N,Gonzalez2017UnbiasedConstraints,Marsh2019StrongConstraints,Zimmermann2025DwarfGalaxies,Teodori2026UltralightDark,Nadler2026WarmFuzzy,Dalal2022ExcludingFuzzy,May2025UpdatedBounds}. While our model assumes that axions are produced before inflation, post-inflationary ULAs are also well-constrained with bounds of $m_\mathrm{a}>10^{-19}$ eV~\cite{Amin2024ALower}.

Studies of the string axiverse predict $\Omega_\mathrm{a}/\Omega_\mathrm{d}\ll 1$ for axion masses below $10^{-22}$ eV~\cite{Stott2017Spectrum}. A combination of the primary cosmic microwave background (CMB) and baryon acoustic oscillation (BAO) surveys strongly constrain the ULA fraction for $10^{-30}\;\mathrm{eV}\lesssim m_\mathrm{a} \lesssim 10^{-27}\;\mathrm{eV}$~\cite{Hlozek2015ASearch}. Studies using the effective field theory of large-scale structure including data from the Sloan Digital Sky Survey (SDSS) further constrain axions for $m_\mathrm{a}\lesssim10^{-25}$ eV~\cite{Lague2022ConstrainingUltralight,Rogers2023UltralightAxions}. The latter notably found a bound of $\Omega_\mathrm{a}/\Omega_\mathrm{d}\lesssim 0.3$ for $m_\mathrm{a} =10^{-25}$ eV. The UVLF of early galaxies constrains the axion fraction to be $\lesssim 20\%$ for masses up to $m_\mathrm{a}\sim 10^{-22}$ eV~\cite{Winch_2024}. The Lyman-$\alpha$ forest constrains $\Omega_\mathrm{a}/\Omega_\mathrm{d}\lesssim 0.16$ at $m_\mathrm{a}=(10^{-23}-10^{-22})$ eV~\cite{Kobayashi2017LymanAlpha}. Further studies using the Lyman-$\alpha$ forest show a strong preference for a small but non-zero fraction of ULAs for $m_\mathrm{a}=(10^{-25}-10^{-23})$ eV~\cite{Rogers20235Sigma}. We will describe existing constraints more quantitatively in Sec.~\ref{sec:other-probes}.

ULAs have been proposed as an explanation for the observed discrepancy in the inferred amplitude of matter fluctuations between galaxy weak lensing surveys and the cosmic microwave background, known as the $S_8$ tension \citep{Rogers2023UltralightAxions}. The parameter combination to which galaxy weak lensing is most sensitive is $S_8 \equiv \sigma_8(\Omega_\mathrm{m}/0.3)^{0.5}$, where $\sigma_8$ is the root-mean-squared amplitude of the linear matter density fluctuations on scales of $8\,\mathrm{Mpc}/h$ at the present day and $\Omega_\mathrm{m}$ is the total matter density. It has been demonstrated that a scale-dependent suppression of the matter power spectrum can reconcile the discrepant measurements~\cite{Perez2025ReconstructingThe}. In particular, an axion with mass around $10^{-25}$ eV introduces the right scale-dependent effects to the matter power spectrum to reduce the amount of lensing measured by galaxy weak lensing surveys while keeping the primary CMB unchanged. Boosting the level of baryonic feedback to smooth out the distribution of matter on nonlinear scales~\cite{amon_2022, Preston2023ANonLinear}, however, could also explain this discrepancy. The latter hypothesis is in agreement with the observed diffuse gas distribution in galaxies, as measured through the kinetic and thermal Sunyaev-Zel'dovich effects~\cite{Bigwood2024WeakLensing,Pandey2025ConstraintsOn,Dalal2026Deciphering} as well as the most recent X-ray measurements~\cite{Popesso2026ERosita}. Likewise, corrections to the redshift estimates from photometric observations have also been shown to lead to an $S_8$ value closer to the value found using the CMB~\cite{Wright2025KiDSLegacy}, possibly indicating that the tension is sourced by observational systematics. ULAs are part of a class of solutions to the $S_8$ tension which involve modifying the shape of the matter power spectrum on small scales at early times. Thus, probing the existence of ULAs is in many ways equivalent to searching for evidence of scale-dependent departures from $\Lambda$CDM in the \textit{linear} regime. While this manuscript was in preparation, the authors of Ref.~\cite{Gaughan2026Ultralight} investigated the impact of nonlinear effects from ULA on the clustering of matter on small scales. As we will discuss in Sec.~\ref{sec:NL_model} and Sec.~\ref{sec:data}, our work complements the findings of Ref.~\cite{Gaughan2026Ultralight} by including a broader data set. This work also builds on the analysis of beyond-$\Lambda$CDM models of Ref.~\cite{Calabrese2025ExtendedModels} which derived constraints on ULAs using the primary CMB. We supplement their work by including observations of gravitational lensing and using the latest nonlinear model for the clustering of ULAs on small scales (which we describe in Sec.~\ref{sec:NL_model}).

We begin by outlining our nonlinear model for mixed ULA-CDM cosmologies in Sec.~\ref{sec:NL_model}. We then list the datasets we fit using this model in Sec.~\ref{sec:data}. In Sec.~\ref{sec:results}, we conduct a combination of Bayesian and frequentist statistical analyses and derive the posterior distributions for the cosmological and axion parameters. We also list the marginalized constraints on axion parameters we obtain and compare our results with existing bounds. Finally, we discuss the implications of our findings in Sec.~\ref{sec:discussion}.

\section{Model}\label{sec:NL_model}

\subsection{Nonlinear matter power spectrum} 
One of the major advancements in the study of ULAs has been the study of their behavior in the nonlinear regime through the use of semi-analytic models, and numerical simulations~\cite{Marsh2014AModel,Dentler2022FuzzyDark,Lague2024CosmologicalSimulations,Vogt2023ImprovedMixed,Dome2025ImprovedHalo}. While the primary CMB is well-modeled by linear theory, it has been shown that nonlinear corrections need to be taken into account when using CMB lensing at high multipoles~\cite{McCarthy_2022,Trendafilova2025TheEnd}. We obtain CMB, lensing convergence, and matter power spectra predictions in the linear regime from the modified Boltzmann code \texttt{axiECAMB}~\cite{Hlozek2015ASearch,Liu2025AccurateMethod}. We include non-linear corrections through the halo model prescription implemented in \texttt{axionHMcode}~\cite{Vogt2023ImprovedMixed}. This model was originally developed as an extension to the non-linear matter power spectrum fitting functions of \texttt{HMcode}~\cite{Mead_2020} and is based on a biased tracer approach originally used to model the effects of massive neutrinos~\cite{Massara2014TheHalo}.

\begin{figure*}[htb]
    \centering
     \begin{subfigure}[t]{0.5\linewidth}
         \includegraphics[width=\linewidth]{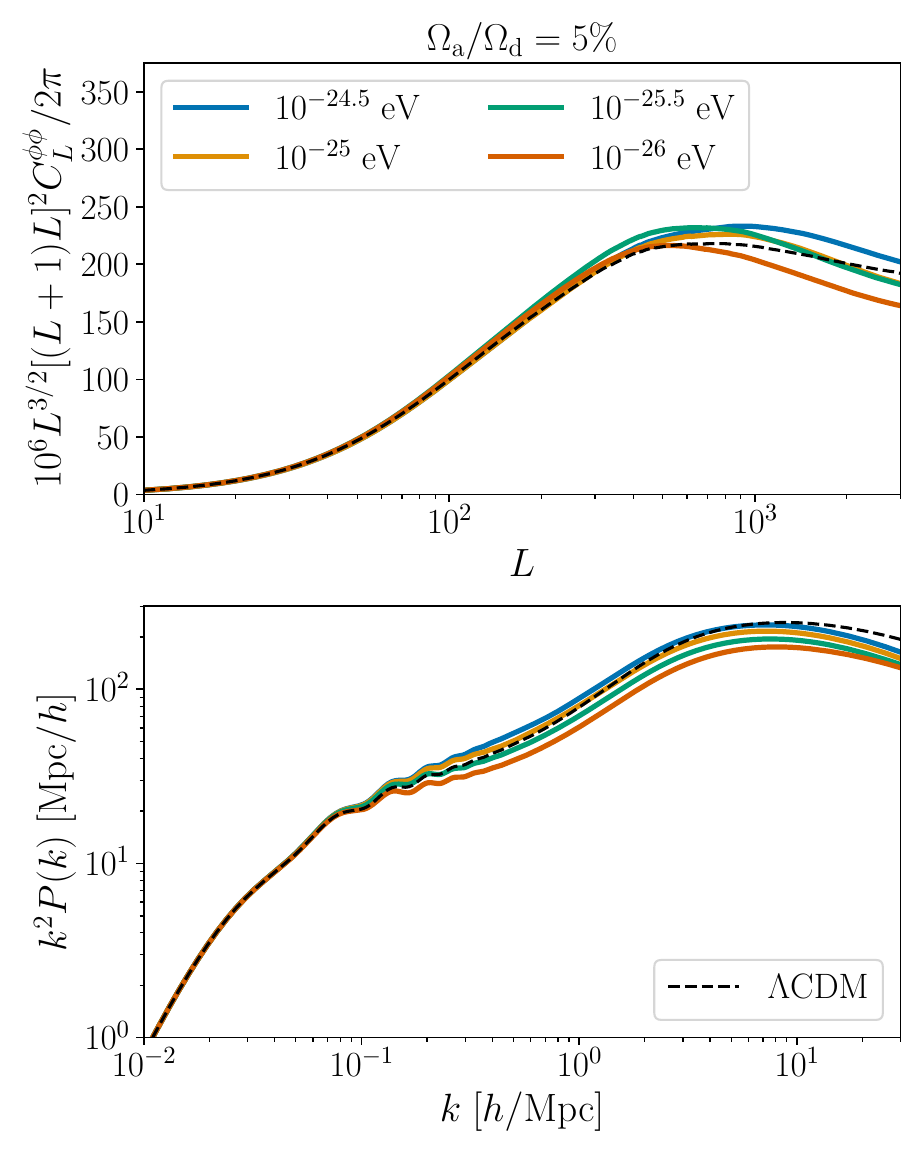}
         \caption{\label{fig:pks_full}}
     \end{subfigure}%
    ~
     \begin{subfigure}[t]{0.5\linewidth}
         \includegraphics[width=\linewidth]{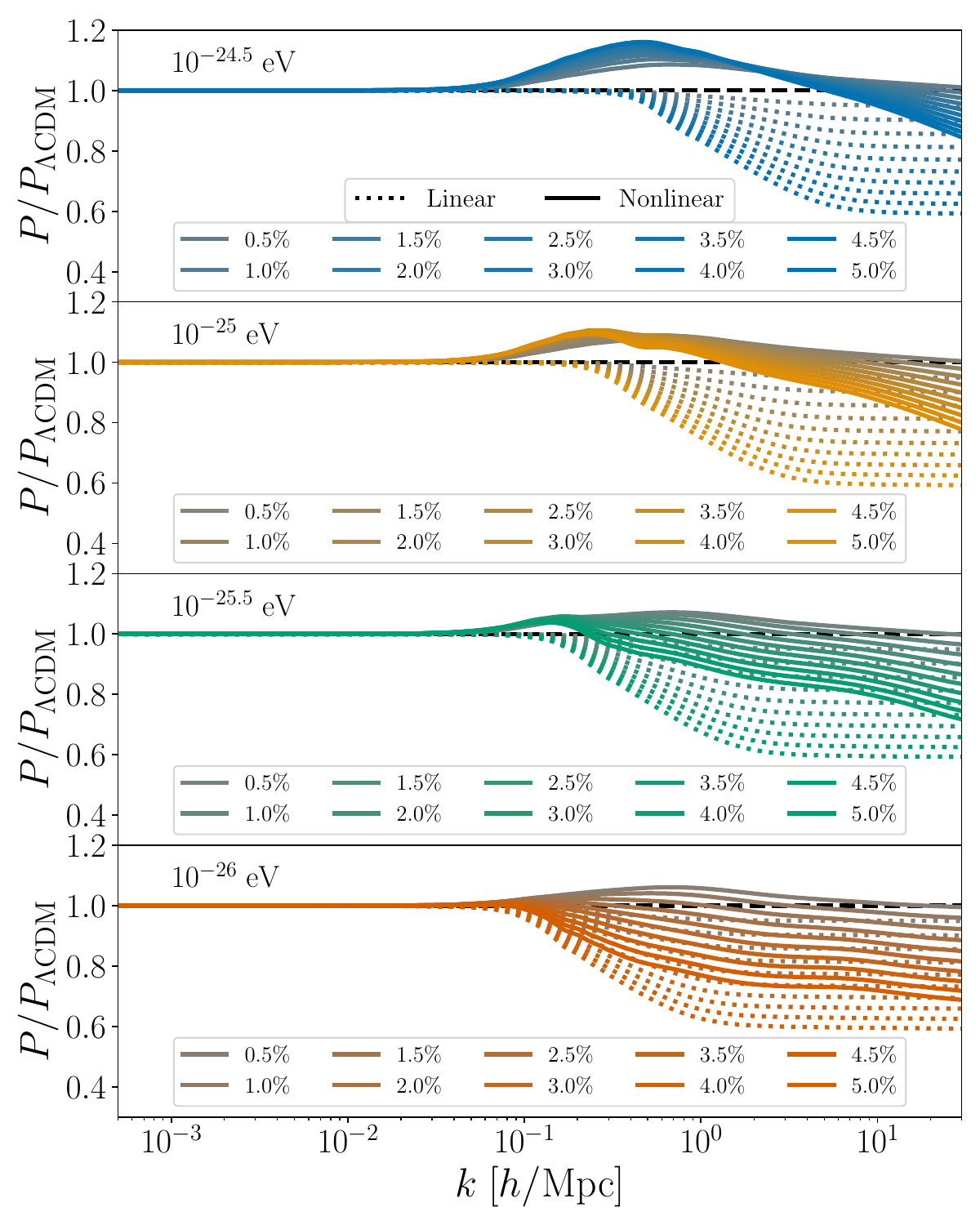}
         \caption{\label{fig:pks_ratios}}
     \end{subfigure}
    \caption{(Top left) CMB lensing convergence angular power spectra as a function of multipole \(L\) for \(\Lambda\)CDM and $\Omega_\mathrm{a}/\Omega_\mathrm{d}=5\%$ for the four axion masses considered in this work. (Bottom left) As top left, except for nonlinear matter power spectra as a function of wavenumber \(k\). (Right) Ratios of matter power spectra to the $\Lambda$CDM reference, varying both \(m_\mathrm{a}\) and \(\Omega_\mathrm{a}/\Omega_\mathrm{d}\). The solid lines denote the nonlinear spectra while the dotted indicate the linear spectra.
    \label{fig:pks}}
\end{figure*}

The model is calibrated on high-resolution mixed ULA-CDM simulations with axion mass ranging from $10^{-25}$ eV to $10^{-21}$ eV~\cite{Dome2025ImprovedHalo} and accounts for the wave behavior of ULAs and the formation of soliton cores and interference fringes in collapsed structures. We use the recently updated version of the model following a re-calibration on a more extensive set of simulations~\cite{Dome2025ImprovedHalo}. This update allows for accurate modeling of the matter power spectrum in mixed DM cosmologies up to $k\sim 10\;h/$Mpc. However, we caution that this model does not include the effects of baryonic feedback on the matter power spectrum as it is based on DM-only simulations. As argued in Ref.~\cite{Popesso2026ERosita}, simulating baryonic processes in a way which matches observations is challenging. Including them in the nonlinear model may require the addition of multiple nuisance parameters quantifying this uncertainty to avoid biasing future axion constraints.  %We leave the development of mixed DM simulations with baryonic effects to future work.

The inclusion of non-linear effects leads to an enhancement in the matter power spectrum when compared to $\Lambda$CDM for some axion masses and fractions. This is due to the formation of coherent structures known as solitonic cores around halos. The existence of an enhancement was predicted in the original version of \texttt{axionHMCode}. In particular, the halo model approach showed such structures create enhancements of up to $\sim$ 30\% in the axion part of the spectrum $P_\mathrm{a, a}(k)$. This was found to occur on scales of $k\sim 0.1-1\;h/$Mpc for $m_\mathrm{a}\sim 10^{-24}-10^{-25}$ eV. Further calibration on mixed DM simulations solving the Schr\"odinger-Poisson system in Ref.~\cite{Dome2025ImprovedHalo} suggests that the (total) matter power spectrum enhancement may have been slightly under-estimated. One possible explanation for this is that the axion-CDM halo model is based on a biased tracer approach, thus assuming that axions follow the CDM gravitational potential. However, axions may induce more severe backreactions on the CDM component than what was originally considered. This statement does not hold if the axion fraction is very large or if its mass is lower than $\sim 10^{-27}$ eV. In that case, the suppression of the linear matter power spectrum dominates even when accounting for non-linear growth of structure. The number density of halos of certain masses was also shown to be higher for small concentrations of ULAs compared to the pure CDM case by Refs.~\cite{Winch_2024,Johnston2026HMF}. Conversely, Ref.~\cite{Gaughan2026Ultralight} found that the modeling of the nonlinear effects in axion cosmologies required tuning to simulation as CDM models applied to a damped linear matter power spectrum can over-estimate the power spectrum on semi-linear scales. We describe the halo model prescription that we use to compute the nonlinear axion effects in more detail in Appendix~\ref{sec:halo-model}. For a full breakdown of the model, we refer the readers to Refs.~\cite{Vogt2023ImprovedMixed,Dome2025ImprovedHalo}.

The matter power spectra for four axion masses are shown in Fig.~\ref{fig:pks}. We compare the non-linear spectra with their linear counterparts to highlight the contribution of the nonlinear effects. On semi-linear scales and for axion masses $m_\mathrm{a}> 10^{-26}$ eV the matter power spectrum shows an enhancement with respect to the $\Lambda$CDM case, but is still suppressed on very small scales ($k\sim 0.5-5\;h/$Mpc, depending on the ULA mass). This contrasts with the linear model which predicts only a suppression of power with respect to the $\Omega_\mathrm{a}\to 0$ limit.

There are some limitations to our nonlinear axion model. First, the set of simulations over which it is calibrated is sparse in the axion parameter space. Indeed, the first simulation suite of Ref.~\cite{Lague2024CosmologicalSimulations} explored the mass range $10^{-25}\,\mathrm{eV}-10^{-21}\,\mathrm{eV}$ at a fixed fraction of 10\% while the simulations of Ref.~\cite{Dome2025ImprovedHalo} explored fractions of $\Omega_\mathrm{a}/\Omega_\mathrm{d}\in [0, 0.01, 0.1, 0.2, 0.3]$, but only for $m_\mathrm{a} = 10^{-24.5}\;\mathrm{eV}$. The model effectively interpolates between these points in parameter space. Nonetheless, the model is designed as a modification of \texttt{HMcode} and recovers the CDM power spectrum in the limits $m_\mathrm{a}\to \infty$ and $\Omega_\mathrm{a}\to 0$. Second, spectral methods for solving the axion Schr\"odinger-Poisson system of partial differential equations must satisfy strict timestep conditions for convergence~\cite{Schwabe2016SimulationsOf}. This makes ULA simulations computationally expensive, particularly at low redshift, where the integration timestep decreases rapidly. Most simulations are therefore limited to redshifts $z \geq 1-2$, with most focusing on high-redshift observables at $z\sim 4$~(see, e.g. Ref.~\cite{Wang2026LymanAlpha}). This means that the matter power spectrum at redshifts below $z=1$ must be extrapolated. In the case of CMB lensing, the kernel in the Limber integral of Eq.~\eqref{eq:limber-phi} peaks around $z=2$, where the model is well calibrated.

\subsection{Gravitational lensing \label{sec:cmb}}
We use the adapted Boltzmann code \texttt{axiECAMB} to obtain the unlensed temperature and polarization angular power spectra. We then compute the lensing corrections using the nonlinear matter power spectrum model (Sec.~\ref{sec:NL_model}). For this calculation, we follow the prescription of Refs.~\cite{Hu2000WeakLensing, Lewis2006WeakGravitational} and use the integral solvers implemented in the \texttt{correlations} package as part of the Boltzmann code \texttt{CAMB}. The expressions we evaluate are in Appendix~\ref{app:lensed-cmb}.

We evaluate the angular power spectrum of the lensing potential \(C_\ell^{\phi\phi}\) given the nonlinear matter power spectrum \(P_\mathrm{m}(k,z)\) (Sec.~\ref{sec:NL_model}). Under the Limber approximation,
\begin{align}
    C_\ell^{\phi\phi} = 4&\int_0^{\chi_\star} d\chi\;\left(\frac{\chi_\star-\chi}{\chi^2\chi_\star}\right)^2 \nonumber \\&\;\;\;\;\times P_\Psi\left[(\ell+1/2)/\chi,z(\chi)\right], \label{eq:limber-phi}
\end{align}
where $\chi$ is the comoving distance, $\chi_\star$ is the comoving distance to the surface of last scattering, and $P_\Psi$ is the power spectrum of the Weyl potential accounting for axion and nonlinear effects. The latter can be obtained from the matter power spectrum through
\begin{align}
    P_\Psi(k, z) = \frac{9\Omega_\mathrm{m}^2(z)H^4(z)}{8\pi^2} \frac{P_\mathrm{m}(k,z)}{k}.
\end{align}
We calculate the nonlinear Weyl potential power spectrum in a two-step process. First, we compute the Weyl potential power spectrum in $\Lambda$CDM assuming linear theory \(P^{\mathrm{L}}_\Psi(\Omega_\mathrm{a}=0)\). We then re-scale the result using a ratio of matter power spectra following the approach of Ref.~\cite{Lewis2006WeakGravitational}, giving
\begin{align}
     P_\Psi = \left[\frac{P_\mathrm{m}}{ P^{\rm L}_\mathrm{m}}\right]\left[\frac{ P_\mathrm{m}^{\rm L} }{P_\mathrm{m}^{\mathrm{L}}(\Omega_\mathrm{a}=0)} \right ]P^{\mathrm{L}}_\Psi(\Omega_\mathrm{a}=0). \label{eq:Weyl-from-camb}
\end{align}
We generate the Weyl power spectrum in the linear regime without axions from \texttt{CAMB} since it is not a standard output of \texttt{axiECAMB} (due to a version difference between the two implementations). We also use \texttt{CAMB} to compute the linear matter power spectrum without axions \(P_\mathrm{m}^{\mathrm{L}}(\Omega_\mathrm{a}=0)\). We use \texttt{axiECAMB} to obtain the linear matter power spectrum accounting for axions \(P_\mathrm{m}^{\mathrm{L}}\) and \texttt{axionHMcode} to compute the matter power spectrum with axions and nonlinear corrections \(P_\mathrm{m}\).

%\keir{Add description of Fig. 1 CMB lensing here.}
The nonlinear model predictions for the CMB lensing angular power spectrum for a few axion masses are shown in Fig.~\ref{fig:pks}. For axions with mass $m_\mathrm{a} \approx 10^{-26}\;\mathrm{eV}$, the scales where the power is suppressed compared to $\Lambda$CDM dominates. We thus expect a strictly lower CMB lensing signal. For masses above this value, we find that the lensing power spectrum amplitude is enhanced for a range of multipoles before exhibiting a scale-dependent suppression.

\subsection{Emulator training and validation}
The numerical evaluation by \texttt{axionHMcode} of the nonlinear matter power spectrum in the presence of ULAs is computationally expensive. This is due to three compounding factors: (1) tracking the evolution of the axion field requires the evaluation of highly oscillatory integrals; (2) the nonlinear model involves gluing together CDM and axion density profiles which must be normalized numerically; and (3) the evaluation of the Limber integral requires computing the nonlinear corrections at a large number of redshifts. For these reasons, we develop a neural network emulator for fast evaluations of the matter and CMB angular power spectra. We follow the training and validation procedures used to produce the \texttt{cosmopower} package~\cite{Mancini2022COSMOPOWER}, modified to account for the effects of axions \texttt{axionEmu}.

\section{Data sets}\label{sec:data}
We constrain the axion model with data from the primary CMB, CMB lensing, and baryon acoustic oscillations. For the primary CMB, we use the combination of Atacama Cosmology Telescope (ACT) Data Release 6 (DR6) and \textit{Planck} measurements of the CMB temperature and polarization angular power spectra \cite{PlanckLikelihood,ACTDR6Likelihood,AtacamaCosmologyTelescope:2025vnj}. Specifically, we use the CMB-only \texttt{act\_dr6\_cmbonly.ACTDR6CMBonly} and \texttt{act\_dr6\_cmbonly.PlanckActCut}. 
At low multipoles, we use the \texttt{Sroll2} likelihood for the \textit{Planck} E-mode polarization~\cite{PlanckSroll2Likelihood} and the \texttt{planck\_2018\_lowl.TT} for the temperature spectra. For the BAO data, we use the second data release from the Dark Energy Spectroscopic Instrument (DESI DR2) likelihood~\cite{DESIDR2BAO}. This primary CMB and BAO data combination is similar to the one used to constrain axions in Ref.~\cite{Calabrese2025ExtendedModels}. Here, we additionally include measurements of the CMB lensing angular power spectra from the South Pole Telescope (SPT-3G) combined with the ACT DR6 and \textit{Planck} CMB lensing measurements~\cite{Madhavacheril_2024, Qu_2024, Planck2018Lensing}. The combination of the three datasets, which we will denote APS, is described in Ref.~\cite{Qu2026APSLensing}. We use the full combination of CMB lensing data up to the maximum angular multipole $L_\mathrm{max}\approx2700$ unless otherwise specified. We use two data set combinations throughout this work. The first is the APS combination described above, including the lensing from three surveys up to $L_\mathrm{max}\approx 2700$ with the BAO data from DESI DR2 and the primary CMB measurements from ACT DR6 and \textit{Planck} 2018. The other, which we refer to as AP, includes the same non-lensing data (BAO and primary CMB), but without the SPT-3G lensing data points, thus going to $L_\mathrm{max}=1250$ in lensing. %When analyzing \textit{Planck} and ACT data, we additionally vary nuisance parameters \(A_\mathrm{Planck}\), \(A_\mathrm{ACT}\) and \(P_\mathrm{ACT}\).

The authors of Ref.~\cite{Gaughan2026Ultralight} conducted a comparative analysis on the choice of nonlinear model for ULAs. Their dataset is similar to ours, with the exception that we include the CMB lensing measurements to smaller scales. They conclude that using the $\Lambda$CDM-based \texttt{HMcode} prescription with a modified linear matter power spectrum leads to biased parameter inference when using CMB lensing. Our internal analysis agrees with this conclusion. This motivates the use of the simulation-calibrated model of Ref.~\cite{Dome2025ImprovedHalo}, as we do here. To recover the result of \texttt{HMcode} to sub-percent accuracy in the $\Omega_\mathrm{a}\to 0$ limit, further corrections must be applied to account for neutrinos. We discuss their implementation in Appendix~\ref{sec:halo-model}. Ref.~\cite{Gaughan2026Ultralight} also introduces additional nuisance parameters, following the weak lensing study of Ref.~\cite{Dentler2022FuzzyDark}, to model higher axion masses. Axions at lower masses have a Jeans instability (see Eq.~\ref{eq:ax-jeans}) on scales where linear theory applies and are simpler to model. We adopt a conservative approach by limiting ourselves to constraining axion masses below the mass at which the model was calibrated ($m_\mathrm{a}=10^{-24.5}$ eV).

\section{Results}\label{sec:results}

\subsection{Parameter inference} \label{sec:param_inference}
We evaluate the posterior distribution of the axion mass and fraction and other cosmological parameters through Markov chain Monte Carlo (MCMC) sampling using the Metropolis-Hastings algorithm. We impose priors as listed in Table~\ref{tab:priors}. Since our nonlinear model does not account for the potential interplay between massive neutrinos and ULAs, we fix the sum of neutrino masses to the fiducial value $\Sigma m_\nu=0.06$ eV. While ULAs with mass $m_\mathrm{a}\sim 10^{-28}$ eV can have effects on large-scale structure degenerate with massive neutrinos~\cite{Lague2022ConstrainingUltralight}, ULAs in the mass range considered suppress the growth of structure on a range of wavelengths far smaller than the typical neutrino free-streaming wavelength. We discuss the effects of massive neutrinos further in Appendix \ref{sec:halo-model}.

\begin{table}[]
    \centering
    \begin{tabular}{c|c}
    Parameter & Prior \\\hline
        $\log( 10^{10} A_\mathrm{s}$) & $\mathcal{U}(2.7, \;3.5)$ \\
        $n_\mathrm{s}$ & $\mathcal{U}(0.85, \;1.1)$ \\
         $H_0/(\mathrm{km/s/Mpc})$ & $\mathcal{U}(50,\; 90)$ \\
         $\Omega_\mathrm{b}h^2$ & $\mathcal{U}(0.017,\;0.027)$ \\
         $\Omega_\mathrm{c}h^2$ & $\mathcal{U}(0.09, \; 0.15)$ \\
         $\tau_\mathrm{reio}$ & $\mathcal{U}(0.02, \; 0.1)$ \\
         $\Omega_\mathrm{a}h^2$ & $\mathcal{U}(10^{-6}, \; 0.15)$\\
         $\log_{10} m_\mathrm{a}/\mathrm{eV}$ & Fixed or $\mathcal{U}(-26, \; -24.5)$
    \end{tabular}
    \caption{$\Lambda$CDM and axion parameters varied in our analysis and their priors (additional calibration and foregrounds parameters are varied following the priors implemented in their respective likelihoods).
    }
    \label{tab:priors}
\end{table}

For part of the analysis, we fix the axion mass and treat the axion density as the only free parameter beyond $\Lambda$CDM (we vary both parameters in Sec.~\ref{sec:nested_sampling}). Fixing the mass is common in the study of ULAs as different axion masses lead to distinct parameter degeneracies and are thus often treated as separate cosmological models. This treatment was employed by Refs.~\cite{Lague2022ConstrainingUltralight,Rogers2023UltralightAxions,Calabrese2025ExtendedModels}. 

Since ULAs with masses below $10^{-27}$ eV are already very well constrained by CMB and large-scale structure observations~\cite{Hlozek2015ASearch,Lague2022ConstrainingUltralight,Rogers2023UltralightAxions,Calabrese2025ExtendedModels}, we focus on higher axion masses. We run four MCMC analyses fixing the axion mass to $10^{-26}\;\mathrm{eV}$, $10^{-25.5}\;\mathrm{eV}$, $10^{-25}\;\mathrm{eV}$, and $10^{-24.5}\;\mathrm{eV}$ respectively.  For all MCMC chains, we remove the first 20\% of samples as burn-in and ensure convergence by imposing that chains run until the Gelman-Rubin statistic achieves $|R-1|<0.01$. We use the \texttt{Cobaya} sampler~\cite{Torrado2021Cobaya} to interface with the likelihoods, the \texttt{iminuit} minimizer~\cite{James:1975dr,iminuit} to identify the best-fit points, and the \texttt{GetDist} package~\cite{Lewis2025GetDist} for postprocessing and visualization. We also conduct a nested sampling analysis which we describe in Sec.~\ref{sec:nested_sampling}.

To ensure reproducibility and that we recover the same results as Ref.~\cite{Calabrese2025ExtendedModels} in the limit $\Omega_\mathrm{a}\to 0$, we compare our emulator to the existing methods for $\Lambda$CDM. We run a full parameter inference analysis using MCMC sampling with the data sets described in Sec.~\ref{sec:data} and fix the value of the axion density to $10^{-6}$ (setting it to exactly zero can lead to numerical instabilities in the numerical integration in the Boltzmann code, and the emulator is never trained with $\Omega_\mathrm{a}=0$). We repeat the analysis using the $\Lambda$CDM \texttt{cosmopower} emulator~\cite{Mancini2022COSMOPOWER,Bolliet2024HighAccuracy}. The marginalized posterior distributions of all parameters given the two emulators agree within $0.3\sigma$. This convergence test ensures that the parameters we recover are unbiased with respect to similar analyses with the same likelihoods. The small remaining differences are due to the two factors mentioned previously: (1) the presence of BAO wiggles in some calculations in the nonlinear halo model, and (2) the modeling of massive neutrinos as a correction. We discuss these in more detail in Appendix~\ref{sec:halo-model}.

\begin{figure*}
    \centering
    \includegraphics[width=\linewidth]{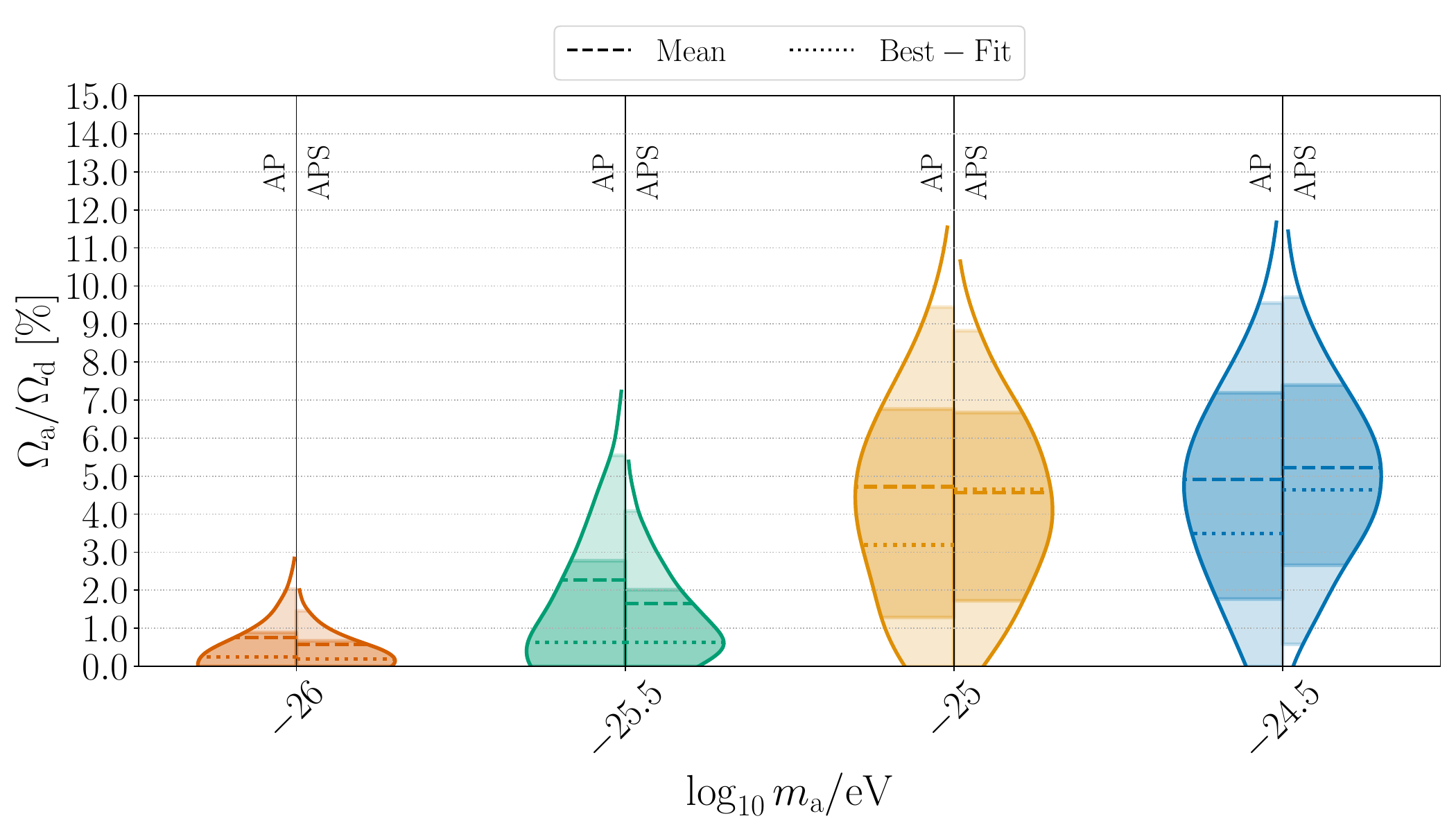}
    \caption{Marginalized posterior distributions of the axion fraction (expressed as a percentage of the total DM content) derived from the APS and AP data combinations (Sec.~\ref{sec:data}). For each violinplot, we fix the axion mass \(m_\mathrm{a} \in \{10^{-24.5}, 10^{-25}, 10^{-25.5}, 10^{-26}\}\,\mathrm{eV}\). The dashed lines denote the posterior mean while the dotted lines denote the best-fit point. The CMB lensing angular power spectra corresponding to these best-fit values are shown in Fig.~\ref{fig:lensing-data}. The darker and lighter shaded regions respectively indicate the 68\% and 95\% credible regions.}
    \label{fig:ap-ext-aps}
\end{figure*}

\begin{figure}
    \centering
    \includegraphics[width=\linewidth]{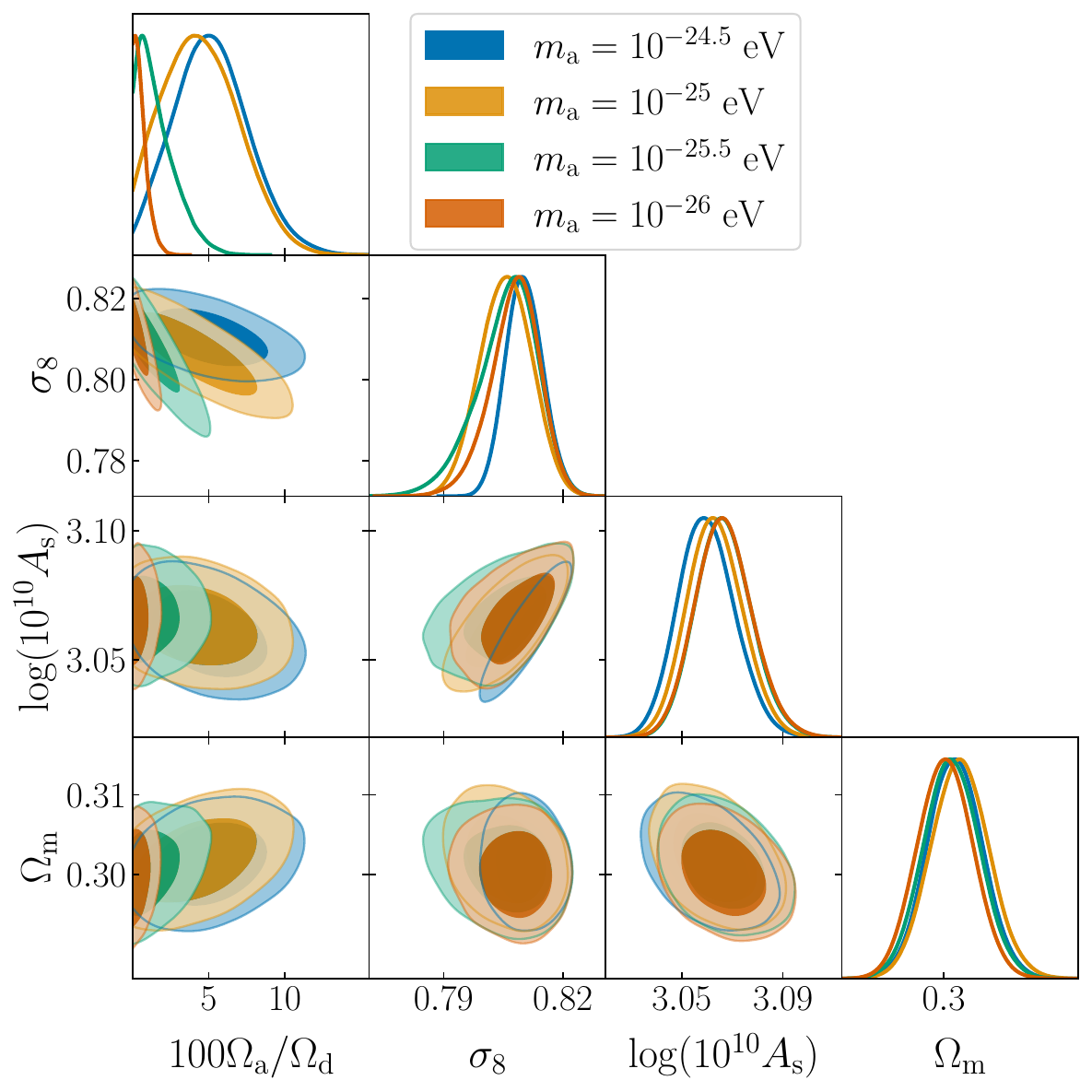}
    \caption{Joint marginalized posterior distributions for axion and cosmological parameters derived from the APS CMB lensing likelihood combined with the primary CMB and BAO described in Sec.~\ref{sec:data}. For each set of contours, we fix the axion mass \(m_\mathrm{a} \in [10^{-24.5}, 10^{-25}, 10^{-25.5}, 10^{-26}]\,\mathrm{eV}\). The darker and lighter shaded regions respectively indicate the 68\% and 95\% credible regions. The full set of parameter degeneracies can be found in App.~\ref{sec:full-results}.}
    \label{fig:four-masses-corner-main-text}
\end{figure}

\begin{table*}[]
    \centering
    \begin{tabular}{c|c|c|c|c|c|c|c}
        Data $C_L^{\phi\phi}$ & $\log_{10}m_\mathrm{a}/\mathrm{eV}$& $\Omega_\mathrm{a}h^2\;(1\sigma)$ & $\Omega_\mathrm{a}h^2\;(2\sigma)$ &  $\Omega_\mathrm{a}/\Omega_\mathrm{d} \;[\%] \;(1\sigma)$ & $\Omega_\mathrm{a}/\Omega_\mathrm{d} \;[\%] \;(2\sigma)$ & $\sigma_8\;(1\sigma)$ & $\Omega_\mathrm{m}\;(1\sigma)$ \\ \hline
        APS & $-24.5$ & $0.0062^{+0.0026}_{-0.0032}$ & $0.0062^{+0.0054}_{-0.0056}$ &  $5.2^{+2.2}_{-2.7}$ & $5.2^{+4.5}_{-4.8}$ & $0.8104\pm 0.0047$ & $0.3015\pm 0.0035$  \\
        & $-25.0$ & $0.0054^{+0.0025}_{-0.0035}$ & $< 0.0105$ &  $4.6^{+2.2}_{-2.9}$ & $< 8.84$ & $0.8057\pm 0.0065$ & $0.3021\pm 0.0037$ \\
        & $-25.5$ & $< 0.00238$ & $< 0.00483$ &   $< 2.02$ & $< 4.09$ & $0.8058^{+0.0088}_{-0.0059}$ & $0.3010\pm 0.0036$ \\
        & $-26.0$ & $<0.000808$ & $< 0.00172$ & $< 0.687$  & $< 1.46$ & $0.8079^{+0.0068}_{-0.0053}$ & $0.3002\pm 0.0035$\\ \hline
        AP & $-24.5$ & $0.0058^{+0.0028}_{-0.0037}$ & $< 0.0113$ & $4.9^{+2.3}_{-3.2}$ & $<9.55$& $0.8104\pm 0.0048$ & $0.3011\pm 0.0037$ \\
        & $-25.0$ & $0.0056^{+0.0025}_{-0.0042}$ & $< 0.0113$ & $4.7^{+2.1}_{-3.5}$ & $<9.52$& $0.8043\pm 0.0068$ & $0.3014\pm 0.0039$ \\
        & $-25.5$ & $< 0.00335$ & $< 0.00657$ & $< 2.85$& $<5.57$& $0.800^{+0.012}_{-0.0070}$ & $0.3003\pm 0.0038$ \\
        & $-26.0$ & $< 0.00105$ & $< 0.00240$ & $< 0.897$& $<2.04$& $0.8039^{+0.0090}_{-0.0056}$ & $0.2993\pm 0.0036$
    \end{tabular}
    \caption{Posterior constraints on the axion density and fraction and cosmological parameters when fixing the axion mass using the full CMB lensing likelihood from ACT DR6 + \textit{Planck} + SPT-3G (APS) or the extended CMB lensing likelihood from ACT DR6 + \textit{Planck} only (AP; Sec.~\ref{sec:data}). We give posterior means and 68\% ($1\sigma$) and 95\% ($2\sigma$) credible intervals.}
    \label{tab:constraints}
\end{table*}

\begin{table}[]
    \centering
    \begin{tabular}{c|c|c|c|c}
        Data $C_L^{\phi\phi}$ & $\log_{10}m_\mathrm{a}/\mathrm{eV}$& $\Omega_\mathrm{a}h^2$ &  $\Omega_\mathrm{a}/\Omega_\mathrm{d}\;[\%] $  & $\Delta \chi^2 (\Lambda\mathrm{CDM})$\\ \hline
        APS & $-24.5$ & $0.0055$ & $4.64$  & $-4.79$ \\
        & $-25.0$ & $0.0055$ & $4.65$& $-3.21$ \\
        & $-25.5$ & $0.00074$ & $0.63$& $-1.44$ \\
        & $-26.0$ & $0.00022$ & $0.18$& $-0.82$
    \end{tabular}
    \caption{Best-fit values of the axion density and fraction when fixing the axion mass using the full CMB lensing likelihood from ACT DR6 + \textit{Planck} + SPT-3G (APS; Sec.~\ref{sec:data}). The $\Delta \chi^2$ denote the improvement of the fit with respect to the case without axions.%or the extended lensing likelihood from ACT DR6 + \textit{Planck} (AP).
    }
    \label{tab:best-fits}
\end{table}

\begin{figure*}[htb!]
    \centering
    \includegraphics[width=\linewidth]{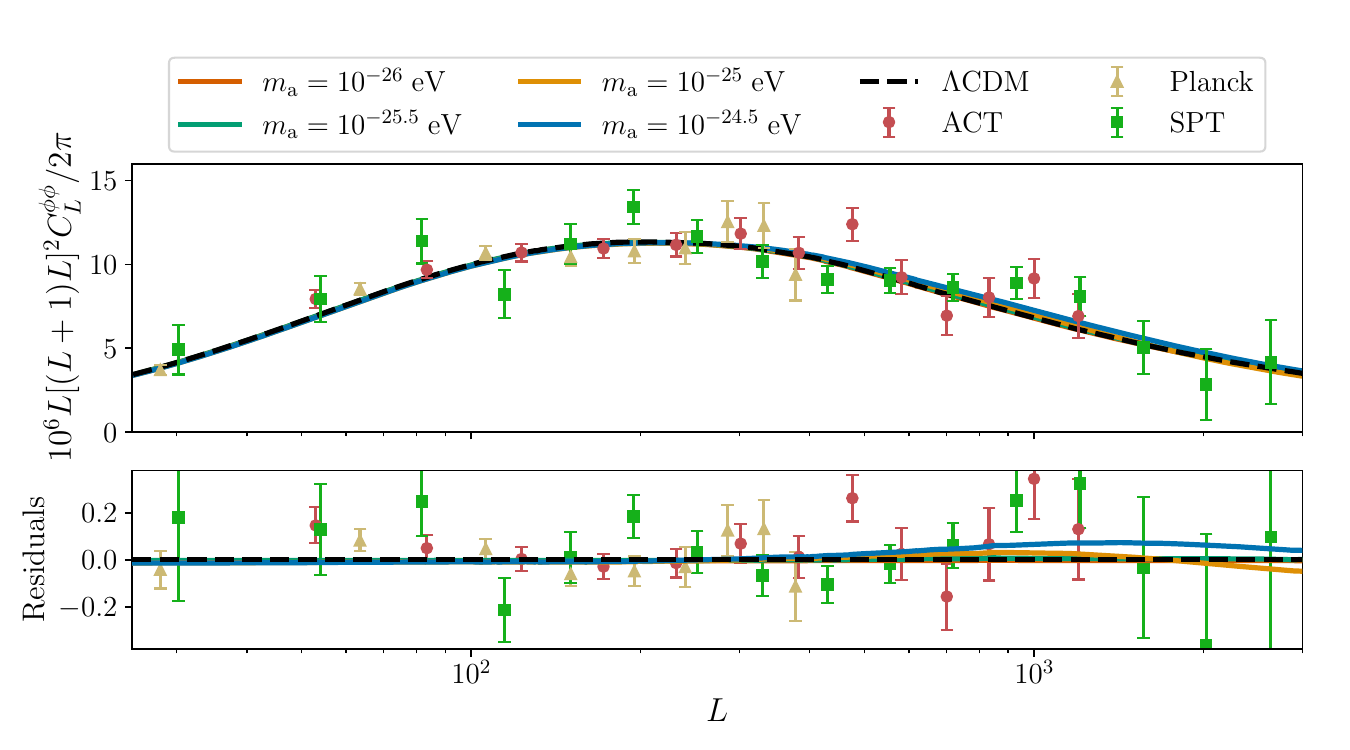}
    \caption{(Top) CMB lensing potential angular power spectra as a function of multipole \(L\) evaluated at best-fit values (Table \ref{tab:best-fits}) given APS bandpowers (data points with errorbars; Sec.~\ref{sec:data}). (Bottom) Residuals are computed with respect to the best-fit $\Lambda$CDM model. We rescale the power spectra by multiple factors of $L$ to enhance visually the small-scale information.
    \label{fig:lensing-data}}
   
\end{figure*}

The posterior contours for all four axion mass bins are shown in Fig.~\ref{fig:four-masses-corner-main-text} and Appendix \ref{sec:full-results}. There is a strong degeneracy between the axion density \(\Omega_\mathrm{a}\) and the linear matter clustering amplitude $\sigma_8$ (and in turn the parameter combination \(S_8\)), since ULAs suppress the linear matter power spectrum. Contrary to $\Lambda$CDM, in cosmologies with ULAs, the parameters $\sigma_8$ and $A_\mathrm{s}$ are no longer equivalent due to the scale-dependent growth of structure induced by axions. There is also a strong degeneracy between \(\Omega_\mathrm{a}\) and \(\Omega_\mathrm{c}\), since the total DM is tightly constrained by the CMB. Otherwise, the remaining cosmological parameters are largely insensitive to the addition of axions to the cosmological model.

In Fig.~\ref{fig:ap-ext-aps}, we show the marginalized posteriors on the axion fraction for two lensing data sets (keeping the primary CMB and BAO the same): the extended range of the ACT-\textit{Planck} combination (with $L_\mathrm{max}=1250$; AP), and the ACT-\textit{Planck}-SPT-3G combination (with $L_\mathrm{max}\approx2700$; APS). We find weaker constraints on the axion fraction when using a lower maximum lensing multipole, since we remove scales that are sensitive to the scale-dependent suppression and enhancement from axions (Sec.~\ref{sec:NL_model}). We summarize the posterior distributions in Table \ref{tab:constraints}. In general, we find that the upper limit on the allowed axion fraction increases as axion mass increases, since the effects manifest on smaller scales that are probed with less precision. We constrain the axion fraction of ULAs of mass $m_\mathrm{a}=10^{-26}$ eV to less than 1.5\% and $m_\mathrm{a}=10^{-25}$ eV to less than 9\% (at a 95\% confidence level). To investigate potential volume effects in the MCMC sampling, we compute the best-fit points for each of the four mass bins (Table \ref{tab:best-fits}). We find that the best-fit points lie below $\Omega_\mathrm{a}/\Omega_\mathrm{d}=1\%$ for the range $ 10^{-26}\;\mathrm{eV} \leq m_\mathrm{a}\leq 10^{-25.5}\;\mathrm{eV}$. The posterior means typically lie above the best-fit values due to the posterior tail and the non-symmetric nature of the prior (i.e., prohibiting $\Omega_\mathrm{a}/\Omega_\mathrm{d}<0$).

For $m_\mathrm{a} = 10^{-24.5}$ eV, the results require further analysis. Indeed, we find a slight preference for axions at around $2\sigma$. Both the mean of the posterior and the best-fit point agree with a preferred axion fraction of $5\%$. We investigate the potential causes for this value. Simulations show that the presence of ultralight axions at a mass around $10^{-24.5}$ eV can cause an increase in the growth of structure compared to $\Lambda$CDM when accounting for nonlinear effects if the axion fraction is sufficiently small (see Sec.~\ref{sec:NL_model}). This excess in the nonlinear matter power spectrum directly translates into an increase in the CMB lensing angular power spectrum. ACT and SPT-3G lensing spectra both have bandpowers which lie above the best-fit $\Lambda$CDM theory $C_L^{\phi\phi}$, as shown in Fig.~\ref{fig:lensing-data}. We thus attribute the mild preference for non-zero axion density to the combination of these two factors: the higher measured lensing bandpowers around the multipole $L=1000$ and the increase in growth of structure for $\Omega_\mathrm{a}/\Omega_\mathrm{d}\approx 5\%$ at $m_\mathrm{a}\approx 10^{-24.5}$ eV. The alignment of these two features is visible in the bottom panel of Fig.~\ref{fig:lensing-data}. As seen in Fig.~\ref{fig:ap-ext-aps}, both the datasets including and excluding the SPT-3G points favor $\Omega_\mathrm{a}>0$ (albeit at $<2\sigma$ for the latter). This suggests that the feature being fitted is not unique to SPT-3G and is also found in the ACT DR6 dataset. We analyze the improvement in $\chi^2$ per datapoint in Appendix~\ref{sec:lensing-best}.

\subsection{Nested sampling} \label{sec:nested_sampling}

\begin{figure}[t]
    \centering
     \begin{subfigure}[t]{\linewidth}
         \includegraphics[width=\linewidth]{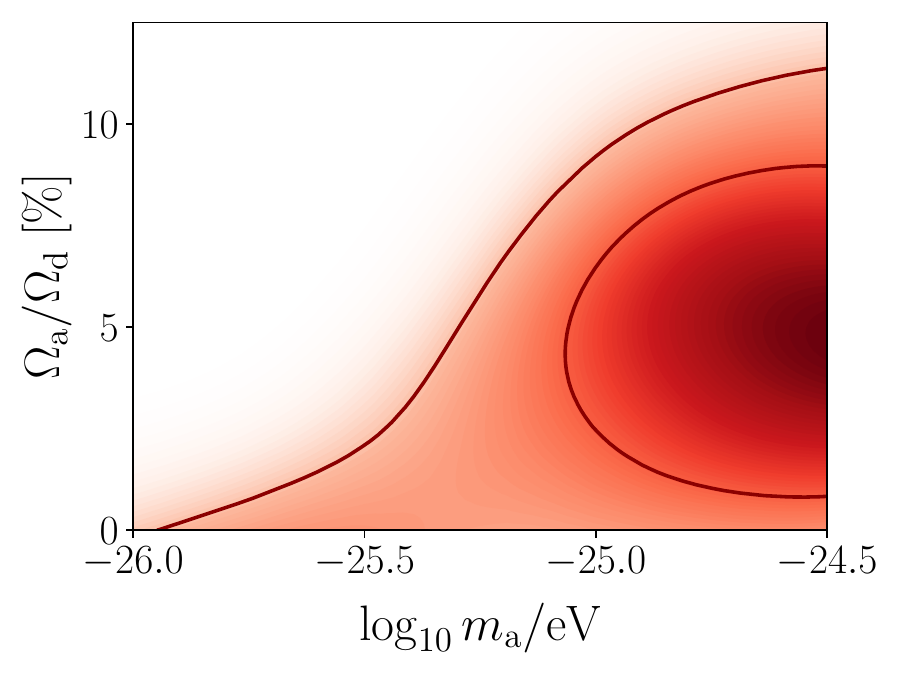}
         \caption{\label{fig:sub1}}
     \end{subfigure}
     \hfill
     \begin{subfigure}[t]{\linewidth}
         \includegraphics[width=\linewidth]{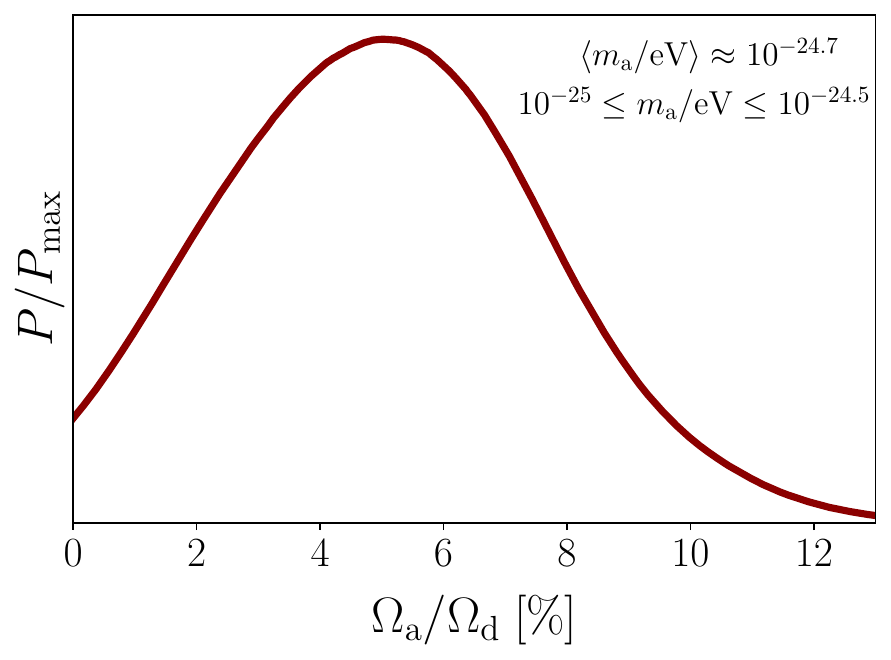}
         \caption{\label{fig:sub2}}
     \end{subfigure}
    \caption{Joint marginalized posterior distributions from the nested sampling analysis varying axion mass and fraction and other cosmological parameters simultaneously given the APS CMB lensing likelihood combined with the primary CMB and BAO (Sec.~\ref{sec:data}). (a) Two-dimensional distribution of axion mass and fraction with darker tones indicating higher probability. The 68\% and 95\% C.L. contours are shown with solid lines. (b) One-dimensional distribution of axion fraction when selecting samples in the range $10^{-25}\;\mathrm{eV}\leq m_\mathrm{a}\leq 10^{-24.5}\;\mathrm{eV}$.\label{fig:nested}}
\end{figure}

To investigate the mild preference for non-zero axion density further, we proceed to vary both the axion mass and fraction simultaneously. When varying both parameters with the other cosmological parameters, the usual Metropolis-Hastings algorithm performs poorly owing to the strong and rapidly-varying parameter degeneracies. Following Refs.~\cite{Hlozek2015ASearch, Winch_2024}, we instead use nested sampling as implemented in the \texttt{dynesty} software \cite{Speagle2020Dynesty}. We use a number of live points equal to 50 times the number of free parameters (eleven) and stop the sampling when the parameter \texttt{dlogz} quantifying the fraction of the Bayesian evidence left to sample falls below $0.001$.

The two-dimensional marginalized posterior in the $\Omega_\mathrm{a}/\Omega_\mathrm{d}-m_\mathrm{a}$ plane is shown in Fig.~\ref{fig:sub1}. We find the same monotonically increasing 95\% C.L. axion fraction limits that were found in the mass range $10^{-27}\;\mathrm{eV}\leq m_\mathrm{a}\leq 10^{-25}\;\mathrm{eV}$ in Ref.~\cite{Hlozek2015ASearch}. The axion fraction for mass around $10^{-26}$ eV in the nested sampling run differs slightly from the fixed-mass analysis of  Sec.~\ref{sec:param_inference}. This is due to the fact that, when varying the mass, we are not only comparing the ULA model to $\Lambda$CDM, but models of different axion masses to each other. In our case, most of the posterior probability is found around $m_\mathrm{a}\approx10^{-24.5}$ eV, strongly disfavoring the part of parameter space near $m_\mathrm{a}\approx 10^{-26}$ eV ULAs with masses differing by orders of magnitude behave very differently and have different degeneracies with $\Lambda$CDM parameters, as shown in Fig.~\ref{fig:four-masses-corner-main-text}. Therefore, naively marginalizing over the axion mass in our nested sampling approach would lead to spurious results that would be hard to interpret. In order to derive a bound on the axion fraction at the higher end of our axion mass range, we combine all samples with mass $m_\mathrm{a}\geq 10^{-25}$ eV since ULA models with $m_\mathrm{a} \sim 10^{-25}$ eV and $m_\mathrm{a} \sim 10^{-24.5}$ eV have similar degeneracies with cosmological parameters such as $\sigma_8$ (as seen in Fig.~\ref{fig:four-masses-corner-main-text}). When marginalizing over the axion mass limited to this range, we find that the preference for $\Omega_\mathrm{a}/\Omega_\mathrm{d}>0$ is exactly $2\sigma$ with a mean value of $5\%$. The marginalized posterior for this range is shown in Fig.~\ref{fig:sub2}, and the mean axion mass in this subsample is $10^{-24.7}$ eV. These results are again consistent with the fixed-mass results.

\subsection{Improvements over existing constraints \label{sec:other-probes}}

\begin{figure*}
    \centering
    \includegraphics[width=0.6\linewidth]{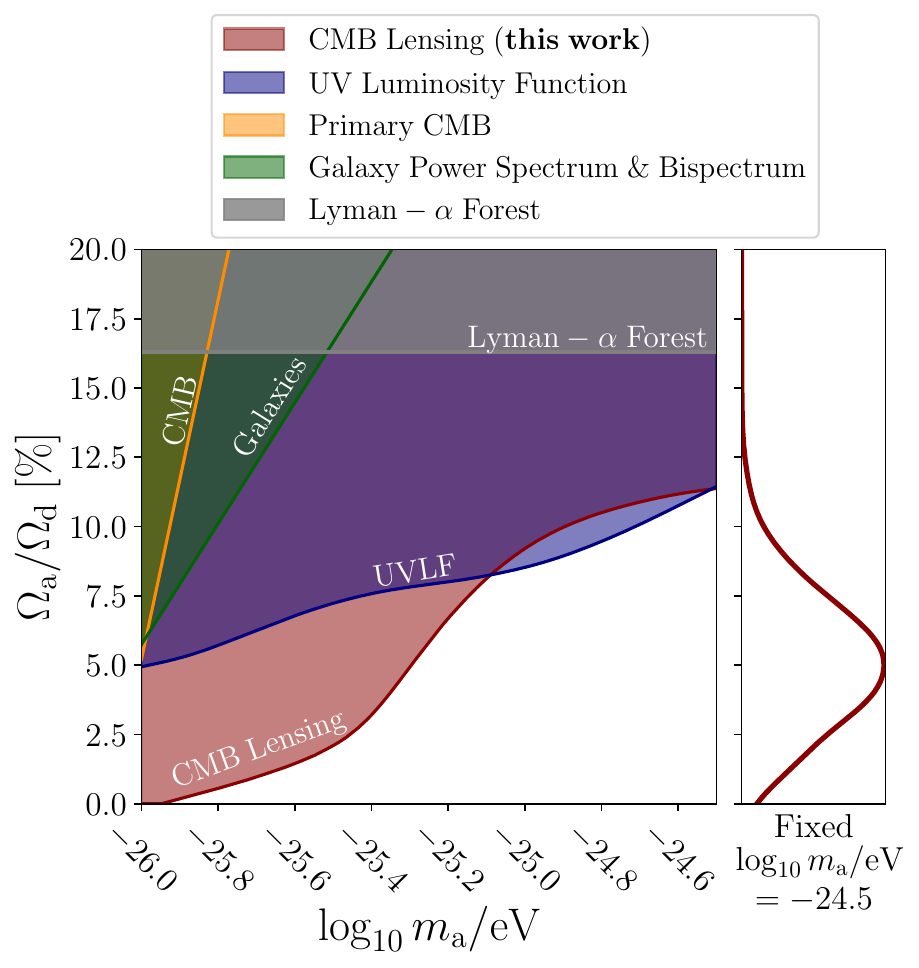}
    \caption{Comparison of constraints on the axion fraction as a function of particle mass for five different probes. The shaded areas denote the regions of parameter space excluded at 95\% C.L. The right-hand inset shows the marginalized posterior on the axion fraction in our fixed-mass analysis for $m_\mathrm{a}=10^{-24.5}$ eV, consistent with $\Omega_\mathrm{a}>0$ at $2.1\sigma$. A detailed description of each probe is in Sec.~\ref{sec:other-probes}.
    %\keir{What's the Lyman-alpha forest bound? How do you calculate the smooth lines?}
    }
    \label{fig:2d_many_probes}
\end{figure*}

ULAs have been the subject of studies involving a wide range of cosmological datasets. For the axion mass range considered in this work, there are five main probes considered in the literature. For the two lower mass bins, we find that CMB lensing improves the constraints on the axion fraction significantly when compared to other probes (Fig.~\ref{fig:2d_many_probes}). For axions at a mass of $10^{-26}\,\mathrm{eV}$, the addition of the lensing information improves the constraints by 72\% compared to the primary CMB without lensing~\cite{Calabrese2025ExtendedModels}. The axion masses above this range were largely unconstrained by the CMB and galaxy surveys. Our bound on ULAs with $m_\mathrm{a}=10^{-25}$ eV is 68\% tighter than the limit found using the Baryon acoustic Oscillation Spectroscopic Survey (BOSS) measurements of the galaxy power spectrum and bispectrum of Ref.~\cite{Rogers2023UltralightAxions}. Some of the strongest bounds on ULA density were found using the UVLF as measured by the Hubble Space Telescope and James Webb Space Telescope~\cite{Winch_2024}. As shown in Fig.~\ref{fig:2d_many_probes}, our constraints improve on these results for axions in the mass range $10^{-26}\;\mathrm{eV}\leq m_\mathrm{a}\lesssim 10^{-25.1}\;\mathrm{eV}$ and matches their constraining power for masses above that range. Modeling of the UVLF is  based on linear theory (in combination with an excursion-set approach), making it a great complement to the nonlinear small-scale lensing of the present work. Both the UVLF and these CMB lensing results include information about the primary CMB and are therefore not fully independent. Finally, the Lyman-$\alpha$ forest probes axion masses up to $\sim 10^{-20}$ eV, but the constraints on lower-mass ULA fractions are about $\Omega_\mathrm{a}/\Omega_\mathrm{d}\approx 16$\% for Ref.~\cite{Kobayashi2017LymanAlpha}. Ref.~\cite{Rogers20235Sigma} finds a preference for an axion fraction of $\sim (1 - 5) \%$ for ULAs in the mass range $10^{-26}\;\mathrm{eV}\lesssim m_\mathrm{a}\lesssim 10^{-24}\;\mathrm{eV}$. Ref.~\cite{Dentler2022FuzzyDark} ruled out axions for all the masses that we consider being all the dark matter using galaxy weak lensing data from the Dark Energy Survey. All of the constraints described above refer to the $2\sigma$ or 95\% C.L. bounds.

\begin{figure}
    \centering
    \includegraphics[width=\linewidth]{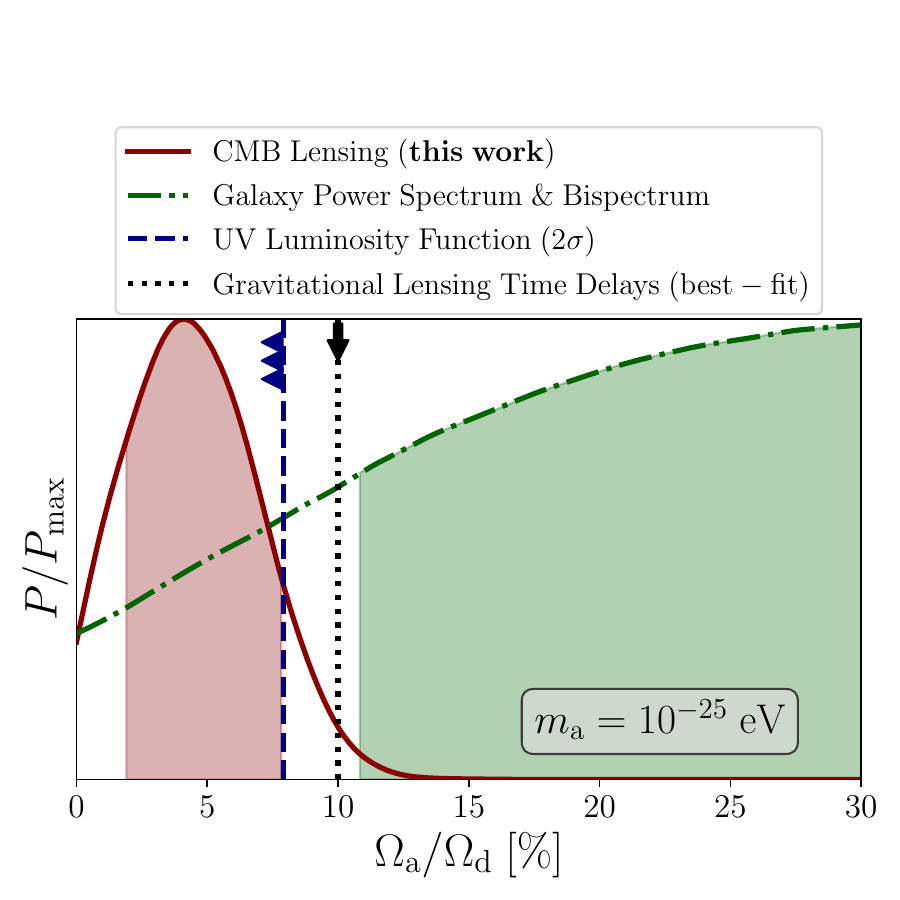}
    \caption{Comparison of constraints on the axion fraction for fixed-mass analyses at $m_\mathrm{a}=10^{-25}$ eV. The shaded areas indicate the 68\% credible intervals; the UVLF 95\% upper limit is indicated by the purple line. A detailed description of each probe is in Sec.~\ref{sec:other-probes}.}
    \label{fig:1e-25_combination}
\end{figure}

In Fig.~\ref{fig:1e-25_combination}, we compare the constraints from different probes on axions at a fixed mass of $10^{-25}$ eV. The UVLF limit derives from the same data as in Fig.~\ref{fig:2d_many_probes}, but with the analysis conducted by fixing \(m_\mathrm{a} = 10^{-25}\,\mathrm{eV}\). The galaxy survey posterior on $\Omega_\mathrm{a}$ is very wide for this mass due to conservative scale cuts that are applied (maximum wavenumbers $k_\mathrm{max} \lesssim 0.4\,h/$Mpc). The distribution is almost uniform, therefore the $1\sigma$ central region is located toward the center of the prior and should not be interpreted as a preference for axions. Finally, the gravitational lensing time delay line refers to the work of Ref.~\cite{Blum2021GravitationalLensing} that found that a ULA at a mass of $\sim 10^{-25}\;\mathrm{eV}$ and axion fraction of $\sim 10\%$ can generate galactic cores leading to an $H_0$ value closer to the one inferred from the CMB.

\subsection{Comparison with cold dark matter}

\begin{figure}
    \centering
    \includegraphics[width=\linewidth]{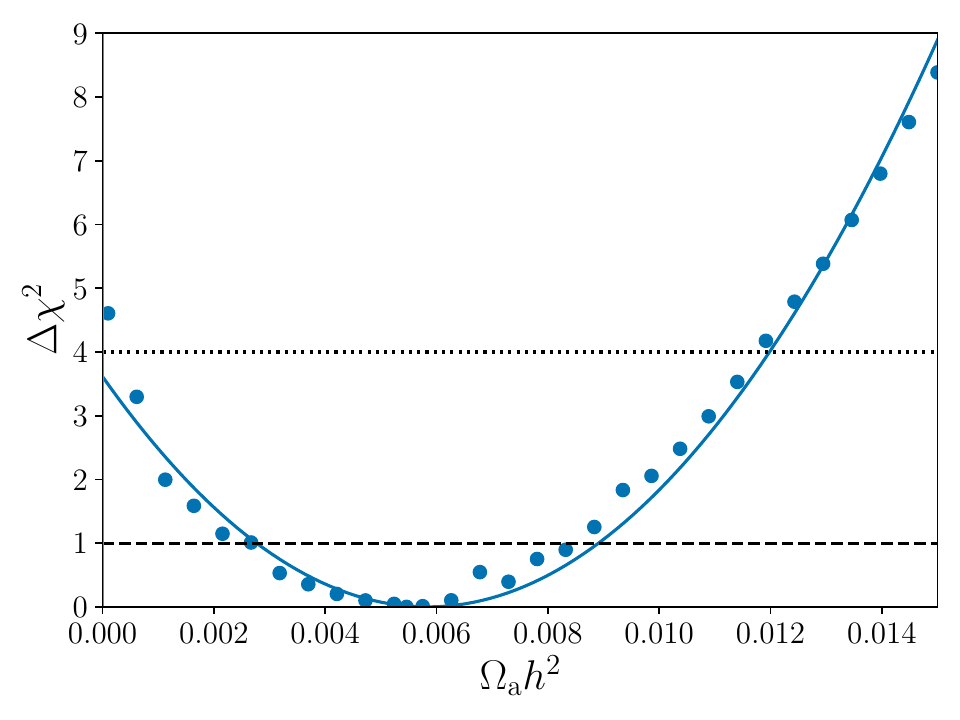}
    \caption{Change in the minimum $\chi^2$ for a fixed axion mass of $10^{-24.5}$ eV and varying axion density. The dashed and dotted lines respectively denote the frequentist thresholds for the 68\% and 95\% confidence limits on $\Omega_\mathrm{a}h^2$.}
    \label{fig:profile_like_curve}
\end{figure}

In order to investigate the preference for the non-zero axion density at $m_\mathrm{a}=10^{-24.5}$ eV, we perform a cross-check with a frequentist profile likelihood evaluation. This method performs well for models with a hard prior boundary such as massive neutrinos or dark-matter-radiation interactions \cite{Straight2026CMBConstraints}. First, we fix the axion mass to $10^{-24.5}$ eV and density close to $0$. We then minimize the $\chi^2$ by varying the $\Lambda$CDM cosmological parameters\footnote{We also minimize the nuisance parameters $A_\mathrm{Planck},\;A_\mathrm{ACT},\;P_\mathrm{ACT}$ and recover values very close to unity (their fiducial values) in our optimization procedure (see Ref.~\cite{ACTDR6Likelihood} for more details on foreground marginalization and nuisance parameters).}. We then repeat the process for linearly-spaced values of the axion density up to $\Omega_\mathrm{a}h^2=0.015$. We run an optimization run varying the axion density as well to obtain a reference minimum $\chi^2_\mathrm{min}$ at fixed axion mass. The change in the minimum chi-squared as a function of axion density $\Delta \chi^2 = \chi^2-\chi^2_\mathrm{min}$ is shown in Fig.~\ref{fig:profile_like_curve}. To this set of points, we fit a Gaussian of the form $e^{-\chi^2(\Omega_\mathrm{a}h^2)/2}$, which appears as a parabola in Fig.~\ref{fig:profile_like_curve}. Using this fit, we can derive the frequentist constraints on the axion density by isolating the values at which $\Delta \chi^2=[1,4]$ for the $[1\sigma,2\sigma]$ bounds. We arrive at similar results from our Bayesian approach with $\Omega_\mathrm{a}h^2 = 0.00584\pm 0.00614$ (95\% confidence uncertainties). It follows that the preference for $\Omega_\mathrm{a}h^2>0$ is $1.9 \sigma$, compared to $2.1\sigma$ from the marginalized posterior. This result is consistent with Ref.~\cite{Straight2026CMBConstraints} that found that the profile likelihood method leads to more conservative constraints on extended model parameters. We still find that the agreement between model and data is improved by the addition of axions composing $\sim 4.7\%$ of the DM. We analyze exactly which data points drive this effect in Appendix~\ref{sec:lensing-best}.

We now calculate the maximum likelihood point when varying all parameters including mass. The minimum $\chi^2$ point is at $m_\mathrm{a}=10^{-24.5}$ eV and has the same $\Delta \chi^2$ to $\Lambda$CDM as the first entry in Table~\ref{tab:best-fits}. Since our axion model introduces additional free parameters to the baseline $\Lambda$CDM, we use model comparison methods which penalize added complexity. The three models we compare are the no-axion case, the fixed axion mass case (at $10^{-24.5}$ eV) and the varying axion mass case. First, we calculate the Aikaike Information Criterion (AIC)~\cite{Aikaike1974AIC} from the best-fit point as 
\begin{align}
    \mathrm{AIC} = \chi^2_\mathrm{min} + 2k,
\end{align}
where $k$ is the number of free parameters. When compared to $\Lambda$CDM, we find $\Delta\mathrm{AIC}=0.79$ for the two-parameter extension and $\Delta\mathrm{AIC}=2.79$ when varying only the axion fraction. This is what we would expect since a lower $\Delta\mathrm{AIC}$ means the model provides a less statistically significant improvement over $\Lambda$CDM, accounting for the number of additional parameters. Given that the best-fit points for the two models are the same, the additional parameter increases the penalty to the model when varying the axion mass. In that case, the preference for axions becomes statistically insignificant when using the AIC. When varying only the axion mass, the AIC falls in the regime of ``weak evidence".

We also compute the Deviance Information Criterion (DIC)~\cite{Spiegelhalter2002DIC}. We obtain the DIC from the Metropolis-Hastings chains for $\Lambda$CDM and $m_\mathrm{a}=10^{-24.5}$ eV models using
\begin{align}
    \mathrm{DIC} = \overline{\chi^2(\theta)} -\chi^2(\overline{\theta}),
\end{align}
where the overbar denotes the mean and $\theta$ represents the sampled parameters. In this case, we find $\Delta\mathrm{DIC}=2.3$ which indicates ``weak evidence" for $\Omega_\mathrm{a}>0$, in agreement with the conclusion from the AIC in the fixed mass analysis.

\section{Discussion}\label{sec:discussion}
We have conducted a search for ULAs with CMB lensing in the nonlinear regime, using observations from \textit{Planck}, ACT, and SPT-3G. Our study extended the latest ACT study on beyond-$\Lambda$CDM models by including nonlinear modeling of ULAs and lensing of the CMB. Calibrated on the largest mixed ULA-CDM simulations in the literature, our model includes the nonlinear wave effects of axions on large-scale structure. It is incorporated into the \texttt{cosmopower} emulator framework allowing for very fast evaluation at the high accuracy settings required by the latest ACT dataset. We derive the strongest constraints on axions with mass $10^{-26}$ eV to $10^{-25}$ eV, and identify a slight preference for the existence of ULAs at a mass of $10^{-24.5}$ eV. This mild preference is driven by the lensing angular power spectrum being slightly higher than the $\Lambda$CDM best-fit around the multipole $L=1000$. This feature aligns with the increase in clustering in the matter power spectrum caused by the nonlinear wave effects, as seen in Fig.~\ref{fig:lensing-data}.

To complement our Metropolis-Hastings MCMC approach, we also run a nested sampling analysis while varying the mass and the fraction simultaneously. Integrating the posterior distribution found in the mass range $m_\mathrm{a}/\mathrm{eV}\in[10^{-25}, 10^{-24.5}]$, we find a $2\sigma$ preference for non-zero axion density with a mean mass in this interval of $10^{-24.7}$ eV. We further test our constraints for the heaviest mass bin by conducting a profile likelihood analysis. The constraints on the axion density in this case are in agreement with the other two approaches, indicating a preference of $1.9\sigma$ for ULAs. For \(10^{-26}\,\mathrm{eV} \leq m_\mathrm{a} \lesssim 10^{-25}\,\mathrm{eV}\), we set the strongest limits on the axion density in the literature.

The simulation set used for calibrating the nonlinear model is limited in some aspects. It includes only redshifts down to $z>1$ and was run for a single mass value of $10^{-24.5}$ eV (although the base \texttt{axionHMcode} also includes information from the first set of mixed DM simulations run over many decades of axion mass). This calibration point nonetheless corresponds exactly with the mass where a preference is identified. The enhancement on semi-linear scales combined with a suppression of the matter power spectrum in the linear regime is a distinct feature of axion models setting them apart from other non-cold DM models like warm DM. %Further, the axion fractions and mass preferred by CMB lensing were also independently identified by Lyman-$\alpha$ forest and weak lensing studies as potential way to reconcile lensing measurements from the CMB and low-redshift cosmic shear~\cite{Rogers2023UltralightAxions,Rogers20235Sigma,Perez2025ReconstructingThe}.

We caution about the possibility of the look elsewhere effect. Given that we repeat our analysis for four axion masses with slightly different signatures, we are effectively testing four models using thirty-eight data points (the combination of \textit{Planck}, ACT, and SPT-3G lensing data points). %The possibility of a random fluctuation driving this result therefore cannot be ruled out. 
However we find that the data points driving the preference (see Fig.~\ref{fig:lensing-data}) are identified in both ACT DR6 and SPT-3G and are uncorrelated.

The axion signature discussed in this work \cite[see also][]{Dome2025ImprovedHalo,Gaughan2026Ultralight} provides a new way to distinguish ULAs from other beyond-$\Lambda$CDM models. We anticipate that future CMB lensing measurements from ACT (including additional daytime CMB measurements), SPT, Simons Observatory \citep{SimonsObservatory:2025wwn}, and CMB-HD will greatly reduce the inferred uncertainties on the axion parameters and will help determine whether the improvement in $\chi^2$ from ULAs is physical or statistical. Combinations with linear probes of the matter power spectrum, such as the Lyman-$\alpha$ forest \citep{Rogers20235Sigma}, the UVLF from the James Webb Space Telescope \citep{Winch_2024} and the void size function \citep{London:2026yiu}, will provide a complementary probe of axions, independent of nonlinear axion effects. On semi-linear scales, full-shape power spectrum and bispectrum analyses of galaxy surveys could also constrain the shape of the matter power spectrum around the axion Jeans scale. In the fully nonlinear regime, studies of galaxy weak lensing will be particularly sensitive to the enhancement and suppression features of ULAs at $m_\mathrm{a}=10^{-24.5}$ eV \citep{Preston:2025tyl}. The analysis of these results will, however, require more advanced simulations for halo model calibration and potentially aggressive scale cuts to avoid contamination from baryonic feedback. In particular, including the suppression of the matter power spectrum caused by baryonic feedback could increase the preference for $\Omega_\mathrm{a}>0$ when combining CMB lensing and cosmic shear measurements. A wider suite of simulations will also prove critical to extend the axion mass range beyond $10^{-24.5}\;\mathrm{eV}$ for CMB lensing. Probes of physics on galactic scales such as strong gravitational lensing, stellar streams \citep{Ma:2025srn}, and Milky Way satellites will also provide an independent check of the wave effects on scales below the de Broglie wavelength for axions around $m_\mathrm{a}=10^{-24}$ eV. Finally, recent developments in optical atomic clocks could provide a laboratory-based test of ultralight bosons~\cite{Kim2024ProbingAn}. Our results help demonstrate the interplay between detector-based particle physics searches and observational cosmology.

\section{Acknowledgments}
We thank Doddy Marsh, Sophie Vogt, Jens Niemeyer, Yourong Frank Wang, Zun Wang, Bruce Partridge, Colin Hill, Edward J. Wollack, and Arthur B. Kosowsky for useful discussions. AL acknowledges support from NASA grant 21-ATP21-0145. KKR is supported by an Ernest Rutherford Fellowship from the UKRI Science and Technology Facilities Council (grant no. ST/Z510191/1). MM acknowledges support from NSF grant AST-2307727 and NASA grant 21-ATP21-0145. The Dunlap Institute is funded through an endowment established by the David Dunlap family and the University of Toronto. RH is supported by the Canadian National Science and Engineering Research Council Discovery Grant RGPIN-2025-06483 and the Arthur B. McDonald Fellowship SMFSU-60768. The authors at the University of Toronto acknowledge that the land on which the University of Toronto is built is the traditional territory of the Haudenosaunee, and most recently, the territory of the Mississaugas of the New Credit First Nation. They are grateful to have the opportunity to work in the community, on this territory. NS acknowledges support from DOE award number DE-SC0025309. Support for ACT was through the U.S.~National Science Foundation through awards AST-0408698, AST-0965625, and AST-1440226 for the ACT project, as well as awards PHY-0355328, PHY-0855887 and PHY-1214379. Funding was also provided by Princeton University, the University of Pennsylvania, and a Canada Foundation for Innovation (CFI) award to UBC. ACT operated in the Parque Astron\'omico Atacama in northern Chile under the auspices of the Agencia Nacional de Investigaci\'on y Desarrollo (ANID). The development of multichroic detectors and lenses was supported by NASA grants NNX13AE56G and NNX14AB58G. Detector research at NIST was supported by the NIST Innovations in Measurement Science program. Computing for ACT was performed using the Princeton Research Computing resources at Princeton University, the National Energy Research Scientific Computing Center (NERSC), and the Niagara supercomputer at the SciNet HPC Consortium. SciNet is funded by the CFI under the auspices of Compute Canada, the Government of Ontario, the Ontario Research Fund–Research Excellence, and the University of Toronto. We thank the Republic of Chile for hosting ACT in the northern Atacama, and the local indigenous Licanantay communities whom we follow in observing and learning from the night sky.

\bibliographystyle{apsrev.bst}
\bibliography{biblio}

@article{amon_2022,
    author = {Amon, Alexandra and Efstathiou, George},
    title = "{A non-linear solution to the S8 tension?}",
    journal = {Monthly Notices of the Royal Astronomical Society},
    volume = {516},
    number = {4},
    pages = {5355-5366},
    year = {2022},
    month = {09},
    issn = {0035-8711},
    doi = {10.1093/mnras/stac2429},
    url = {https://doi.org/10.1093/mnras/stac2429},
    eprint = {https://academic.oup.com/mnras/article-pdf/516/4/5355/46130148/stac2429.pdf},
}

@ARTICLE{MacFadden2026DNA,
       author = {{MacFadden}, Nate and {Schachner}, Andreas and {Sheridan}, Elijah},
        title = "{The DNA of Calabi─Yau Hypersurfaces: A Genetic Algorithm for Polytope Triangulations}",
      journal = {Fortschritte der Physik},
         year = 2026,
        month = feb,
       volume = {74},
        pages = {70060},
          doi = {10.1002/prop.70060},
archivePrefix = {arXiv},
       eprint = {2405.08871},
 primaryClass = {hep-th},
       adsurl = {https://ui.adsabs.harvard.edu/abs/2026ForPh..7470060M}
}

@article{Qu_2024,
doi = {10.3847/1538-4357/acfe06},
url = {https://dx.doi.org/10.3847/1538-4357/acfe06},
year = {2024},
month = {feb},
publisher = {The American Astronomical Society},
volume = {962},
number = {2},
pages = {112},
author = {Frank J. Qu and Blake D. Sherwin and Mathew S. Madhavacheril and Dongwon Han and Kevin T. Crowley and Irene Abril-Cabezas and Peter A. R. Ade and Simone Aiola and Tommy Alford and Mandana Amiri and Stefania Amodeo and Rui An and Zachary Atkins and Jason E. Austermann and Nicholas Battaglia and Elia Stefano Battistelli and James A. Beall and Rachel Bean and Benjamin Beringue and Tanay Bhandarkar and Emily Biermann and Boris Bolliet and J Richard Bond and Hongbo Cai and Erminia Calabrese and Victoria Calafut and Valentina Capalbo and Felipe Carrero and Julien Carron and Anthony Challinor and Grace E. Chesmore and Hsiao-mei Cho and Steve K. Choi and Susan E. Clark and Rodrigo Córdova Rosado and Nicholas F. Cothard and Kevin Coughlin and William Coulton and Roohi Dalal and Omar Darwish and Mark J. Devlin and Simon Dicker and Peter Doze and Cody J. Duell and Shannon M. Duff and Adriaan J. Duivenvoorden and Jo Dunkley and Rolando Dünner and Valentina Fanfani and Max Fankhanel and Gerrit Farren and Simone Ferraro and Rodrigo Freundt and Brittany Fuzia and Patricio A. Gallardo and Xavier Garrido and Vera Gluscevic and Joseph E. Golec and Yilun Guan and Mark Halpern and Ian Harrison and Matthew Hasselfield and Erin Healy and Shawn Henderson and Brandon Hensley and Carlos Hervías-Caimapo and J. Colin Hill and Gene C. Hilton and Matt Hilton and Adam D. Hincks and Renée Hložek and Shuay-Pwu Patty Ho and Zachary B. Huber and Johannes Hubmayr and Kevin M. Huffenberger and John P. Hughes and Kent Irwin and Giovanni Isopi and Hidde T. Jense and Ben Keller and Joshua Kim and Kenda Knowles and Brian J. Koopman and Arthur Kosowsky and Darby Kramer and Aleksandra Kusiak and Adrien La Posta and Alex Lague and Victoria Lakey and Eunseong Lee and Zack Li and Yaqiong Li and Michele Limon and Martine Lokken and Thibaut Louis and Marius Lungu and Niall MacCrann and Amanda MacInnis and Diego Maldonado and Felipe Maldonado and Maya Mallaby-Kay and Gabriela A. Marques and Jeff McMahon and Yogesh Mehta and Felipe Menanteau and Kavilan Moodley and Thomas W. Morris and Tony Mroczkowski and Sigurd Naess and Toshiya Namikawa and Federico Nati and Laura Newburgh and Andrina Nicola and Michael D. Niemack and Michael R. Nolta and John Orlowski-Scherer and Lyman A. Page and Shivam Pandey and Bruce Partridge and Heather Prince and Roberto Puddu and Federico Radiconi and Naomi Robertson and Felipe Rojas and Tai Sakuma and Maria Salatino and Emmanuel Schaan and Benjamin L. Schmitt and Neelima Sehgal and Shabbir Shaikh and Carlos Sierra and Jon Sievers and Cristóbal Sifón and Sara Simon and Rita Sonka and David N. Spergel and Suzanne T. Staggs and Emilie Storer and Eric R. Switzer and Niklas Tampier and Robert Thornton and Hy Trac and Jesse Treu and Carole Tucker and Joel Ullom and Leila R. Vale and Alexander Van Engelen and Jeff Van Lanen and Joshiwa van Marrewijk and Cristian Vargas and Eve M. Vavagiakis and Kasey Wagoner and Yuhan Wang and Lukas Wenzl and Edward J. Wollack and Zhilei Xu and Fernando Zago and Kaiwen Zheng},
title = {The Atacama Cosmology Telescope: A Measurement of the DR6 CMB Lensing Power Spectrum and Its Implications for Structure Growth},
journal = {The Astrophysical Journal}
}

@article{Madhavacheril_2024,
doi = {10.3847/1538-4357/acff5f},
url = {https://dx.doi.org/10.3847/1538-4357/acff5f},
year = {2024},
month = {feb},
publisher = {The American Astronomical Society},
volume = {962},
number = {2},
pages = {113},
author = {Mathew S. Madhavacheril and Frank J. Qu and Blake D. Sherwin and Niall MacCrann and Yaqiong Li and Irene Abril-Cabezas and Peter A. R. Ade and Simone Aiola and Tommy Alford and Mandana Amiri and Stefania Amodeo and Rui An and Zachary Atkins and Jason E. Austermann and Nicholas Battaglia and Elia Stefano Battistelli and James A. Beall and Rachel Bean and Benjamin Beringue and Tanay Bhandarkar and Emily Biermann and Boris Bolliet and J Richard Bond and Hongbo Cai and Erminia Calabrese and Victoria Calafut and Valentina Capalbo and Felipe Carrero and Anthony Challinor and Grace E. Chesmore and Hsiao-mei Cho and Steve K. Choi and Susan E. Clark and Rodrigo Córdova Rosado and Nicholas F. Cothard and Kevin Coughlin and William Coulton and Kevin T. Crowley and Roohi Dalal and Omar Darwish and Mark J. Devlin and Simon Dicker and Peter Doze and Cody J. Duell and Shannon M. Duff and Adriaan J. Duivenvoorden and Jo Dunkley and Rolando Dünner and Valentina Fanfani and Max Fankhanel and Gerrit Farren and Simone Ferraro and Rodrigo Freundt and Brittany Fuzia and Patricio A. Gallardo and Xavier Garrido and Jahmour Givans and Vera Gluscevic and Joseph E. Golec and Yilun Guan and Kirsten R. Hall and Mark Halpern and Dongwon Han and Ian Harrison and Matthew Hasselfield and Erin Healy and Shawn Henderson and Brandon Hensley and Carlos Hervías-Caimapo and J. Colin Hill and Gene C. Hilton and Matt Hilton and Adam D. Hincks and Renée Hložek and Shuay-Pwu Patty Ho and Zachary B. Huber and Johannes Hubmayr and Kevin M. Huffenberger and John P. Hughes and Kent Irwin and Giovanni Isopi and Hidde T. Jense and Ben Keller and Joshua Kim and Kenda Knowles and Brian J. Koopman and Arthur Kosowsky and Darby Kramer and Aleksandra Kusiak and Adrien La Posta and Alex Lague and Victoria Lakey and Eunseong Lee and Zack Li and Michele Limon and Martine Lokken and Thibaut Louis and Marius Lungu and Amanda MacInnis and Diego Maldonado and Felipe Maldonado and Maya Mallaby-Kay and Gabriela A. Marques and Jeff McMahon and Yogesh Mehta and Felipe Menanteau and Kavilan Moodley and Thomas W. Morris and Tony Mroczkowski and Sigurd Naess and Toshiya Namikawa and Federico Nati and Laura Newburgh and Andrina Nicola and Michael D. Niemack and Michael R. Nolta and John Orlowski-Scherer and Lyman A. Page and Shivam Pandey and Bruce Partridge and Heather Prince and Roberto Puddu and Federico Radiconi and Naomi Robertson and Felipe Rojas and Tai Sakuma and Maria Salatino and Emmanuel Schaan and Benjamin L. Schmitt and Neelima Sehgal and Shabbir Shaikh and Carlos Sierra and Jon Sievers and Cristóbal Sifón and Sara Simon and Rita Sonka and David N. Spergel and Suzanne T. Staggs and Emilie Storer and Eric R. Switzer and Niklas Tampier and Robert Thornton and Hy Trac and Jesse Treu and Carole Tucker and Joel Ullom and Leila R. Vale and Alexander Van Engelen and Jeff Van Lanen and Joshiwa van Marrewijk and Cristian Vargas and Eve M. Vavagiakis and Kasey Wagoner and Yuhan Wang and Lukas Wenzl and Edward J. Wollack and Zhilei Xu and Fernando Zago and Kaiwen Zheng},
title = {The Atacama Cosmology Telescope: DR6 Gravitational Lensing Map and Cosmological Parameters},
journal = {The Astrophysical Journal}
}

@article{CAMB,
doi = {10.1086/309179},
url = {https://dx.doi.org/10.1086/309179},
year = {2000},
month = {aug},
publisher = {},
volume = {538},
number = {2},
pages = {473},
author = {Antony Lewis and Anthony Challinor and Anthony Lasenby},
title = {Efficient Computation of Cosmic Microwave Background Anisotropies in
Closed Friedmann-Robertson-Walker Models},
journal = {The Astrophysical Journal}
}

@article{Mead_2020,
   title={<scp>hmcode-2020</scp>: improved modelling of non-linear cosmological power spectra with baryonic feedback},
   volume={502},
   ISSN={1365-2966},
   url={http://dx.doi.org/10.1093/mnras/stab082},
   DOI={10.1093/mnras/stab082},
   number={1},
   journal={Monthly Notices of the Royal Astronomical Society},
   publisher={Oxford University Press (OUP)},
   author={Mead, A J and Brieden, S and Tröster, T and Heymans, C},
   year={2021},
   month=jan, pages={1401–1422} }

@article{Rogers20235Sigma,
    author = "Rogers, Keir K. and Poulin, Vivian",
    title = "{5{\ensuremath{\sigma}} tension between Planck cosmic microwave background and eBOSS Lyman-alpha forest and constraints on physics beyond {\ensuremath{\Lambda}}CDM}",
    eprint = "2311.16377",
    archivePrefix = "arXiv",
    primaryClass = "astro-ph.CO",
    doi = "10.1103/PhysRevResearch.7.L012018",
    journal = "Phys. Rev. Res.",
    volume = "7",
    number = "1",
    pages = "L012018",
    year = "2025"
}

@ARTICLE{Gendler2024QCDAxion,
       author = {{Gendler}, Naomi and {Marsh}, David J.~E.},
        title = "{QCD Axion Dark Matter in String Theory: Haloscopes and Helioscopes as Probes of the Landscape}",
      journal = {arXiv e-prints},
         year = 2024,
        month = jul,
          eid = {arXiv:2407.07143},
        pages = {arXiv:2407.07143},
          doi = {10.48550/arXiv.2407.07143},
archivePrefix = {arXiv},
       eprint = {2407.07143},
 primaryClass = {hep-th},
       adsurl = {https://ui.adsabs.harvard.edu/abs/2024arXiv240707143G}
}

@ARTICLE{Vogt2023ImprovedMixed,
       author = {{Vogt}, Sophie M.~L. and {Marsh}, David J.~E. and {Lagu{\"e}}, Alex},
        title = "{Improved mixed dark matter halo model for ultralight axions}",
      journal = {\prd},
         year = 2023,
        month = mar,
       volume = {107},
       number = {6},
          eid = {063526},
        pages = {063526},
          doi = {10.1103/PhysRevD.107.063526},
archivePrefix = {arXiv},
       eprint = {2209.13445},
 primaryClass = {astro-ph.CO},
       adsurl = {https://ui.adsabs.harvard.edu/abs/2023PhRvD.107f3526V}
}

@ARTICLE{Hlozek2015ASearch,
       author = {{Hlozek}, Ren{\'e}e and {Grin}, Daniel and {Marsh}, David J.~E. and {Ferreira}, Pedro G.},
        title = "{A search for ultralight axions using precision cosmological data}",
      journal = {\prd},
         year = 2015,
        month = may,
       volume = {91},
       number = {10},
          eid = {103512},
        pages = {103512},
          doi = {10.1103/PhysRevD.91.103512},
archivePrefix = {arXiv},
       eprint = {1410.2896},
 primaryClass = {astro-ph.CO},
       adsurl = {https://ui.adsabs.harvard.edu/abs/2015PhRvD..91j3512H}
}

@ARTICLE{Lague2022ConstrainingUltralight,
       author = {{Lagu{\"e}}, A. and {Bond}, J.~R. and {Hlo{\v{z}}ek}, R. and {Rogers}, K.~K. and {Marsh}, D.~J.~E. and {Grin}, D.},
        title = "{Constraining ultralight axions with galaxy surveys}",
      journal = {\jcap},
         year = 2022,
        month = jan,
       volume = {2022},
       number = {1},
          eid = {049},
        pages = {049},
          doi = {10.1088/1475-7516/2022/01/049},
archivePrefix = {arXiv},
       eprint = {2104.07802},
 primaryClass = {astro-ph.CO},
       adsurl = {https://ui.adsabs.harvard.edu/abs/2022JCAP...01..049L}
}

@ARTICLE{Rogers2023UltralightAxions,
       author = {{Rogers}, Keir K. and {Hlo{\v{z}}ek}, Ren{\'e}e and {Lagu{\"e}}, Alex and {Ivanov}, Mikhail M. and {Philcox}, Oliver H.~E. and {Cabass}, Giovanni and {Akitsu}, Kazuyuki and {Marsh}, David J.~E.},
        title = "{Ultra-light axions and the S $_{8}$ tension: joint constraints from the cosmic microwave background and galaxy clustering}",
      journal = {\jcap},
         year = 2023,
        month = jun,
       volume = {2023},
       number = {6},
          eid = {023},
        pages = {023},
          doi = {10.1088/1475-7516/2023/06/023},
archivePrefix = {arXiv},
       eprint = {2301.08361},
 primaryClass = {astro-ph.CO},
       adsurl = {https://ui.adsabs.harvard.edu/abs/2023JCAP...06..023R}
}

@ARTICLE{Planck2018Lensing,
       author = {{Planck Collaboration} and {Aghanim}, N. and {Akrami}, Y. and {Ashdown}, M. and {Aumont}, J. and {Baccigalupi}, C. and {Ballardini}, M. and {Banday}, A.~J. and {Barreiro}, R.~B. and {Bartolo}, N. and {Basak}, S. and {Benabed}, K. and {Bernard}, J. -P. and {Bersanelli}, M. and {Bielewicz}, P. and {Bock}, J.~J. and {Bond}, J.~R. and {Borrill}, J. and {Bouchet}, F.~R. and {Boulanger}, F. and {Bucher}, M. and {Burigana}, C. and {Calabrese}, E. and {Cardoso}, J. -F. and {Carron}, J. and {Challinor}, A. and {Chiang}, H.~C. and {Colombo}, L.~P.~L. and {Combet}, C. and {Crill}, B.~P. and {Cuttaia}, F. and {de Bernardis}, P. and {de Zotti}, G. and {Delabrouille}, J. and {Di Valentino}, E. and {Diego}, J.~M. and {Dor{\'e}}, O. and {Douspis}, M. and {Ducout}, A. and {Dupac}, X. and {Efstathiou}, G. and {Elsner}, F. and {En{\ss}lin}, T.~A. and {Eriksen}, H.~K. and {Fantaye}, Y. and {Fernandez-Cobos}, R. and {Finelli}, F. and {Forastieri}, F. and {Frailis}, M. and {Fraisse}, A.~A. and {Franceschi}, E. and {Frolov}, A. and {Galeotta}, S. and {Galli}, S. and {Ganga}, K. and {G{\'e}nova-Santos}, R.~T. and {Gerbino}, M. and {Ghosh}, T. and {Gonz{\'a}lez-Nuevo}, J. and {G{\'o}rski}, K.~M. and {Gratton}, S. and {Gruppuso}, A. and {Gudmundsson}, J.~E. and {Hamann}, J. and {Handley}, W. and {Hansen}, F.~K. and {Herranz}, D. and {Hivon}, E. and {Huang}, Z. and {Jaffe}, A.~H. and {Jones}, W.~C. and {Karakci}, A. and {Keih{\"a}nen}, E. and {Keskitalo}, R. and {Kiiveri}, K. and {Kim}, J. and {Knox}, L. and {Krachmalnicoff}, N. and {Kunz}, M. and {Kurki-Suonio}, H. and {Lagache}, G. and {Lamarre}, J. -M. and {Lasenby}, A. and {Lattanzi}, M. and {Lawrence}, C.~R. and {Le Jeune}, M. and {Levrier}, F. and {Lewis}, A. and {Liguori}, M. and {Lilje}, P.~B. and {Lindholm}, V. and {L{\'o}pez-Caniego}, M. and {Lubin}, P.~M. and {Ma}, Y. -Z. and {Mac{\'\i}as-P{\'e}rez}, J.~F. and {Maggio}, G. and {Maino}, D. and {Mandolesi}, N. and {Mangilli}, A. and {Marcos-Caballero}, A. and {Maris}, M. and {Martin}, P.~G. and {Mart{\'\i}nez-Gonz{\'a}lez}, E. and {Matarrese}, S. and {Mauri}, N. and {McEwen}, J.~D. and {Melchiorri}, A. and {Mennella}, A. and {Migliaccio}, M. and {Miville-Desch{\^e}nes}, M. -A. and {Molinari}, D. and {Moneti}, A. and {Montier}, L. and {Morgante}, G. and {Moss}, A. and {Natoli}, P. and {Pagano}, L. and {Paoletti}, D. and {Partridge}, B. and {Patanchon}, G. and {Perrotta}, F. and {Pettorino}, V. and {Piacentini}, F. and {Polastri}, L. and {Polenta}, G. and {Puget}, J. -L. and {Rachen}, J.~P. and {Reinecke}, M. and {Remazeilles}, M. and {Renzi}, A. and {Rocha}, G. and {Rosset}, C. and {Roudier}, G. and {Rubi{\~n}o-Mart{\'\i}n}, J.~A. and {Ruiz-Granados}, B. and {Salvati}, L. and {Sandri}, M. and {Savelainen}, M. and {Scott}, D. and {Sirignano}, C. and {Sunyaev}, R. and {Suur-Uski}, A. -S. and {Tauber}, J.~A. and {Tavagnacco}, D. and {Tenti}, M. and {Toffolatti}, L. and {Tomasi}, M. and {Trombetti}, T. and {Valiviita}, J. and {Van Tent}, B. and {Vielva}, P. and {Villa}, F. and {Vittorio}, N. and {Wandelt}, B.~D. and {Wehus}, I.~K. and {White}, M. and {White}, S.~D.~M. and {Zacchei}, A. and {Zonca}, A.},
        title = "{Planck 2018 results. VIII. Gravitational lensing}",
      journal = {\aap},
         year = 2020,
        month = sep,
       volume = {641},
          eid = {A8},
        pages = {A8},
          doi = {10.1051/0004-6361/201833886},
archivePrefix = {arXiv},
       eprint = {1807.06210},
 primaryClass = {astro-ph.CO},
       adsurl = {https://ui.adsabs.harvard.edu/abs/2020A&A...641A...8P}
}

@ARTICLE{Dome2025ImprovedHalo,
       author = {{Dome}, Tibor and {May}, Simon and {Lagu{\"e}}, Alex and {Marsh}, David J.~E. and {Johnston}, Sarah and {Bose}, Sownak and {Tocher}, Alex and {Fialkov}, Anastasia},
        title = "{Improved halo model calibrations for mixed dark matter models of ultralight axions}",
      journal = {\mnras},
         year = 2025,
        month = feb,
       volume = {537},
       number = {1},
        pages = {252-271},
          doi = {10.1093/mnras/staf005},
archivePrefix = {arXiv},
       eprint = {2409.11469},
 primaryClass = {astro-ph.CO},
       adsurl = {https://ui.adsabs.harvard.edu/abs/2025MNRAS.537..252D}
}

@ARTICLE{Preskill_1983,
       author = {{Preskill}, John and {Wise}, Mark B. and {Wilczek}, Frank},
        title = "{Cosmology of the invisible axion}",
      journal = {Physics Letters B},
         year = 1983,
        month = jan,
       volume = {120},
       number = {1-3},
        pages = {127-132},
          doi = {10.1016/0370-2693(83)90637-8},
       adsurl = {https://ui.adsabs.harvard.edu/abs/1983PhLB..120..127P}
}

@ARTICLE{Dine_1983,
       author = {{Dine}, Michael and {Fischler}, Willy},
        title = "{The not-so-harmless axion}",
      journal = {Physics Letters B},
         year = 1983,
        month = jan,
       volume = {120},
       number = {1-3},
        pages = {137-141},
          doi = {10.1016/0370-2693(83)90639-1},
       adsurl = {https://ui.adsabs.harvard.edu/abs/1983PhLB..120..137D}
}

@ARTICLE{Abbott_1983,
       author = {{Abbott}, L.~F. and {Sikivie}, P.},
        title = "{A cosmological bound on the invisible axion}",
      journal = {Physics Letters B},
         year = 1983,
        month = jan,
       volume = {120},
       number = {1-3},
        pages = {133-136},
          doi = {10.1016/0370-2693(83)90638-X},
       adsurl = {https://ui.adsabs.harvard.edu/abs/1983PhLB..120..133A}
}

@article{Peccei1977CP,
    author = "Peccei, R. D. and Quinn, Helen R.",
    title = "{CP Conservation in the Presence of Instantons}",
    reportNumber = "ITP-568-STANFORD",
    doi = "10.1103/PhysRevLett.38.1440",
    journal = "Phys. Rev. Lett.",
    volume = "38",
    pages = "1440--1443",
    year = "1977"
}

@ARTICLE{Svrcek_2006,
       author = {{Svrcek}, Peter and {Witten}, Edward},
        title = "{Axions in string theory}",
      journal = {Journal of High Energy Physics},
         year = 2006,
        month = jun,
       volume = {2006},
       number = {6},
          eid = {051},
        pages = {051},
          doi = {10.1088/1126-6708/2006/06/051},
archivePrefix = {arXiv},
       eprint = {hep-th/0605206},
 primaryClass = {hep-th},
       adsurl = {https://ui.adsabs.harvard.edu/abs/2006JHEP...06..051S}
}

@ARTICLE{Arvanitaki_2010,
       author = {{Arvanitaki}, Asimina and {Dimopoulos}, Savas and {Dubovsky}, Sergei and {Kaloper}, Nemanja and {March-Russell}, John},
        title = "{String axiverse}",
      journal = {\prd},
         year = 2010,
        month = jun,
       volume = {81},
       number = {12},
          eid = {123530},
        pages = {123530},
          doi = {10.1103/PhysRevD.81.123530},
archivePrefix = {arXiv},
       eprint = {0905.4720},
 primaryClass = {hep-th},
       adsurl = {https://ui.adsabs.harvard.edu/abs/2010PhRvD..81l3530A}
}

@article{Sheridan2024FuzzyAxions,
    author = "Sheridan, Elijah and Carta, Federico and Gendler, Naomi and Jain, Mudit and Marsh, David J. E. and McAllister, Liam and Righi, Nicole and Rogers, Keir K. and Schachner, Andreas",
    title = "{Fuzzy axions and associated relics}",
    eprint = "2412.12012",
    archivePrefix = "arXiv",
    primaryClass = "hep-th",
    reportNumber = "KCL-PH-TH/2024-75, KCL-PH-TH/2024-75",
    doi = "10.1007/JHEP09(2025)016",
    journal = "JHEP",
    volume = "09",
    pages = "016",
    year = "2025"
}

@ARTICLE{Gendler2024GlimmersOf,
       author = {{Gendler}, Naomi and {Marsh}, David J.~E. and {McAllister}, Liam and {Moritz}, Jakob},
        title = "{Glimmers from the axiverse}",
      journal = {\jcap},
         year = 2024,
        month = sep,
       volume = {2024},
       number = {9},
          eid = {071},
        pages = {071},
          doi = {10.1088/1475-7516/2024/09/071},
archivePrefix = {arXiv},
       eprint = {2309.13145},
 primaryClass = {hep-th},
       adsurl = {https://ui.adsabs.harvard.edu/abs/2024JCAP...09..071G}
}

@ARTICLE{Mehta_2021,
       author = {{Mehta}, Viraf M. and {Demirtas}, Mehmet and {Long}, Cody and {Marsh}, David J.~E. and {McAllister}, Liam and {Stott}, Matthew J.},
        title = "{Superradiance in string theory}",
      journal = {\jcap},
         year = 2021,
        month = jul,
       volume = {2021},
       number = {7},
          eid = {033},
        pages = {033},
          doi = {10.1088/1475-7516/2021/07/033},
archivePrefix = {arXiv},
       eprint = {2103.06812},
 primaryClass = {hep-th},
       adsurl = {https://ui.adsabs.harvard.edu/abs/2021JCAP...07..033M}
}

@ARTICLE{Bachlechner_2017,
       author = {{Bachlechner}, Thomas C. and {Eckerle}, Kate and {Janssen}, Oliver and {Kleban}, Matthew},
        title = "{Multiple-axion framework}",
      journal = {\prd},
         year = 2018,
        month = sep,
       volume = {98},
       number = {6},
          eid = {061301},
        pages = {061301},
          doi = {10.1103/PhysRevD.98.061301},
       adsurl = {https://ui.adsabs.harvard.edu/abs/2018PhRvD..98f1301B}
}

@article{Winch_2024,
    author = "Winch, Harrison and Rogers, Keir K. and Hlo{\v{z}}ek, Ren{\'e}e and Marsh, David J. E.",
    title = "{High-redshift, Small-scale Tests of Ultralight Axion Dark Matter Using Hubble and Webb Galaxy UV Luminosities}",
    eprint = "2404.11071",
    archivePrefix = "arXiv",
    primaryClass = "astro-ph.CO",
    doi = "10.3847/1538-4357/ad7a73",
    journal = "Astrophys. J.",
    volume = "976",
    number = "1",
    pages = "40",
    year = "2024"
}

@ARTICLE{Massara2014TheHalo,
       author = {{Massara}, Elena and {Villaescusa-Navarro}, Francisco and {Viel}, Matteo},
        title = "{The halo model in a massive neutrino cosmology}",
      journal = {\jcap},
         year = 2014,
        month = dec,
       volume = {2014},
       number = {12},
        pages = {053-053},
          doi = {10.1088/1475-7516/2014/12/053},
archivePrefix = {arXiv},
       eprint = {1410.6813},
 primaryClass = {astro-ph.CO},
       adsurl = {https://ui.adsabs.harvard.edu/abs/2014JCAP...12..053M}
}

@ARTICLE{Trendafilova2025TheEnd,
       author = {{Trendafilova}, Cynthia and {Khalife}, Ali Rida and {Galli}, Silvia},
        title = "{The end of easy phenomenology for CMB experiments: A case study in the dark sector}",
      journal = {\jcap},
         year = 2025,
        month = may,
       volume = {2025},
       number = {5},
          eid = {094},
        pages = {094},
          doi = {10.1088/1475-7516/2025/05/094},
archivePrefix = {arXiv},
       eprint = {2502.19383},
 primaryClass = {astro-ph.CO},
       adsurl = {https://ui.adsabs.harvard.edu/abs/2025JCAP...05..094T}
}

@ARTICLE{McCarthy_2022,
       author = {{McCarthy}, Fiona and {Hill}, J. Colin and {Madhavacheril}, Mathew S.},
        title = "{Baryonic feedback biases on fundamental physics from lensed CMB power spectra}",
      journal = {\prd},
         year = 2022,
        month = jan,
       volume = {105},
       number = {2},
          eid = {023517},
        pages = {023517},
          doi = {10.1103/PhysRevD.105.023517},
archivePrefix = {arXiv},
       eprint = {2103.05582},
 primaryClass = {astro-ph.CO},
       adsurl = {https://ui.adsabs.harvard.edu/abs/2022PhRvD.105b3517M}
}

@ARTICLE{Mancini2022COSMOPOWER,
       author = {{Spurio Mancini}, Alessio and {Piras}, Davide and {Alsing}, Justin and {Joachimi}, Benjamin and {Hobson}, Michael P.},
        title = "{COSMOPOWER: emulating cosmological power spectra for accelerated Bayesian inference from next-generation surveys}",
      journal = {\mnras},
         year = 2022,
        month = apr,
       volume = {511},
       number = {2},
        pages = {1771-1788},
          doi = {10.1093/mnras/stac064},
archivePrefix = {arXiv},
       eprint = {2106.03846},
 primaryClass = {astro-ph.CO},
       adsurl = {https://ui.adsabs.harvard.edu/abs/2022MNRAS.511.1771S}
}

@ARTICLE{Hu2000WeakLensing,
       author = {{Hu}, Wayne},
        title = "{Weak lensing of the CMB: A harmonic approach}",
      journal = {\prd},
         year = 2000,
        month = aug,
       volume = {62},
       number = {4},
          eid = {043007},
        pages = {043007},
          doi = {10.1103/PhysRevD.62.043007},
archivePrefix = {arXiv},
       eprint = {astro-ph/0001303},
 primaryClass = {astro-ph},
       adsurl = {https://ui.adsabs.harvard.edu/abs/2000PhRvD..62d3007H}
}

@ARTICLE{Lewis2006WeakGravitational,
       author = {{Lewis}, Antony and {Challinor}, Anthony},
        title = "{Weak gravitational lensing of the CMB}",
      journal = {\physrep},
         year = 2006,
        month = jun,
       volume = {429},
       number = {1},
        pages = {1-65},
          doi = {10.1016/j.physrep.2006.03.002},
archivePrefix = {arXiv},
       eprint = {astro-ph/0601594},
 primaryClass = {astro-ph},
       adsurl = {https://ui.adsabs.harvard.edu/abs/2006PhR...429....1L}
}

@ARTICLE{Stott2017Spectrum,
       author = {{Stott}, Matthew J. and {Marsh}, David J.~E. and {Pongkitivanichkul}, Chakrit and {Price}, Layne C. and {Acharya}, Bobby S.},
        title = "{Spectrum of the axion dark sector}",
      journal = {\prd},
         year = 2017,
        month = oct,
       volume = {96},
       number = {8},
          eid = {083510},
        pages = {083510},
          doi = {10.1103/PhysRevD.96.083510},
archivePrefix = {arXiv},
       eprint = {1706.03236},
 primaryClass = {astro-ph.CO},
       adsurl = {https://ui.adsabs.harvard.edu/abs/2017PhRvD..96h3510S}
}

@ARTICLE{Kobayashi2017LymanAlpha,
       author = {{Kobayashi}, Takeshi and {Murgia}, Riccardo and {De Simone}, Andrea and {Ir{\v{s}}i{\v{c}}}, Vid and {Viel}, Matteo},
        title = "{Lyman-{\ensuremath{\alpha}} constraints on ultralight scalar dark matter: Implications for the early and late universe}",
      journal = {\prd},
         year = 2017,
        month = dec,
       volume = {96},
       number = {12},
          eid = {123514},
        pages = {123514},
          doi = {10.1103/PhysRevD.96.123514},
archivePrefix = {arXiv},
       eprint = {1708.00015},
 primaryClass = {astro-ph.CO},
       adsurl = {https://ui.adsabs.harvard.edu/abs/2017PhRvD..96l3514K}
}

@ARTICLE{Armengaud2017ConstrainingThe,
       author = {{Armengaud}, Eric and {Palanque-Delabrouille}, Nathalie and {Y{\`e}che}, Christophe and {Marsh}, David J.~E. and {Baur}, Julien},
        title = "{Constraining the mass of light bosonic dark matter using SDSS Lyman-{\ensuremath{\alpha}} forest}",
      journal = {\mnras},
         year = 2017,
        month = nov,
       volume = {471},
       number = {4},
        pages = {4606-4614},
          doi = {10.1093/mnras/stx1870},
archivePrefix = {arXiv},
       eprint = {1703.09126},
 primaryClass = {astro-ph.CO},
       adsurl = {https://ui.adsabs.harvard.edu/abs/2017MNRAS.471.4606A}
}

@ARTICLE{Irsic2017FirstConstraints,
       author = {{Ir{\v{s}}i{\v{c}}}, Vid and {Viel}, Matteo and {Haehnelt}, Martin G. and {Bolton}, James S. and {Becker}, George D.},
        title = "{First Constraints on Fuzzy Dark Matter from Lyman-{\ensuremath{\alpha}} Forest Data and Hydrodynamical Simulations}",
      journal = {\prl},
         year = 2017,
        month = jul,
       volume = {119},
       number = {3},
          eid = {031302},
        pages = {031302},
          doi = {10.1103/PhysRevLett.119.031302},
archivePrefix = {arXiv},
       eprint = {1703.04683},
 primaryClass = {astro-ph.CO},
       adsurl = {https://ui.adsabs.harvard.edu/abs/2017PhRvL.119c1302I}
}

@ARTICLE{Rogers2021StrongBound,
       author = {{Rogers}, Keir K. and {Peiris}, Hiranya V.},
        title = "{Strong Bound on Canonical Ultralight Axion Dark Matter from the Lyman-Alpha Forest}",
      journal = {\prl},
         year = 2021,
        month = feb,
       volume = {126},
       number = {7},
          eid = {071302},
        pages = {071302},
          doi = {10.1103/PhysRevLett.126.071302},
archivePrefix = {arXiv},
       eprint = {2007.12705},
 primaryClass = {astro-ph.CO},
       adsurl = {https://ui.adsabs.harvard.edu/abs/2021PhRvL.126g1302R}
}

@ARTICLE{Lague2024CosmologicalSimulations,
       author = {{Lagu{\"e}}, Alex and {Schwabe}, Bodo and {Hlo{\v{z}}ek}, Ren{\'e}e and {Marsh}, David J.~E. and {Rogers}, Keir K.},
        title = "{Cosmological simulations of mixed ultralight dark matter}",
      journal = {\prd},
         year = 2024,
        month = feb,
       volume = {109},
       number = {4},
          eid = {043507},
        pages = {043507},
          doi = {10.1103/PhysRevD.109.043507},
archivePrefix = {arXiv},
       eprint = {2310.20000},
 primaryClass = {astro-ph.CO},
       adsurl = {https://ui.adsabs.harvard.edu/abs/2024PhRvD.109d3507L}
}

@ARTICLE{Bolliet2024HighAccuracy,
       author = {{Bolliet}, Boris and {Spurio Mancini}, Alessio and {Hill}, J. Colin and {Madhavacheril}, Mathew and {Jense}, Hidde T. and {Calabrese}, Erminia and {Dunkley}, Jo},
        title = "{High-accuracy emulators for observables in {\ensuremath{\Lambda}}CDM, N$_{eff}$, {\ensuremath{\Sigma}}m$_{{\ensuremath{\nu}}}$, and w cosmologies}",
      journal = {\mnras},
         year = 2024,
        month = jun,
       volume = {531},
       number = {1},
        pages = {1351-1370},
          doi = {10.1093/mnras/stae1201},
archivePrefix = {arXiv},
       eprint = {2303.01591},
 primaryClass = {astro-ph.CO},
       adsurl = {https://ui.adsabs.harvard.edu/abs/2024MNRAS.531.1351B}
}

@ARTICLE{Dentler2022FuzzyDark,
       author = {{Dentler}, Mona and {Marsh}, David J.~E. and {Hlo{\v{z}}ek}, Ren{\'e}e and {Lagu{\"e}}, Alex and {Rogers}, Keir K. and {Grin}, Daniel},
        title = "{Fuzzy dark matter and the Dark Energy Survey Year 1 data}",
      journal = {\mnras},
         year = 2022,
        month = oct,
       volume = {515},
       number = {4},
        pages = {5646-5664},
          doi = {10.1093/mnras/stac1946},
archivePrefix = {arXiv},
       eprint = {2111.01199},
 primaryClass = {astro-ph.CO},
       adsurl = {https://ui.adsabs.harvard.edu/abs/2022MNRAS.515.5646D}
}

@ARTICLE{Demirtas2023PQAxiverse,
       author = {{Demirtas}, Mehmet and {Gendler}, Naomi and {Long}, Cody and {McAllister}, Liam and {Moritz}, Jakob},
        title = "{PQ axiverse}",
      journal = {Journal of High Energy Physics},
         year = 2023,
        month = jun,
       volume = {2023},
       number = {6},
          eid = {92},
        pages = {92},
          doi = {10.1007/JHEP06(2023)092},
archivePrefix = {arXiv},
       eprint = {2112.04503},
 primaryClass = {hep-th},
       adsurl = {https://ui.adsabs.harvard.edu/abs/2023JHEP...06..092D}
}

@ARTICLE{Marsh2014AModel,
       author = {{Marsh}, David J.~E. and {Silk}, Joseph},
        title = "{A model for halo formation with axion mixed dark matter}",
      journal = {\mnras},
         year = 2014,
        month = jan,
       volume = {437},
       number = {3},
        pages = {2652-2663},
          doi = {10.1093/mnras/stt2079},
archivePrefix = {arXiv},
       eprint = {1307.1705},
 primaryClass = {astro-ph.CO},
       adsurl = {https://ui.adsabs.harvard.edu/abs/2014MNRAS.437.2652M}
}

@ARTICLE{Wright2025KiDSLegacy,
       author = {{Wright}, Angus H. and {St{\"o}lzner}, Benjamin and {Asgari}, Marika and {Bilicki}, Maciej and {Giblin}, Benjamin and {Heymans}, Catherine and {Hildebrandt}, Hendrik and {Hoekstra}, Henk and {Joachimi}, Benjamin and {Kuijken}, Konrad and {Li}, Shun-Sheng and {Reischke}, Robert and {von Wietersheim-Kramsta}, Maximilian and {Yoon}, Mijin and {Burger}, Pierre and {Chisari}, Nora Elisa and {de Jong}, Jelte and {Dvornik}, Andrej and {Georgiou}, Christos and {Harnois-D{\'e}raps}, Joachim and {Jalan}, Priyanka and {William}, Anjitha John and {Joudaki}, Shahab and {Lesci}, Giorgio Francesco and {Linke}, Laila and {Loureiro}, Arthur and {Mahony}, Constance and {Maturi}, Matteo and {Miller}, Lance and {Moscardini}, Lauro and {Napolitano}, Nicola R. and {Porth}, Lucas and {Radovich}, Mario and {Schneider}, Peter and {Tr{\"o}ster}, Tilman and {Valentijn}, Edwin and {Wittje}, Anna and {Yan}, Ziang and {Zhang}, Yun-Hao},
        title = "{KiDS-Legacy: Cosmological constraints from cosmic shear with the complete Kilo-Degree Survey}",
      journal = {\aap},
         year = 2025,
        month = nov,
       volume = {703},
          eid = {A158},
        pages = {A158},
          doi = {10.1051/0004-6361/202554908},
archivePrefix = {arXiv},
       eprint = {2503.19441},
 primaryClass = {astro-ph.CO},
       adsurl = {https://ui.adsabs.harvard.edu/abs/2025A&A...703A.158W}
}

@ARTICLE{Benabou2025QCDAxion,
       author = {{Benabou}, Joshua N. and {Fraser}, Katherine and {Reig}, Mario and {Safdi}, Benjamin R.},
        title = "{String theory and grand unification suggest a submicroelectronvolt QCD axion}",
      journal = {\prd},
         year = 2025,
        month = sep,
       volume = {112},
       number = {6},
          eid = {066003},
        pages = {066003},
          doi = {10.1103/lthr-97lm},
archivePrefix = {arXiv},
       eprint = {2505.15884},
 primaryClass = {hep-ph},
       adsurl = {https://ui.adsabs.harvard.edu/abs/2025PhRvD.112f6003B}
}

@ARTICLE{Jain2025BayesianCY,
       author = {{Jain}, Mudit and {Sheridan}, Elijah and {Marsh}, David J.~E. and {Heyes}, Elli and {Rogers}, Keir K. and {Schachner}, Andreas},
        title = "{Bayesian inference on Calabi--Yau moduli spaces and the axiverse: experimental data meets string theory}",
      journal = {arXiv e-prints},
         year = 2025,
        month = nov,
          eid = {arXiv:2512.00144},
        pages = {arXiv:2512.00144},
          doi = {10.48550/arXiv.2512.00144},
archivePrefix = {arXiv},
       eprint = {2512.00144},
 primaryClass = {hep-th},
       adsurl = {https://ui.adsabs.harvard.edu/abs/2025arXiv251200144J}
}

@ARTICLE{Vander2025FTheory,
       author = {{Vander Ploeg Fallon}, Sebastian and {Halverson}, James and {McAllister}, Liam and {Zhu}, Yunhao},
        title = "{F-theory Axiverse}",
      journal = {arXiv e-prints},
         year = 2025,
        month = nov,
          eid = {arXiv:2511.20458},
        pages = {arXiv:2511.20458},
          doi = {10.48550/arXiv.2511.20458},
archivePrefix = {arXiv},
       eprint = {2511.20458},
 primaryClass = {hep-th},
       adsurl = {https://ui.adsabs.harvard.edu/abs/2025arXiv251120458V}
}

@ARTICLE{Cheng2025Universality,
       author = {{Cheng}, Junyi and {Gendler}, Naomi},
        title = "{Universality in the axiverse}",
      journal = {Journal of High Energy Physics},
         year = 2025,
        month = nov,
       volume = {2025},
       number = {11},
          eid = {12},
        pages = {12},
          doi = {10.1007/JHEP11(2025)012},
archivePrefix = {arXiv},
       eprint = {2507.12516},
 primaryClass = {hep-th},
       adsurl = {https://ui.adsabs.harvard.edu/abs/2025JHEP...11..012C}
}

@ARTICLE{Yin2025AxiverseReio,
       author = {{Yin}, Ziwen and {Cheng}, Hanyu and {Di Valentino}, Eleonora and {Gendler}, Naomi and {Marsh}, David J.~E. and {Visinelli}, Luca},
        title = "{Constraining the axiverse with reionization}",
      journal = {arXiv e-prints},
         year = 2025,
        month = jul,
          eid = {arXiv:2507.03535},
        pages = {arXiv:2507.03535},
          doi = {10.48550/arXiv.2507.03535},
archivePrefix = {arXiv},
       eprint = {2507.03535},
 primaryClass = {hep-ph},
       adsurl = {https://ui.adsabs.harvard.edu/abs/2025arXiv250703535Y}
}

@ARTICLE{Perez2025ReconstructingThe,
       author = {{Perez Sarmiento}, Karen and {Lagu{\"e}}, Alex and {Madhavacheril}, Mathew S. and {Jain}, Bhuvnesh and {Sherwin}, Blake},
        title = "{Reconstructing the shape of the nonlinear matter power spectrum using CMB lensing and cosmic shear}",
      journal = {\prd},
         year = 2025,
        month = sep,
       volume = {112},
       number = {6},
          eid = {063510},
        pages = {063510},
          doi = {10.1103/bzrj-76sn},
archivePrefix = {arXiv},
       eprint = {2502.06687},
 primaryClass = {astro-ph.CO},
       adsurl = {https://ui.adsabs.harvard.edu/abs/2025PhRvD.112f3510P}
}

@ARTICLE{Marsh2019StrongConstraints,
       author = {{Marsh}, David J.~E. and {Niemeyer}, Jens C.},
        title = "{Strong Constraints on Fuzzy Dark Matter from Ultrafaint Dwarf Galaxy Eridanus II}",
      journal = {\prl},
         year = 2019,
        month = aug,
       volume = {123},
       number = {5},
          eid = {051103},
        pages = {051103},
          doi = {10.1103/PhysRevLett.123.051103},
archivePrefix = {arXiv},
       eprint = {1810.08543},
 primaryClass = {astro-ph.CO},
       adsurl = {https://ui.adsabs.harvard.edu/abs/2019PhRvL.123e1103M}
}

@ARTICLE{Dalal2022ExcludingFuzzy,
       author = {{Dalal}, Neal and {Kravtsov}, Andrey},
        title = "{Excluding fuzzy dark matter with sizes and stellar kinematics of ultrafaint dwarf galaxies}",
      journal = {\prd},
         year = 2022,
        month = sep,
       volume = {106},
       number = {6},
          eid = {063517},
        pages = {063517},
          doi = {10.1103/PhysRevD.106.063517},
archivePrefix = {arXiv},
       eprint = {2203.05750},
 primaryClass = {astro-ph.CO},
       adsurl = {https://ui.adsabs.harvard.edu/abs/2022PhRvD.106f3517D}
}

@ARTICLE{May2025UpdatedBounds,
       author = {{May}, Simon and {Dalal}, Neal and {Kravtsov}, Andrey},
        title = "{Updated bounds on ultra-light dark matter from the tiniest galaxies}",
      journal = {arXiv e-prints},
         year = 2025,
        month = sep,
          eid = {arXiv:2509.02781},
        pages = {arXiv:2509.02781},
          doi = {10.48550/arXiv.2509.02781},
archivePrefix = {arXiv},
       eprint = {2509.02781},
 primaryClass = {astro-ph.CO},
       adsurl = {https://ui.adsabs.harvard.edu/abs/2025arXiv250902781M}
}

@ARTICLE{Schwabe2016SimulationsOf,
       author = {{Schwabe}, Bodo and {Niemeyer}, Jens C. and {Engels}, Jan F.},
        title = "{Simulations of solitonic core mergers in ultralight axion dark matter cosmologies}",
      journal = {\prd},
         year = 2016,
        month = aug,
       volume = {94},
       number = {4},
          eid = {043513},
        pages = {043513},
          doi = {10.1103/PhysRevD.94.043513},
archivePrefix = {arXiv},
       eprint = {1606.05151},
 primaryClass = {astro-ph.CO},
       adsurl = {https://ui.adsabs.harvard.edu/abs/2016PhRvD..94d3513S}
}

@ARTICLE{Wang2026LymanAlpha,
       author = {{Wang}, Yourong Frank},
        title = "{Lyman-$α$ Forest Signatures of Mixed Fuzzy and Cold Dark Matter}",
      journal = {arXiv e-prints},
         year = 2026,
        month = apr,
          eid = {arXiv:2604.06038},
        pages = {arXiv:2604.06038},
          doi = {10.48550/arXiv.2604.06038},
archivePrefix = {arXiv},
       eprint = {2604.06038},
 primaryClass = {astro-ph.CO},
       adsurl = {https://ui.adsabs.harvard.edu/abs/2026arXiv260406038W}
}

@ARTICLE{Witten1984Superstrings,
       author = {{Witten}, Edward},
        title = "{Some properties of O(32) superstrings}",
      journal = {Physics Letters B},
         year = 1984,
        month = dec,
       volume = {149},
       number = {4-5},
        pages = {351-356},
          doi = {10.1016/0370-2693(84)90422-2},
       adsurl = {https://ui.adsabs.harvard.edu/abs/1984PhLB..149..351W}
}

@ARTICLE{Conlon2006ModuliStabilisation,
       author = {{Conlon}, Joseph P.},
        title = "{The QCD axion and moduli stabilisation}",
      journal = {Journal of High Energy Physics},
         year = 2006,
        month = may,
       volume = {2006},
       number = {5},
          eid = {078},
        pages = {078},
          doi = {10.1088/1126-6708/2006/05/078},
archivePrefix = {arXiv},
       eprint = {hep-th/0602233},
 primaryClass = {hep-th},
       adsurl = {https://ui.adsabs.harvard.edu/abs/2006JHEP...05..078C}
}

@ARTICLE{Acharya2010MTheory,
       author = {{Samir Acharya}, Bobby and {Bobkov}, Konstantin and {Kumar}, Piyush},
        title = "{An M theory solution to the strong CP-problem, and constraints on the axiverse}",
      journal = {Journal of High Energy Physics},
         year = 2010,
        month = nov,
       volume = {2010},
          eid = {105},
        pages = {105},
          doi = {10.1007/JHEP11(2010)105},
archivePrefix = {arXiv},
       eprint = {1004.5138},
 primaryClass = {hep-th},
       adsurl = {https://ui.adsabs.harvard.edu/abs/2010JHEP...11..105S}
}

@ARTICLE{Cicoli2012TypeIIB,
       author = {{Cicoli}, Michele and {Goodsell}, Mark D. and {Ringwald}, Andreas},
        title = "{The type IIB string axiverse and its low-energy phenomenology}",
      journal = {Journal of High Energy Physics},
         year = 2012,
        month = oct,
       volume = {2012},
          eid = {146},
        pages = {146},
          doi = {10.1007/JHEP10(2012)146},
archivePrefix = {arXiv},
       eprint = {1206.0819},
 primaryClass = {hep-th},
       adsurl = {https://ui.adsabs.harvard.edu/abs/2012JHEP...10..146C}
}

@ARTICLE{Liu2025AccurateMethod,
       author = {{Liu}, Rayne and {Hu}, Wayne and {Grin}, Daniel},
        title = "{Accurate method for ultralight axion CMB and matter power spectra}",
      journal = {\prd},
         year = 2025,
        month = jul,
       volume = {112},
       number = {2},
          eid = {023513},
        pages = {023513},
          doi = {10.1103/1z4c-1w7f},
archivePrefix = {arXiv},
       eprint = {2412.15192},
 primaryClass = {astro-ph.CO},
       adsurl = {https://ui.adsabs.harvard.edu/abs/2025PhRvD.112b3513L}
}

@ARTICLE{DESIDR2BAO,
       author = {{Abdul Karim}, M. and {Aguilar}, J. and {Ahlen}, S. and {Alam}, S. and {Allen}, L. and {Allende Prieto}, C. and {Alves}, O. and {Anand}, A. and {Andrade}, U. and {Armengaud}, E. and {Aviles}, A. and {Bailey}, S. and {Baltay}, C. and {Bansal}, P. and {Bault}, A. and {Behera}, J. and {BenZvi}, S. and {Bianchi}, D. and {Blake}, C. and {Brieden}, S. and {Brodzeller}, A. and {Brooks}, D. and {Buckley-Geer}, E. and {Burtin}, E. and {Calderon}, R. and {Canning}, R. and {Rosell}, A. Carnero and {Carrilho}, P. and {Casas}, L. and {Castander}, F.~J. and {Charles}, M. and {Chaussidon}, E. and {Chaves-Montero}, J. and {Chebat}, D. and {Chen}, X. and {Claybaugh}, T. and {Cole}, S. and {Cooper}, A.~P. and {Cuceu}, A. and {Dawson}, K.~S. and {de la Macorra}, A. and {de Mattia}, A. and {Deiosso}, N. and {Della Costa}, J. and {Demina}, R. and {Dey}, A. and {Dey}, B. and {Ding}, Z. and {Doel}, P. and {Edelstein}, J. and {Eisenstein}, D.~J. and {Elbers}, W. and {Fagrelius}, P. and {Fanning}, K. and {Fern{\'a}ndez-Garc{\'\i}a}, E. and {Ferraro}, S. and {Font-Ribera}, A. and {Forero-Romero}, J.~E. and {Frenk}, C.~S. and {Garcia-Quintero}, C. and {Garrison}, L.~H. and {Gazta{\~n}aga}, E. and {Gil-Mar{\'\i}n}, H. and {Gontcho A Gontcho}, S. and {Gonzalez}, D. and {Gonzalez-Morales}, A.~X. and {Gordon}, C. and {Green}, D. and {Gutierrez}, G. and {Guy}, J. and {Hadzhiyska}, B. and {Hahn}, C. and {He}, S. and {Herbold}, M. and {Herrera-Alcantar}, H.~K. and {Ho}, M.-F. and {Honscheid}, K. and {Howlett}, C. and {Huterer}, D. and {Ishak}, M. and {Juneau}, S. and {Kamble}, N.~V. and {Kara{\c{c}}ayl{\i}}, N.~G. and {Kehoe}, R. and {Kent}, S. and {Kim}, A.~G. and {Kirkby}, D. and {Kisner}, T. and {Koposov}, S.~E. and {Kremin}, A. and {Krolewski}, A. and {Lahav}, O. and {Lamman}, C. and {Landriau}, M. and {Lang}, D. and {Lasker}, J. and {Le Goff}, J.~M. and {Le Guillou}, L. and {Leauthaud}, A. and {Levi}, M.~E. and {Li}, Q. and {Li}, T.~S. and {Lodha}, K. and {Lokken}, M. and {Lozano-Rodr{\'\i}guez}, F. and {Magneville}, C. and {Manera}, M. and {Martini}, P. and {Matthewson}, W.~L. and {Meisner}, A. and {Mena-Fern{\'a}ndez}, J. and {Menegas}, A. and {Mergulh{\~a}o}, T. and {Miquel}, R. and {Moustakas}, J. and {Mu{\~n}oz-Guti{\'e}rrez}, A. and {Mu{\~n}oz-Santos}, D. and {Myers}, A.~D. and {Nadathur}, S. and {Naidoo}, K. and {Napolitano}, L. and {Newman}, J.~A. and {Niz}, G. and {Noriega}, H.~E. and {Paillas}, E. and {Palanque-Delabrouille}, N. and {Pan}, J. and {Peacock}, J.~A. and {Pellejero Ibanez}, M. and {Percival}, W.~J. and {P{\'e}rez-Fern{\'a}ndez}, A. and {P{\'e}rez-R{\`a}fols}, I. and {Pieri}, M.~M. and {Poppett}, C. and {Prada}, F. and {Rabinowitz}, D. and {Raichoor}, A. and {Ram{\'\i}rez-P{\'e}rez}, C. and {Rashkovetskyi}, M. and {Ravoux}, C. and {Rich}, J. and {Rocher}, A. and {Rockosi}, C. and {Rohlf}, J. and {Rom{\'a}n-Herrera}, J.~O. and {Ross}, A.~J. and {Rossi}, G. and {Ruggeri}, R. and {Ruhlmann-Kleider}, V. and {Samushia}, L. and {Sanchez}, E. and {Sanders}, N. and {Schlegel}, D. and {Schubnell}, M. and {Seo}, H. and {Shafieloo}, A. and {Sharples}, R. and {Silber}, J. and {Sinigaglia}, F. and {Sprayberry}, D. and {Tan}, T. and {Tarl{\'e}}, G. and {Taylor}, P. and {Turner}, W. and {Ure{\~n}a-L{\'o}pez}, L.~A. and {Vaisakh}, R. and {Valdes}, F. and {Valogiannis}, G. and {Vargas-Maga{\~n}a}, M. and {Verde}, L. and {Walther}, M. and {Weaver}, B.~A. and {Weinberg}, D.~H. and {White}, M. and {Wolfson}, M. and {Y{\`e}che}, C. and {Yu}, J. and {Zaborowski}, E.~A. and {Zarrouk}, P. and {Zhai}, Z. and {Zhang}, H. and {Zhao}, C. and {Zhao}, G.~B. and {Zhou}, R. and {Zou}, H. and {DESI Collaboration}},
        title = "{DESI DR2 results. II. Measurements of baryon acoustic oscillations and cosmological constraints}",
      journal = {\prd},
         year = 2025,
        month = oct,
       volume = {112},
       number = {8},
          eid = {083515},
        pages = {083515},
          doi = {10.1103/tr6y-kpc6},
archivePrefix = {arXiv},
       eprint = {2503.14738},
 primaryClass = {astro-ph.CO},
       adsurl = {https://ui.adsabs.harvard.edu/abs/2025PhRvD.112h3515A}
}

@ARTICLE{PlanckLikelihood,
       author = {{Planck Collaboration} and {Aghanim}, N. and {Akrami}, Y. and {Ashdown}, M. and {Aumont}, J. and {Baccigalupi}, C. and {Ballardini}, M. and {Banday}, A.~J. and {Barreiro}, R.~B. and {Bartolo}, N. and {Basak}, S. and {Benabed}, K. and {Bernard}, J.-P. and {Bersanelli}, M. and {Bielewicz}, P. and {Bock}, J.~J. and {Bond}, J.~R. and {Borrill}, J. and {Bouchet}, F.~R. and {Boulanger}, F. and {Bucher}, M. and {Burigana}, C. and {Butler}, R.~C. and {Calabrese}, E. and {Cardoso}, J.-F. and {Carron}, J. and {Casaponsa}, B. and {Challinor}, A. and {Chiang}, H.~C. and {Colombo}, L.~P.~L. and {Combet}, C. and {Crill}, B.~P. and {Cuttaia}, F. and {de Bernardis}, P. and {de Rosa}, A. and {de Zotti}, G. and {Delabrouille}, J. and {Delouis}, J.-M. and {Di Valentino}, E. and {Diego}, J.~M. and {Dor{\'e}}, O. and {Douspis}, M. and {Ducout}, A. and {Dupac}, X. and {Dusini}, S. and {Efstathiou}, G. and {Elsner}, F. and {En{\ss}lin}, T.~A. and {Eriksen}, H.~K. and {Fantaye}, Y. and {Fernandez-Cobos}, R. and {Finelli}, F. and {Frailis}, M. and {Fraisse}, A.~A. and {Franceschi}, E. and {Frolov}, A. and {Galeotta}, S. and {Galli}, S. and {Ganga}, K. and {G{\'e}nova-Santos}, R.~T. and {Gerbino}, M. and {Ghosh}, T. and {Giraud-H{\'e}raud}, Y. and {Gonz{\'a}lez-Nuevo}, J. and {G{\'o}rski}, K.~M. and {Gratton}, S. and {Gruppuso}, A. and {Gudmundsson}, J.~E. and {Hamann}, J. and {Handley}, W. and {Hansen}, F.~K. and {Herranz}, D. and {Hivon}, E. and {Huang}, Z. and {Jaffe}, A.~H. and {Jones}, W.~C. and {Keih{\"a}nen}, E. and {Keskitalo}, R. and {Kiiveri}, K. and {Kim}, J. and {Kisner}, T.~S. and {Krachmalnicoff}, N. and {Kunz}, M. and {Kurki-Suonio}, H. and {Lagache}, G. and {Lamarre}, J.-M. and {Lasenby}, A. and {Lattanzi}, M. and {Lawrence}, C.~R. and {Le Jeune}, M. and {Levrier}, F. and {Lewis}, A. and {Liguori}, M. and {Lilje}, P.~B. and {Lilley}, M. and {Lindholm}, V. and {L{\'o}pez-Caniego}, M. and {Lubin}, P.~M. and {Ma}, Y.-Z. and {Mac{\'\i}as-P{\'e}rez}, J.~F. and {Maggio}, G. and {Maino}, D. and {Mandolesi}, N. and {Mangilli}, A. and {Marcos-Caballero}, A. and {Maris}, M. and {Martin}, P.~G. and {Mart{\'\i}nez-Gonz{\'a}lez}, E. and {Matarrese}, S. and {Mauri}, N. and {McEwen}, J.~D. and {Meinhold}, P.~R. and {Melchiorri}, A. and {Mennella}, A. and {Migliaccio}, M. and {Millea}, M. and {Miville-Desch{\^e}nes}, M.-A. and {Molinari}, D. and {Moneti}, A. and {Montier}, L. and {Morgante}, G. and {Moss}, A. and {Natoli}, P. and {N{\o}rgaard-Nielsen}, H.~U. and {Pagano}, L. and {Paoletti}, D. and {Partridge}, B. and {Patanchon}, G. and {Peiris}, H.~V. and {Perrotta}, F. and {Pettorino}, V. and {Piacentini}, F. and {Polenta}, G. and {Puget}, J.-L. and {Rachen}, J.~P. and {Reinecke}, M. and {Remazeilles}, M. and {Renzi}, A. and {Rocha}, G. and {Rosset}, C. and {Roudier}, G. and {Rubi{\~n}o-Mart{\'\i}n}, J.~A. and {Ruiz-Granados}, B. and {Salvati}, L. and {Sandri}, M. and {Savelainen}, M. and {Scott}, D. and {Shellard}, E.~P.~S. and {Sirignano}, C. and {Sirri}, G. and {Spencer}, L.~D. and {Sunyaev}, R. and {Suur-Uski}, A.-S. and {Tauber}, J.~A. and {Tavagnacco}, D. and {Tenti}, M. and {Toffolatti}, L. and {Tomasi}, M. and {Trombetti}, T. and {Valiviita}, J. and {Van Tent}, B. and {Vielva}, P. and {Villa}, F. and {Vittorio}, N. and {Wandelt}, B.~D. and {Wehus}, I.~K. and {Zacchei}, A. and {Zonca}, A.},
        title = "{Planck 2018 results. V. CMB power spectra and likelihoods}",
      journal = {\aap},
         year = 2020,
        month = sep,
       volume = {641},
          eid = {A5},
        pages = {A5},
          doi = {10.1051/0004-6361/201936386},
archivePrefix = {arXiv},
       eprint = {1907.12875},
 primaryClass = {astro-ph.CO},
       adsurl = {https://ui.adsabs.harvard.edu/abs/2020A&A...641A...5P}
}

@ARTICLE{PlanckSroll2Likelihood,
       author = {{Pagano}, L. and {Delouis}, J.-M. and {Mottet}, S. and {Puget}, J.-L. and {Vibert}, L.},
        title = "{Reionization optical depth determination from Planck HFI data with ten percent accuracy}",
      journal = {\aap},
         year = 2020,
        month = mar,
       volume = {635},
          eid = {A99},
        pages = {A99},
          doi = {10.1051/0004-6361/201936630},
archivePrefix = {arXiv},
       eprint = {1908.09856},
 primaryClass = {astro-ph.CO},
       adsurl = {https://ui.adsabs.harvard.edu/abs/2020A&A...635A..99P}
}

@ARTICLE{ACTDR6Likelihood,
       author = {{Louis}, Thibaut and {La Posta}, Adrien and {Atkins}, Zachary and {Jense}, Hidde T. and {Abril-Cabezas}, Irene and {Addison}, Graeme E. and {Ade}, Peter A.~R. and {Aiola}, Simone and {Alford}, Tommy and {Alonso}, David and {Amiri}, Mandana and {An}, Rui and {Austermann}, Jason E. and {Barbavara}, Eleonora and {Battaglia}, Nicholas and {Battistelli}, Elia Stefano and {Beall}, James A. and {Bean}, Rachel and {Beheshti}, Ali and {Beringue}, Benjamin and {Bhandarkar}, Tanay and {Biermann}, Emily and {Bolliet}, Boris and {Bond}, J. Richard and {Calabrese}, Erminia and {Capalbo}, Valentina and {Carrero}, Felipe and {Chen}, Shi-Fan and {Chesmore}, Grace and {Cho}, Hsiao-mei and {Choi}, Steve K. and {Clark}, Susan E. and {Cothard}, Nicholas F. and {Coughlin}, Kevin and {Coulton}, William and {Crichton}, Devin and {Crowley}, Kevin T. and {Darwish}, Omar and {Devlin}, Mark J. and {Dicker}, Simon and {Duell}, Cody J. and {Duff}, Shannon M. and {Duivenvoorden}, Adriaan J. and {Dunkley}, Jo and {Dunner}, Rolando and {Embil Villagra}, Carmen and {Fankhanel}, Max and {Farren}, Gerrit S. and {Ferraro}, Simone and {Foster}, Allen and {Freundt}, Rodrigo and {Fuzia}, Brittany and {Gallardo}, Patricio A. and {Garrido}, Xavier and {Gerbino}, Martina and {Giardiello}, Serena and {Gill}, Ajay and {Givans}, Jahmour and {Gluscevic}, Vera and {Goldstein}, Samuel and {Golec}, Joseph E. and {Gong}, Yulin and {Guan}, Yilun and {Halpern}, Mark and {Harrison}, Ian and {Hasselfield}, Matthew and {Healy}, Erin and {Henderson}, Shawn and {Hensley}, Brandon and {Herv{\'\i}as-Caimapo}, Carlos and {Hill}, J. Colin and {Hilton}, Gene C. and {Hilton}, Matt and {Hincks}, Adam D. and {Hlo{\v{z}}ek}, Ren{\'e}e and {Ho}, Shuay-Pwu Patty and {Hood}, John and {Hornecker}, Erika and {Huber}, Zachary B. and {Hubmayr}, Johannes and {Huffenberger}, Kevin M. and {Hughes}, John P. and {Ikape}, Margaret and {Irwin}, Kent and {Isopi}, Giovanni and {Joshi}, Neha and {Keller}, Ben and {Kim}, Joshua and {Knowles}, Kenda and {Koopman}, Brian J. and {Kosowsky}, Arthur and {Kramer}, Darby and {Kusiak}, Aleksandra and {Lagu{\"e}}, Alex and {Lakey}, Victoria and {Lee}, Eunseong and {Li}, Yaqiong and {Li}, Zack and {Limon}, Michele and {Lokken}, Martine and {Lungu}, Marius and {MacCrann}, Niall and {MacInnis}, Amanda and {Madhavacheril}, Mathew S. and {Maldonado}, Diego and {Maldonado}, Felipe and {Mallaby-Kay}, Maya and {Marques}, Gabriela A. and {van Marrewijk}, Joshiwa and {McCarthy}, Fiona and {McMahon}, Jeff and {Mehta}, Yogesh and {Menanteau}, Felipe and {Moodley}, Kavilan and {Morris}, Thomas W. and {Mroczkowski}, Tony and {Naess}, Sigurd and {Namikawa}, Toshiya and {Nati}, Federico and {Nerval}, Simran K. and {Newburgh}, Laura and {Nicola}, Andrina and {Niemack}, Michael D. and {Nolta}, Michael R. and {Orlowski-Scherer}, John and {Pagano}, Luca and {Page}, Lyman A. and {Pandey}, Shivam and {Partridge}, Bruce and {Perez Sarmiento}, Karen and {Prince}, Heather and {Puddu}, Roberto and {Qu}, Frank J. and {Ragavan}, Damien C. and {Ried Guachalla}, Bernardita and {Rogers}, Keir K. and {Rojas}, Felipe and {Sakuma}, Tai and {Schaan}, Emmanuel and {Schmitt}, Benjamin L. and {Sehgal}, Neelima and {Shaikh}, Shabbir and {Sherwin}, Blake D. and {Sierra}, Carlos and {Sievers}, Jon and {Sif{\'o}n}, Crist{\'o}bal and {Simon}, Sara and {Sonka}, Rita and {Spergel}, David N. and {Staggs}, Suzanne T. and {Storer}, Emilie and {Surrao}, Kristen and {Switzer}, Eric R. and {Tampier}, Niklas and {Thornton}, Robert and {Trac}, Hy and {Tucker}, Carole and {Ullom}, Joel and {Vale}, Leila R. and {Van Engelen}, Alexander and {Van Lanen}, Jeff and {Vargas}, Cristian and {Vavagiakis}, Eve M. and {Wagoner}, Kasey and {Wang}, Yuhan and {Wenzl}, Lukas and {Wollack}, Edward J. and {Zheng}, Kaiwen and {The Atacama Cosmology Telescope collaboration}},
        title = "{The Atacama Cosmology Telescope: DR6 power spectra, likelihoods and {\ensuremath{\Lambda}}CDM parameters}",
      journal = {\jcap},
         year = 2025,
        month = nov,
       volume = {2025},
       number = {11},
          eid = {062},
        pages = {062},
          doi = {10.1088/1475-7516/2025/11/062},
archivePrefix = {arXiv},
       eprint = {2503.14452},
 primaryClass = {astro-ph.CO},
       adsurl = {https://ui.adsabs.harvard.edu/abs/2025JCAP...11..062L}
}

@ARTICLE{Qu2026APSLensing,
       author = {{Qu}, Frank J. and {Ge}, Fei and {Wu}, W.~L. Kimmy and {Abril-Cabezas}, Irene and {Madhavacheril}, Mathew S. and {Millea}, Marius and {Ahmed}, Zeeshan and {Anderes}, Ethan and {Anderson}, Adam J. and {Ansarinejad}, Behzad and {Archipley}, Melanie and {Atkins}, Zachary and {Balkenhol}, Lennart and {Battaglia}, Nicholas and {Benabed}, Karim and {Bender}, Amy N. and {Benson}, Bradford A. and {Bianchini}, Federico and {Bleem}, Lindsey. E. and {Bolliet}, Boris and {Bond}, J. Richard and {Bouchet}, Fran{\c{c}}ois. R. and {Bryant}, Lincoln and {Calabrese}, Erminia and {Camphuis}, Etienne and {Carlstrom}, John E. and {Carron}, Julien and {Challinor}, Anthony and {Chang}, Clarence L. and {Chaubal}, Prakrut and {Chen}, Geoff and {Chichura}, Paul M. and {Choi}, Steve K. and {Chokshi}, Aman and {Chou}, Ti-Lin and {Coerver}, Anna and {Coulton}, William and {Crawford}, Thomas M. and {Daley}, Cail and {Darwish}, Omar and {de Haan}, Tijmen and {Devlin}, Mark J. and {Dibert}, Karia R. and {Dobbs}, Matthew A. and {Doohan}, Michael and {Doussot}, Aristide and {Duivenvoorden}, Adriaan J. and {Dunkley}, Jo and {Dunner}, Rolando and {Dutcher}, Daniel and {Villagra}, Carmen Embil and {Everett}, Wendy and {Farren}, Gerrit S. and {Feng}, Chang and {Ferraro}, Simone and {Ferguson}, Kyle R. and {Fichman}, Kyra and {Finson}, Emily and {Foster}, Allen and {Gallardo}, Patricio A. and {Galli}, Silvia and {Gambrel}, Anne E. and {Gardner}, Rob W. and {Goeckner-Wald}, Neil and {Gualtieri}, Riccardo and {Guidi}, Federica and {Guns}, Sam and {Halpern}, Mark and {Halverson}, Nils W. and {Hill}, J. Colin and {Hilton}, Matt and {Hivon}, Eric and {Holder}, Gilbert P. and {Holzapfel}, William L. and {Hood}, John C. and {Howe}, Doug and {Hryciuk}, Alec and {Huang}, Nicholas and {Hubmayr}, Johannes and {K{\'e}ruzor{\'e}}, Florian and {Khalife}, Ali R. and {Kim}, Joshua and {Knox}, Lloyd and {Korman}, Milo and {Kornoelje}, Kayla and {Kosowsky}, Arthur and {Kuo}, Chao-Lin and {Jense}, Hidde T. and {La Posta}, Adrien and {Levy}, Kevin and {Lowitz}, Amy E. and {Louis}, Thibaut and {Lu}, Chunyu and {Lynch}, Gabriel P. and {MacCrann}, Niall and {Maniyar}, Abhishek and {Martsen}, Emily S. and {McMahon}, Jeff and {Menanteau}, Felipe and {Montgomery}, Joshua and {Nakato}, Yuka and {Moodley}, Kavilan and {Namikawa}, Toshiya and {Natoli}, Tyler and {Niemack}, Michael D. and {Noble}, Gavin I. and {Omori}, Yuuki and {Ouellette}, Aaron and {Page}, Lyman A. and {Pan}, Zhaodi and {Paschos}, Pascal and {Phadke}, Kedar A. and {Pollak}, Alexander W. and {Prabhu}, Karthik and {Quan}, Wei and {Raghunathan}, Srinivasan and {Rahimi}, Mahsa and {Rahlin}, Alexandra and {Reichardt}, Christian L. and {Riebel}, Dave and {Rouble}, Maclean and {Ruhl}, John E. and {Schaan}, Emmanuel and {Schiappucci}, Eduardo and {Sehgal}, Neelima and {Sierra}, Carlos E. and {Simpson}, Aidan and {Sherwin}, Blake D. and {Sif{\'o}n}, Crist{\'o}bal and {Spergel}, David N. and {Staggs}, Suzanne T. and {Sobrin}, Joshua A. and {Stark}, Antony A. and {Stephen}, Judith and {Tandoi}, Chris and {Thorne}, Ben and {Trendafilova}, Cynthia and {Umilta}, Caterina and {Van Engelen}, Alexander and {Vieira}, Joaquin D. and {Vitrier}, Aline and {Wan}, Yujie and {Whitehorn}, Nathan and {Wollack}, Edward J. and {Young}, Matthew R. and {Zebrowski}, Jessica A. and {(ACT + SPT-3G Collaborations)}},
        title = "{Unified and Consistent Structure Growth Measurements from Joint ACT, SPT, and Planck CMB Lensing}",
      journal = {\prl},
         year = 2026,
        month = jan,
       volume = {136},
       number = {2},
          eid = {021001},
        pages = {021001},
          doi = {10.1103/k5yr-3h6d},
archivePrefix = {arXiv},
       eprint = {2504.20038},
 primaryClass = {astro-ph.CO},
       adsurl = {https://ui.adsabs.harvard.edu/abs/2026PhRvL.136b1001Q}
}

@ARTICLE{Calabrese2025ExtendedModels,
       author = {{Calabrese}, Erminia and {Hill}, J. Colin and {Jense}, Hidde T. and {La Posta}, Adrien and {Abril-Cabezas}, Irene and {Addison}, Graeme E. and {Ade}, Peter A.~R. and {Aiola}, Simone and {Alford}, Tommy and {Alonso}, David and {Amiri}, Mandana and {An}, Rui and {Atkins}, Zachary and {Austermann}, Jason E. and {Barbavara}, Eleonora and {Barbieri}, Nicola and {Battaglia}, Nicholas and {Battistelli}, Elia Stefano and {Beall}, James A. and {Bean}, Rachel and {Beheshti}, Ali and {Beringue}, Benjamin and {Bhandarkar}, Tanay and {Biermann}, Emily and {Bolliet}, Boris and {Bond}, J. Richard and {Capalbo}, Valentina and {Carrero}, Felipe and {Chen}, Shi-Fan and {Chesmore}, Grace and {Cho}, Hsiao-mei and {Choi}, Steve K. and {Clark}, Susan E. and {Cothard}, Nicholas F. and {Coughlin}, Kevin and {Coulton}, William and {Crichton}, Devin and {Crowley}, Kevin T. and {Darwish}, Omar and {Devlin}, Mark J. and {Dicker}, Simon and {Duell}, Cody J. and {Duff}, Shannon M. and {Duivenvoorden}, Adriaan J. and {Dunkley}, Jo and {Dunner}, Rolando and {Embil Villagra}, Carmen and {Fankhanel}, Max and {Farren}, Gerrit S. and {Ferraro}, Simone and {Foster}, Allen and {Freundt}, Rodrigo and {Fuzia}, Brittany and {Gallardo}, Patricio A. and {Garrido}, Xavier and {Gerbino}, Martina and {Giardiello}, Serena and {Gill}, Ajay and {Givans}, Jahmour and {Gluscevic}, Vera and {Goldstein}, Samuel and {Golec}, Joseph E. and {Gong}, Yulin and {Guan}, Yilun and {Halpern}, Mark and {Harrison}, Ian and {Hasselfield}, Matthew and {He}, Adam and {Healy}, Erin and {Henderson}, Shawn and {Hensley}, Brandon and {Herv{\'\i}as-Caimapo}, Carlos and {Hilton}, Gene C. and {Hilton}, Matt and {Hincks}, Adam D. and {Hlo{\v{z}}ek}, Ren{\'e}e and {Ho}, Shuay-Pwu Patty and {Hood}, John and {Hornecker}, Erika and {Huber}, Zachary B. and {Hubmayr}, Johannes and {Huffenberger}, Kevin M. and {Hughes}, John P. and {Ikape}, Margaret and {Irwin}, Kent and {Isopi}, Giovanni and {Joshi}, Neha and {Keller}, Ben and {Kim}, Joshua and {Knowles}, Kenda and {Koopman}, Brian J. and {Kosowsky}, Arthur and {Kramer}, Darby and {Kusiak}, Aleksandra and {Lagu{\"e}}, Alex and {Lakey}, Victoria and {Lattanzi}, Massimiliano and {Lee}, Eunseong and {Li}, Yaqiong and {Li}, Zack and {Limon}, Michele and {Lokken}, Martine and {Louis}, Thibaut and {Lungu}, Marius and {MacCrann}, Niall and {MacInnis}, Amanda and {Madhavacheril}, Mathew S. and {Maldonado}, Diego and {Maldonado}, Felipe and {Mallaby-Kay}, Maya and {Marques}, Gabriela A. and {van Marrewijk}, Joshiwa and {McCarthy}, Fiona and {McMahon}, Jeff and {Mehta}, Yogesh and {Menanteau}, Felipe and {Moodley}, Kavilan and {Morris}, Thomas W. and {Mroczkowski}, Tony and {Naess}, Sigurd and {Namikawa}, Toshiya and {Nati}, Federico and {Nerval}, Simran K. and {Newburgh}, Laura and {Nicola}, Andrina and {Niemack}, Michael D. and {Nolta}, Michael R. and {Orlowski-Scherer}, John and {Pagano}, Luca and {Page}, Lyman A. and {Pandey}, Shivam and {Partridge}, Bruce and {Perez Sarmiento}, Karen and {Prince}, Heather and {Puddu}, Roberto and {Qu}, Frank J. and {Ragavan}, Damien C. and {Ried Guachalla}, Bernardita and {Rogers}, Keir K. and {Rojas}, Felipe and {Sakuma}, Tai and {Schaan}, Emmanuel and {Schmitt}, Benjamin L. and {Sehgal}, Neelima and {Shaikh}, Shabbir and {Sherwin}, Blake D. and {Sierra}, Carlos and {Sievers}, Jon and {Sif{\'o}n}, Crist{\'o}bal and {Simon}, Sara and {Sonka}, Rita and {Spergel}, David N. and {Staggs}, Suzanne T. and {Storer}, Emilie and {Surrao}, Kristen and {Switzer}, Eric R. and {Tampier}, Niklas and {Thiele}, Leander and {Thornton}, Robert and {Trac}, Hy and {Tucker}, Carole and {Ullom}, Joel and {Vale}, Leila R. and {Van Engelen}, Alexander and {Van Lanen}, Jeff and {Vargas}, Cristian and {Vavagiakis}, Eve M. and {Wagoner}, Kasey and {Wang}, Yuhan and {Wenzl}, Lukas and {Wollack}, Edward J. and {Zheng}, Kaiwen and {The Atacama Cosmology Telescope collaboration}},
        title = "{The Atacama Cosmology Telescope: DR6 constraints on extended cosmological models}",
      journal = {\jcap},
         year = 2025,
        month = nov,
       volume = {2025},
       number = {11},
          eid = {063},
        pages = {063},
          doi = {10.1088/1475-7516/2025/11/063},
archivePrefix = {arXiv},
       eprint = {2503.14454},
 primaryClass = {astro-ph.CO},
       adsurl = {https://ui.adsabs.harvard.edu/abs/2025JCAP...11..063C}
}

@ARTICLE{Speagle2020Dynesty,
       author = {{Speagle}, Joshua S.},
        title = "{DYNESTY: a dynamic nested sampling package for estimating Bayesian posteriors and evidences}",
      journal = {\mnras},
         year = 2020,
        month = apr,
       volume = {493},
       number = {3},
        pages = {3132-3158},
          doi = {10.1093/mnras/staa278},
archivePrefix = {arXiv},
       eprint = {1904.02180},
 primaryClass = {astro-ph.IM},
       adsurl = {https://ui.adsabs.harvard.edu/abs/2020MNRAS.493.3132S}
}

@ARTICLE{Torrado2021Cobaya,
       author = {{Torrado}, Jes{\'u}s and {Lewis}, Antony},
        title = "{Cobaya: code for Bayesian analysis of hierarchical physical models}",
      journal = {\jcap},
         year = 2021,
        month = may,
       volume = {2021},
       number = {5},
          eid = {057},
        pages = {057},
          doi = {10.1088/1475-7516/2021/05/057},
archivePrefix = {arXiv},
       eprint = {2005.05290},
 primaryClass = {astro-ph.IM},
       adsurl = {https://ui.adsabs.harvard.edu/abs/2021JCAP...05..057T}
}

@article{iminuit,
  author={Hans Dembinski and Piti Ongmongkolkul et al.},
  title={scikit-hep/iminuit},
  DOI={10.5281/zenodo.3949207},
  publisher={Zenodo},
  year={2020},
  month={Dec},
  url={https://doi.org/10.5281/zenodo.3949207}
}

@article{James:1975dr,
    author = "James, F. and Roos, M.",
    title = "{Minuit: A System for Function Minimization and Analysis of the Parameter Errors and Correlations}",
    reportNumber = "CERN-DD-75-20",
    doi = "10.1016/0010-4655(75)90039-9",
    journal = "Comput. Phys. Commun.",
    volume = "10",
    pages = "343--367",
    year = "1975"
}

@ARTICLE{Lewis2025GetDist,
       author = {{Lewis}, Antony},
        title = "{GetDist: a Python package for analysing Monte Carlo samples}",
      journal = {\jcap},
         year = 2025,
        month = aug,
       volume = {2025},
       number = {8},
          eid = {025},
        pages = {025},
          doi = {10.1088/1475-7516/2025/08/025},
archivePrefix = {arXiv},
       eprint = {1910.13970},
 primaryClass = {astro-ph.IM},
       adsurl = {https://ui.adsabs.harvard.edu/abs/2025JCAP...08..025L}
}

@ARTICLE{Demirtas2018Kreuzer-Skarke,
       author = {{Demirtas}, Mehmet and {Long}, Cody and {McAllister}, Liam and {Stillman}, Mike},
        title = "{The Kreuzer-Skarke Axiverse}",
      journal = {arXiv e-prints},
         year = 2018,
        month = aug,
          eid = {arXiv:1808.01282},
        pages = {arXiv:1808.01282},
          doi = {10.48550/arXiv.1808.01282},
archivePrefix = {arXiv},
       eprint = {1808.01282},
 primaryClass = {hep-th},
       adsurl = {https://ui.adsabs.harvard.edu/abs/2018arXiv180801282D}
}

@ARTICLE{Demirtas2022CYTools,
       author = {{Demirtas}, Mehmet and {Rios-Tascon}, Andres and {McAllister}, Liam},
        title = "{CYTools: A Software Package for Analyzing Calabi-Yau Manifolds}",
      journal = {arXiv e-prints},
         year = 2022,
        month = nov,
          eid = {arXiv:2211.03823},
        pages = {arXiv:2211.03823},
          doi = {10.48550/arXiv.2211.03823},
archivePrefix = {arXiv},
       eprint = {2211.03823},
 primaryClass = {hep-th},
       adsurl = {https://ui.adsabs.harvard.edu/abs/2022arXiv221103823D}
}

@ARTICLE{Weinberg1978LightBoson,
       author = {{Weinberg}, Steven},
        title = "{A new light boson?}",
      journal = {\prl},
         year = 1978,
        month = jan,
       volume = {40},
       number = {4},
        pages = {223-226},
          doi = {10.1103/PhysRevLett.40.223},
       adsurl = {https://ui.adsabs.harvard.edu/abs/1978PhRvL..40..223W}
}

@ARTICLE{Wilczek1978ProblemOf,
       author = {{Wilczek}, F.},
        title = "{Problem of strong P and T invariance in the presence of instantons}",
      journal = {\prl},
         year = 1978,
        month = jan,
       volume = {40},
       number = {5},
        pages = {279-282},
          doi = {10.1103/PhysRevLett.40.279},
       adsurl = {https://ui.adsabs.harvard.edu/abs/1978PhRvL..40..279W}
}

@unpublished{Johnston2026HMF,
  author = {Johnston, Sarah},
  title = {Title of the Paper in Preparation},
  note = {In preparation},
  year = {2026}
}

@ARTICLE{Dalal2026Deciphering,
       author = {{Dalal}, Nihar and {To}, Chun-Hao and {Hirata}, Chris and {Hyeon-Shin}, Tae and {Hilton}, Matt and {Pandey}, Shivam and {Richard Bond}, J.},
        title = "{Deciphering baryonic feedback from ACT tSZ galaxy clusters}",
      journal = {\jcap},
         year = 2026,
        month = mar,
       volume = {2026},
       number = {3},
          eid = {036},
        pages = {036},
          doi = {10.1088/1475-7516/2026/03/036},
archivePrefix = {arXiv},
       eprint = {2507.04476},
 primaryClass = {astro-ph.CO},
       adsurl = {https://ui.adsabs.harvard.edu/abs/2026JCAP...03..036D}
}

@ARTICLE{Bigwood2024WeakLensing,
       author = {{Bigwood}, L. and {Amon}, A. and {Schneider}, A. and {Salcido}, J. and {McCarthy}, I.~G. and {Preston}, C. and {Sanchez}, D. and {Sijacki}, D. and {Schaan}, E. and {Ferraro}, S. and {Battaglia}, N. and {Chen}, A. and {Dodelson}, S. and {Roodman}, A. and {Pieres}, A. and {Fert{\'e}}, A. and {Alarcon}, A. and {Drlica-Wagner}, A. and {Choi}, A. and {Navarro-Alsina}, A. and {Campos}, A. and {Ross}, A.~J. and {Carnero Rosell}, A. and {Yin}, B. and {Yanny}, B. and {S{\'a}nchez}, C. and {Chang}, C. and {Davis}, C. and {Doux}, C. and {Gruen}, D. and {Rykoff}, E.~S. and {Huff}, E.~M. and {Sheldon}, E. and {Tarsitano}, F. and {Andrade-Oliveira}, F. and {Bernstein}, G.~M. and {Giannini}, G. and {Diehl}, H.~T. and {Huang}, H. and {Harrison}, I. and {Sevilla-Noarbe}, I. and {Tutusaus}, I. and {Elvin-Poole}, J. and {McCullough}, J. and {Zuntz}, J. and {Blazek}, J. and {DeRose}, J. and {Cordero}, J. and {Prat}, J. and {Myles}, J. and {Eckert}, K. and {Bechtol}, K. and {Herner}, K. and {Secco}, L.~F. and {Gatti}, M. and {Raveri}, M. and {Kind}, M. Carrasco and {Becker}, M.~R. and {Troxel}, M.~A. and {Jarvis}, M. and {MacCrann}, N. and {Friedrich}, O. and {Alves}, O. and {Leget}, P.-F. and {Chen}, R. and {Rollins}, R.~P. and {Wechsler}, R.~H. and {Gruendl}, R.~A. and {Cawthon}, R. and {Allam}, S. and {Bridle}, S.~L. and {Pandey}, S. and {Everett}, S. and {Shin}, T. and {Hartley}, W.~G. and {Fang}, X. and {Zhang}, Y. and {Aguena}, M. and {Annis}, J. and {Bacon}, D. and {Bertin}, E. and {Bocquet}, S. and {Brooks}, D. and {Carretero}, J. and {Castander}, F.~J. and {da Costa}, L.~N. and {Pereira}, M.~E.~S. and {De Vicente}, J. and {Desai}, S. and {Doel}, P. and {Ferrero}, I. and {Flaugher}, B. and {Frieman}, J. and {Garc{\'\i}a-Bellido}, J. and {Gaztanaga}, E. and {Gutierrez}, G. and {Hinton}, S.~R. and {Hollowood}, D.~L. and {Honscheid}, K. and {Huterer}, D. and {James}, D.~J. and {Kuehn}, K. and {Lahav}, O. and {Lee}, S. and {Marshall}, J.~L. and {Mena-Fern{\'a}ndez}, J. and {Miquel}, R. and {Muir}, J. and {Paterno}, M. and {Plazas Malag{\'o}n}, A.~A. and {Porredon}, A. and {Romer}, A.~K. and {Samuroff}, S. and {Sanchez}, E. and {Sanchez Cid}, D. and {Smith}, M. and {Soares-Santos}, M. and {Suchyta}, E. and {Swanson}, M.~E.~C. and {Tarle}, G. and {To}, C. and {Weaverdyck}, N. and {Weller}, J. and {Wiseman}, P. and {Yamamoto}, M.},
        title = "{Weak lensing combined with the kinetic Sunyaev-Zel'dovich effect: a study of baryonic feedback}",
      journal = {\mnras},
         year = 2024,
        month = oct,
       volume = {534},
       number = {1},
        pages = {655-682},
          doi = {10.1093/mnras/stae2100},
archivePrefix = {arXiv},
       eprint = {2404.06098},
 primaryClass = {astro-ph.CO},
       adsurl = {https://ui.adsabs.harvard.edu/abs/2024MNRAS.534..655B}
}

@ARTICLE{Pandey2025ConstraintsOn,
       author = {{Pandey}, S. and {Hill}, J.~C. and {Alarcon}, A. and {Alves}, O. and {Amon}, A. and {Anbajagane}, D. and {Andrade-Oliveira}, F. and {Battaglia}, N. and {Baxter}, E. and {Bechtol}, K. and {Becker}, M.~R. and {Bernstein}, G.~M. and {Blazek}, J. and {Bridle}, S.~L. and {Calabrese}, E. and {Camacho}, H. and {Campos}, A. and {Carnero Rosell}, A. and {Carrasco Kind}, M. and {Cawthon}, R. and {Chang}, C. and {Chen}, R. and {Chintalapati}, P. and {Choi}, A. and {Cordero}, J. and {Coulton}, W. and {Crocce}, M. and {Davis}, C. and {DeRose}, J. and {Devlin}, M. and {Diehl}, H.~T. and {Dodelson}, S. and {Doux}, C. and {Drlica-Wagner}, A. and {Eckert}, K. and {Eifler}, T.~F. and {Elvin-Poole}, J. and {Everett}, S. and {Fang}, X. and {Fert{\'e}}, A. and {Fosalba}, P. and {Friedrich}, O. and {Gatti}, M. and {Gaztanaga}, E. and {Giannini}, G. and {Gluscevic}, V. and {Gruen}, D. and {Gruendl}, R.~A. and {Ried Guachalla}, B. and {Harrison}, I. and {Hartley}, W.~G. and {Herner}, K. and {Huang}, H. and {Huff}, E.~M. and {Huterer}, D. and {Jain}, B. and {Jarvis}, M. and {Krause}, E. and {Kuropatkin}, N. and {Kusiak}, A. and {Leget}, P. and {Lemos}, P. and {Liddle}, A.~R. and {Lokken}, M. and {MacCrann}, N. and {McCullough}, J. and {Moodley}, K. and {Muir}, J. and {Myles}, J. and {Navarro-Alsina}, A. and {Omori}, Y. and {Park}, Y. and {Partridge}, B. and {Porredon}, A. and {Prat}, J. and {Raveri}, M. and {Refregier}, A. and {Rollins}, R.~P. and {Roodman}, A. and {Rosenfeld}, R. and {Ross}, A.~J. and {Rykoff}, E.~S. and {Samuroff}, S. and {Sanchez}, J. and {S{\'a}nchez}, C. and {Secco}, L.~F. and {Sevilla-Noarbe}, I. and {Shaikh}, S. and {Sheldon}, E. and {Shin}, T. and {Sif{\'o}n}, Crist{\'o}bal and {To}, C. and {Troja}, A. and {Troxel}, M.~A. and {Tutusaus}, I. and {Varga}, T.~N. and {Weaverdyck}, N. and {Wechsler}, R.~H. and {Wollack}, E.~J. and {Yanny}, B. and {Yin}, B. and {Zhang}, Y. and {Zuntz}, J. and {Allam}, S.~S. and {Bacon}, D. and {Bocquet}, S. and {Brooks}, D. and {Burke}, D.~L. and {Carretero}, J. and {Cawthon}, R. and {Costanzi}, M. and {da Costa}, L.~N. and {da Silva Pereira}, M.~E. and {Davis}, T.~M. and {Desai}, S. and {Frieman}, J. and {Garc{\'\i}a-Bellido}, J. and {Gutierrez}, G. and {Hinton}, S.~R. and {Hollowood}, D.~L. and {Honscheid}, K. and {James}, D.~J. and {Jeffrey}, N. and {Lee}, S. and {Marshall}, J.~L. and {Mena-Fern{\'a}ndez}, J. and {Miquel}, R. and {Mohr}, J.~J. and {Ogando}, R.~L.~C. and {Plazas Malag'on}, A.~A. and {Romer}, A.~K. and {Sanchez}, E. and {Santiago}, B. and {Smith}, M. and {Suchyta}, E. and {Swanson}, M.~E.~C. and {Thomas}, D. and {Vikram}, V. and {Walker}, A.~R. and {Weller}, J. and {Wiseman}, P.},
        title = "{Constraints on cosmology and baryonic feedback with joint analysis of Dark Energy Survey Year 3 lensing data and ACT DR6 thermal Sunyaev-Zel'dovich effect observations}",
      journal = {arXiv e-prints},
         year = 2025,
        month = jun,
          eid = {arXiv:2506.07432},
        pages = {arXiv:2506.07432},
          doi = {10.48550/arXiv.2506.07432},
archivePrefix = {arXiv},
       eprint = {2506.07432},
 primaryClass = {astro-ph.CO},
       adsurl = {https://ui.adsabs.harvard.edu/abs/2025arXiv250607432P}
}

@ARTICLE{Straight2026CMBConstraints,
       author = {{Straight}, Maria C. and {Karwal}, Tanvi and {Bernal}, Jos{\'e} Luis and {Boddy}, Kimberly K.},
        title = "{CMB constraints on dark matter-proton scattering: investigating prior-volume effects using profile likelihoods}",
      journal = {arXiv e-prints},
         year = 2026,
        month = mar,
          eid = {arXiv:2603.25731},
        pages = {arXiv:2603.25731},
          doi = {10.48550/arXiv.2603.25731},
archivePrefix = {arXiv},
       eprint = {2603.25731},
 primaryClass = {astro-ph.CO},
       adsurl = {https://ui.adsabs.harvard.edu/abs/2026arXiv260325731S}
}

@ARTICLE{Blum2021GravitationalLensing,
       author = {{Blum}, Kfir and {Teodori}, Luca},
        title = "{Gravitational lensing H$_{0}$ tension from ultralight axion galactic cores}",
      journal = {\prd},
         year = 2021,
        month = dec,
       volume = {104},
       number = {12},
          eid = {123011},
        pages = {123011},
          doi = {10.1103/PhysRevD.104.123011},
archivePrefix = {arXiv},
       eprint = {2105.10873},
 primaryClass = {astro-ph.CO},
       adsurl = {https://ui.adsabs.harvard.edu/abs/2021PhRvD.104l3011B}
}

@ARTICLE{Eisenstein1998BaryonicFeatures,
       author = {{Eisenstein}, Daniel J. and {Hu}, Wayne},
        title = "{Baryonic Features in the Matter Transfer Function}",
      journal = {\apj},
         year = 1998,
        month = mar,
       volume = {496},
       number = {2},
        pages = {605-614},
          doi = {10.1086/305424},
archivePrefix = {arXiv},
       eprint = {astro-ph/9709112},
 primaryClass = {astro-ph},
       adsurl = {https://ui.adsabs.harvard.edu/abs/1998ApJ...496..605E}
}

@article{Spiegelhalter2002DIC,
author = {Spiegelhalter, David J. and Best, Nicola G. and Carlin, Bradley P. and Van Der Linde, Angelika},
title = {Bayesian measures of model complexity and fit},
journal = {Journal of the Royal Statistical Society: Series B (Statistical Methodology)},
volume = {64},
number = {4},
pages = {583-639},
doi = {https://doi.org/10.1111/1467-9868.00353},
url = {https://rss.onlinelibrary.wiley.com/doi/abs/10.1111/1467-9868.00353},
eprint = {https://rss.onlinelibrary.wiley.com/doi/pdf/10.1111/1467-9868.00353},
year = {2002}
}

@ARTICLE{Winch2024Extreme,
       author = {{Winch}, Harrison and {Hlo{\v{z}}ek}, Ren{\'e}e and {Marsh}, David J.~E. and {Grin}, Daniel and {Rogers}, Keir K.},
        title = "{Extreme axions unveiled: A novel fluid approach for cosmological modeling}",
      journal = {\prd},
         year = 2024,
        month = aug,
       volume = {110},
       number = {4},
          eid = {043517},
        pages = {043517},
          doi = {10.1103/PhysRevD.110.043517},
archivePrefix = {arXiv},
       eprint = {2311.02052},
 primaryClass = {astro-ph.CO},
       adsurl = {https://ui.adsabs.harvard.edu/abs/2024PhRvD.110d3517W}
}

@ARTICLE{Gaughan2026Ultralight,
       author = {{Gaughan}, Lauren and {Green}, Anne M. and {Moss}, Adam},
        title = "{Ultra-light axion constraints from Planck and ACT: the role of nonlinear modelling}",
      journal = {arXiv e-prints},
         year = 2026,
        month = may,
          eid = {arXiv:2605.12054},
        pages = {arXiv:2605.12054},
archivePrefix = {arXiv},
       eprint = {2605.12054},
 primaryClass = {astro-ph.CO},
       adsurl = {https://ui.adsabs.harvard.edu/abs/2026arXiv260512054G}
}

\appendix

\section{Lensing of the primary cosmic microwave background anisotropies \label{app:lensed-cmb}}

The lensed temperature T angular power spectrum is obtained by the following transformation:
\begin{align}
    C_\ell^{TT} =\,&(1-\ell^2 R)C_\ell^{TT,\rm unl} \nonumber \\ &+ \int \frac{d^2\boldsymbol{\ell}^\prime}{(2\pi)^2} \left[\boldsymbol{\ell}^\prime \cdot (\boldsymbol{\ell}-\boldsymbol{\ell}^\prime)\right]^2 C^{\phi\phi}_{|\boldsymbol{\ell}-\boldsymbol{\ell}^\prime|} C^{TT,\rm unl}_{\ell^\prime},
\end{align}
where
\begin{align}
    R\equiv \frac{1}{4\pi}\int d\ell\,\ell^3 C_\ell^{\phi\phi},
\end{align}
and $\rm unl$ denotes the unlensed power spectrum. Lensing also affects polarization by rotating the E modes and B modes as shown in Ref.~\cite{Hu2000WeakLensing}:
\begin{align}
    C_\ell^{EE} =\,&(1-\ell^2R) C_\ell^{EE, \rm unl} \nonumber \\ &+ \int \frac{d^2\boldsymbol{\ell}^\prime}{(2\pi)^2} \left[\boldsymbol{\ell}^\prime \cdot (\boldsymbol{\ell}-\boldsymbol{\ell}^\prime)\right]^2 C^{\phi\phi}_{|\boldsymbol{\ell}-\boldsymbol{\ell}^\prime|} C_\ell^{EE, \rm unl}\left( 1-\cos 4\theta^\prime \right); \\
    C_\ell^{TE} =\,&(1-\ell^2R) C_\ell^{TE, \rm unl} \nonumber \\ &+ \int \frac{d^2\boldsymbol{\ell}^\prime}{(2\pi)^2} \left[\boldsymbol{\ell}^\prime \cdot (\boldsymbol{\ell}-\boldsymbol{\ell}^\prime)\right]^2 C^{\phi\phi}_{|\boldsymbol{\ell}-\boldsymbol{\ell}^\prime|}  C_\ell^{TE, \rm unl} \cos 2\theta^\prime,
\end{align}
where $\theta^\prime$ is the azimutal angle of $\boldsymbol{\ell}^\prime$.

\section{Ultralight axion halo model}\label{sec:halo-model}

ULAs do not cluster in the same way as CDM. It was found in Ref.~\cite{Lague2024CosmologicalSimulations} that, in mixed DM cosmologies, halos comprising only CDM can form even if their radius is smaller than the ULA wavelength. In our halo model, we account for this using the ULA halo fraction. The axion perturbation
\begin{align}
    \delta_\mathrm{a} = F_{\rm h}\delta_\mathrm{a}^{\rm h} + (1-F_{\rm h}) \delta_\mathrm{a}^{\rm L},
\end{align}
where $F_{\rm h}$ is the ULA halo fraction (the proportion of axion perturbations that clusters into bound structures), $\delta_\mathrm{a}^{\rm h}$ denotes the axion density fluctuations in halos and $\delta_\mathrm{a}^{\rm L}$ denotes linearly-evolving axion fluctuations outside of halos. The total nonlinear matter power spectrum is calculated using the following prescription:
\begin{align}
    P_\mathrm{m}(k) = \left(\frac{\Omega_\mathrm{a}}{\Omega_\mathrm{m}} \right)^2 &P_{\rm a, a}(k) + \frac{2\Omega_\mathrm{a}(1-\Omega_\mathrm{a})}{\Omega_\mathrm{m}^2} P_{\rm a, cb}(k) \nonumber \\  &+ \left(1-\frac{\Omega_\mathrm{a}}{\Omega_\mathrm{m}} \right)^2 P_{\rm cb, cb}(k), \label{eq:main-nl-pk}
\end{align}
where the $\rm a, cb$ denote respectively the axion and CDM + baryons components and all components are nonlinear. The cold and baryon power spectrum is calculated using a modified \texttt{HMcode} procedure while the axion auto- and cross-spectra are adjusted to account for changes in the halo mass function, density profiles, and clustering of axions into halos.

\begin{figure}
    \centering
    \includegraphics[width=\linewidth]{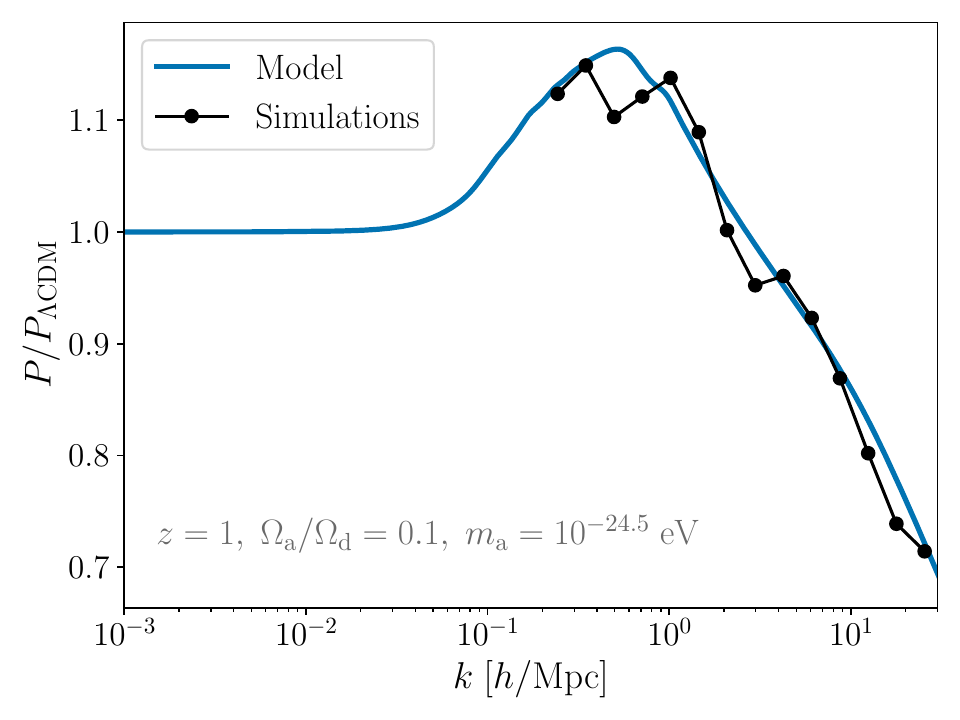}
    \caption{Comparison between the matter power spectrum ratio to $\Lambda$CDM between the nonlinear simulations of Ref.~\cite{Dome2025ImprovedHalo} for $m_\mathrm{a}=10^{-24.5}$ eV. We find good agreement between the excess seen in the simulations and the halo model approach. (The data were digitized from the original figure.)}
    \label{fig:dome_compare}
\end{figure}

%We also confirm that our model captures the correct enhancement due to nonlinear structure growth on semi-linear scales. Since this is the main feature driving the preference for $\Omega_\mathrm{a}>0$ in the CMB lensing data. 
We compare the ratio of the nonlinear matter power spectrum with respect to $\Lambda$CDM measured in the simulations of Ref.~\cite{Dome2025ImprovedHalo}. The result is shown in Fig.~\ref{fig:dome_compare}

\begin{figure}
    \centering
    \includegraphics[width=\linewidth]{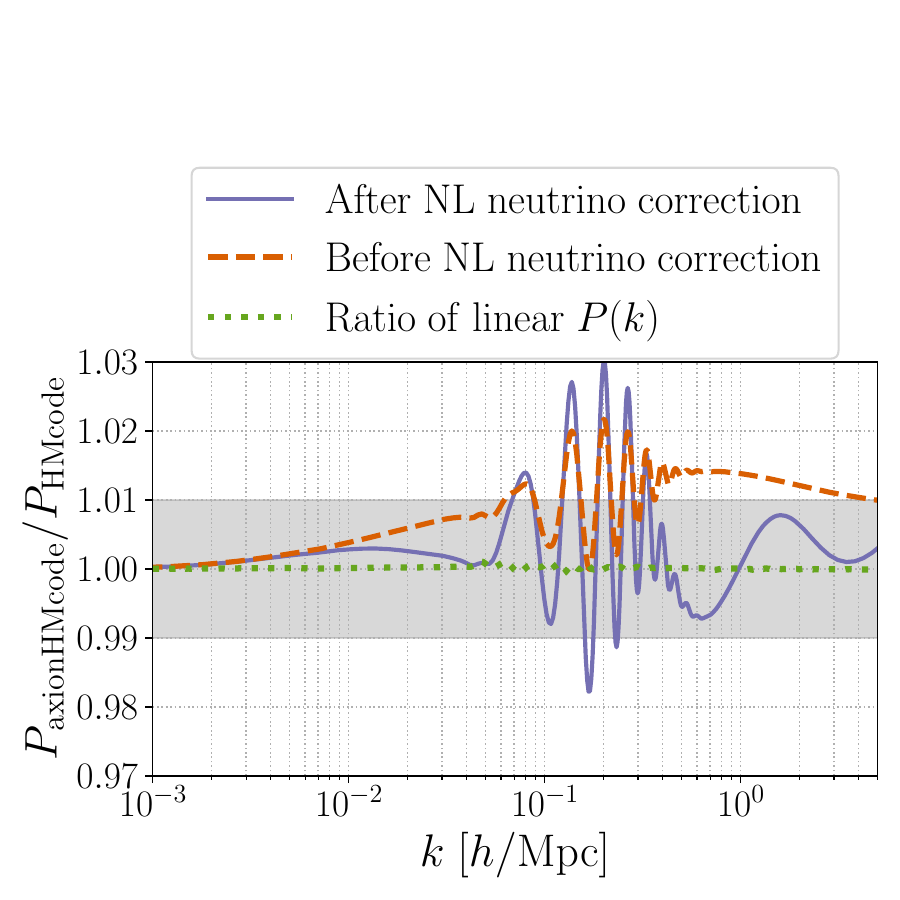}
    \caption{(Purple and orange lines) Nonlinear matter power spectrum ratios of \texttt{axionHMcode} to \texttt{HMCode}. The inclusion of the nonlinear neutrino correction (orange to purple) ensures that the power spectrum does not have spuriously enhanced power on small scales. The oscillatory features are due to different treatments of the BAO scales between the two codes. (Green line) Linear matter power spectrum ratio of \texttt{AxiECAMB} to \texttt{CAMB}, which is negligible. The axion mass used is $10^{-25}$ eV at redshift $z=1$ with $\Omega_\mathrm{a}=10^{-6}$.}
    \label{fig:neutrino-corr}
\end{figure}

A subtle distinction between our nonlinear model and \texttt{HMcode} is that we use the full matter power spectrum when calculating the variance of the matter power spectrum to obtain the halo mass function in the halo model. This approach is in contrast with \texttt{HMcode} which uses the matter power spectrum without its baryon acoustic oscillations (often referred to as the ``no-wiggle" power spectrum) to obtain these quantities. The latter is more numerically stable as it does not involve the oscillatory part of the spectrum. However, ULAs introduce scale-dependent features which can overlap with these oscillations. In this case, the traditional methods to derive the smooth part of the matter power spectrum (such as the Eisenstein-Hu fitting function~\cite{Eisenstein1998BaryonicFeatures}) no longer work as intended and introduce spurious features in the smoothed spectrum. This choice of numerical implementation leads to a percent-level increase in the nonlinear matter power spectrum on scales around $k=0.1\;h/$Mpc. Our convergence tests involve comparison with $\Lambda$CDM emulators which are based on the Mead-2020 nonlinear prescription. To ensure consistency with these emulators, we rescale the lensing power spectrum $C_L^{\phi\phi}$ by $1.25\%$. This value is determined by the amplitude difference between the lensing potential power spectrum from the $\Lambda$CDM emulator of the \texttt{cosmopower-organization}\footnote{\url{https://github.com/cosmopower-organization}.} in the $\Omega_\mathrm{a}/\Omega_\mathrm{d}\to 0$ limit at the reference multipole of $L=100$. This constant rescaling is not degenerate with scale-dependent axion effects and makes our constraints slightly more conservative.

The model does not explore the nonlinear interactions between massive neutrinos and ULAs. It computes the nonlinear terms following \texttt{HMcode} (Mead2020 version) in the limit of $\Omega_\nu\to 0$. However, the fiducial cosmology for \texttt{cosmopower} to which we compare assumes one massive neutrino species and a sum of neutrino masses $\Sigma m_\nu =0.06$ eV with a total number of effective degrees of freedom $N_{\rm eff}=3.044$~\cite{Bolliet2024HighAccuracy}. The input linear calculations from \texttt{AxiECAMB} include the contribution of massive neutrinos while the halo model calculations from \texttt{AxionHMcode} do not. %The version of \texttt{HMcode} is also based on the Mead2016 version with $c_{\rm min}=3.13$ and $\eta_0=0.603$. %To account for these differences consistently with our model, we apply a nonlinear correction denoted $P_m(k) \to \mathcal{C}_{\rm NL}(k) P_m(k)$ defined by
%\begin{align}
     %\mathcal{C}_{\rm NL}(k) &\equiv \frac{P^{\rm cosmopower}_m(k)}{ P_m(k;f_{\rm ax}=0)}. \label{eq:nl-correction}
%\end{align}
To account for these differences, we define a multiplicative nonlinear neutrino correction
\begin{align}
    \mathcal{C}^\nu_{\rm NL} &\equiv \left[\frac{P^{\rm }_m( \Omega_\mathrm{a}=0)}{ P_m(\Omega_{\nu}=\Omega_\mathrm{a}=0)}\right]\left[\frac{P^{\rm L}_m(\Omega_{\nu}=\Omega_\mathrm{a}=0)}{ P_m^{\rm L}(\Omega_\mathrm{a}=0)}\right], \label{eq:nl-correction}
\end{align}
%(pk_nonlin[0] / pk_nonlin_nonu[0]) / (pk[0] / pk_nonu[0])
where the power spectra with neutrinos (without the explicit $\Omega_\nu=0$) are set assuming $\Sigma m_\nu =0.06$ eV and the superscript ``$\rm L$" denotes linear spectra. In order to reflect the differences in nonlinear implementations between \texttt{cosmopower} and \texttt{axionHMcode}, we generate the nonlinear spectrum without neutrinos using the 2020 version of \texttt{HMcode}. This choice is justified by the fact that \texttt{axionHMcode} reproduces the nonlinear spectra of \texttt{HMcode} in the limit $(f_{\rm ax}, f_{\nu})\to (0, 0)$.
The correction in Eq.~(\ref{eq:nl-correction}) thus rescales the matter power spectrum to account for the assumptions of the nonlinear calculations through the change $P_\mathrm{m}(k) \to \mathcal{C}^\nu_{\rm NL}(k) P_\mathrm{m}(k)$. The residual effect on the power spectrum of using the full (non-smoothed) power spectrum and the nonlinear neutrino correction is in Fig.~\ref{fig:neutrino-corr}. This change is minor and amounts to a variation of less than one percent in the CMB lensing potential power spectrum (we describe this observable in greater detail in \S\ref{sec:cmb}). 

\section{Lensing data best fit analysis}\label{sec:lensing-best}
\begin{figure}[!htb]
    \centering
    \includegraphics[width=\linewidth]{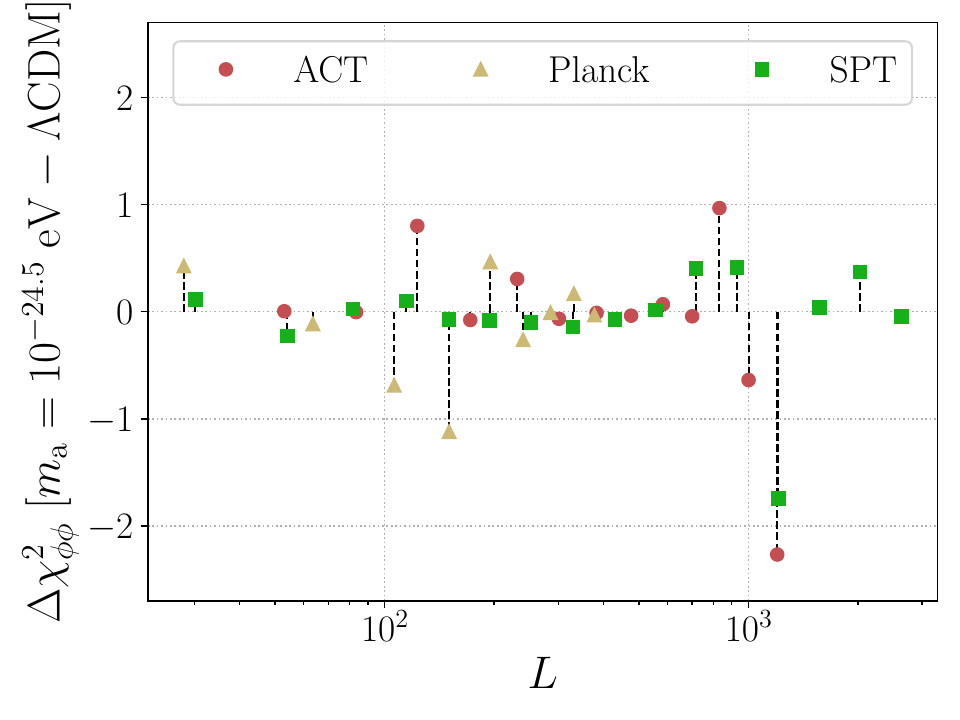}
    \caption{$\chi^2$ contribution for individual data points in the CMB lensing likelihood. The reduction in the total $\chi^2$ from the axion model at $10^{-24.5}$ eV is largely driven by ACT and SPT data points near $L=1200$.}
    \label{fig:individual-lensing}
\end{figure}
We study the improvement in fit from individual data points in the lensing likelihood. We compare the best-fit models for $\Lambda$CDM and the axion model at fixed mass $m_\mathrm{a}=10^{-24.5}$ eV and examine the change in $\chi^2$ for individual data points. The results of this analysis are shown in Fig.~\ref{fig:individual-lensing}. We attribute the slight preference for $\Omega_\mathrm{a}>0$ to the two data points around the multipole $L\approx 1200$. These points correspond to an enhancement in lensing power compared to the best-fit \(\Lambda\)CDM model, as shown in Fig.~\ref{fig:lensing-data}. The increases are present in both the ACT and SPT lensing data and are largely uncorrelated, with a correlation coefficient $\rho=0.0037$.

\section{Full parameter constraints}\label{sec:full-results}

\begin{figure*}
    \centering
    \includegraphics[width=\linewidth]{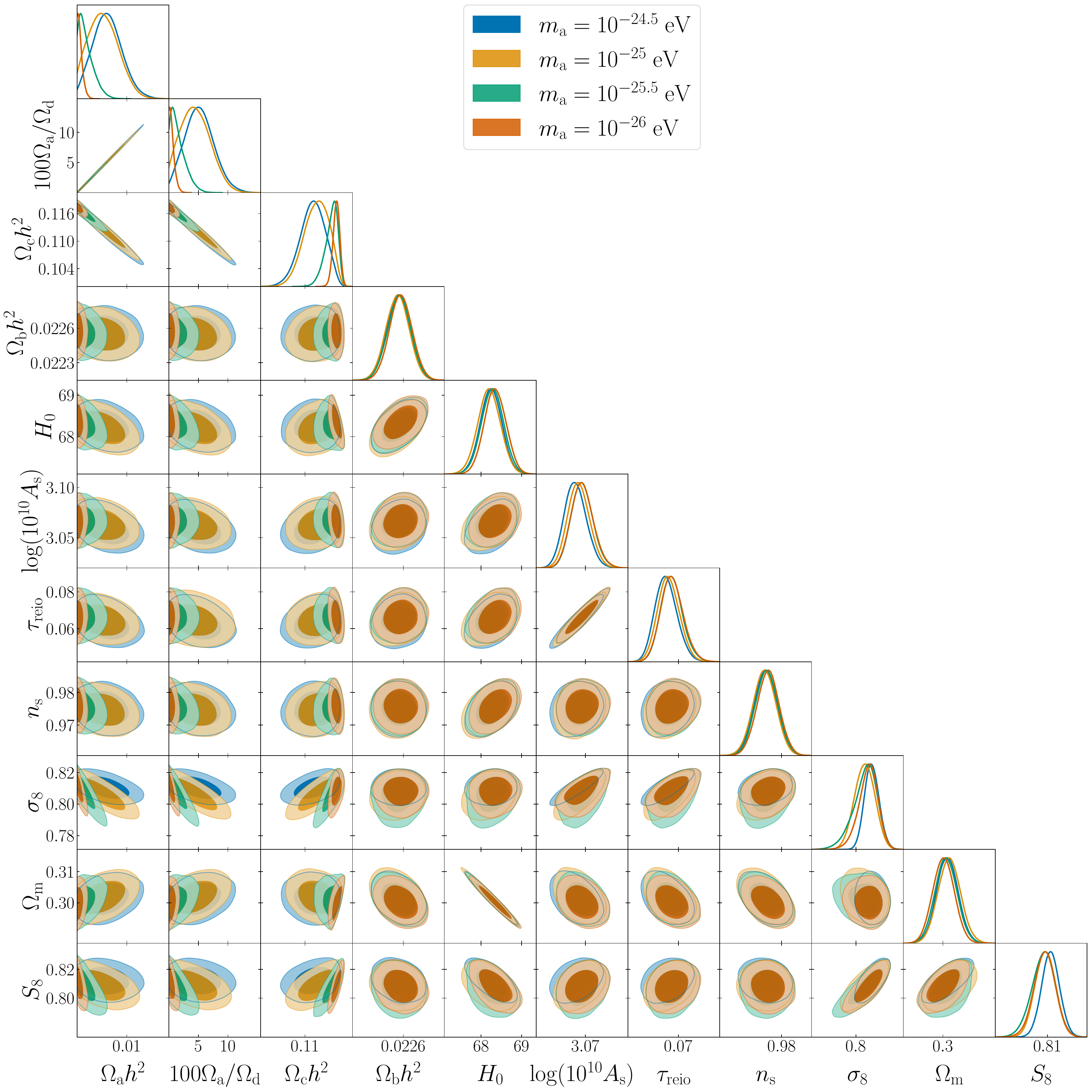}
    \caption{As Fig.~\ref{fig:four-masses-corner-main-text}, but for all the axion and cosmological parameters that we vary.}
    \label{fig:four-masses-corner}
\end{figure*}

Here, we show the full set of 1D and 2D marginalized posterior distributions from the analysis in Sec.~\ref{sec:param_inference} (Fig.~\ref{fig:four-masses-corner}).

\end{document}